\documentclass[11pt]{article}
\pdfoutput=1 

\usepackage{jheppub_qcdNf} 
\def\@preprint{MOO}

\usepackage[T1]{fontenc} 

\usepackage{graphicx}
\usepackage{bm}
\usepackage{amsfonts}
\usepackage{booktabs}

\usepackage{xcolor}

\usepackage{dsfont}
\def\openone{\mathds{1}}

\def\b{{\bm b}}

\def\p{{\bm p}}

\def\B{{\bm B}}
\def\C{{\bm C}}
\def\P{{\bm P}}

\def\Ybar{\overline{Y}}
\def\eps{\epsilon}

\def\Nc{N_{\rm c}}
\def\Nf{N_{\rm f}}
\def\CA{C_{\rm A}}
\def\CF{C_{\rm F}}
\def\tF{t_{\rm F}}
\def\TF{T_{\rm F}}
\def\dF{d_{\rm F}}
\def\dA{d_{\rm A}}
\def\mD{m_{\rm D}}
\def\alphas{\alpha_{\rm s}}
\def\alphaqed{\alpha_{\scriptscriptstyle\rm EM}}
\def\Re{\operatorname{Re}}

\def\tr{\operatorname{tr}}
\def\sgn{\operatorname{sgn}}

\def\grad{{\bm\nabla}}

\def\yfrak{{\mathfrak y}}
\def\yfrakQ{\yfrak_\ssQ}
\def\yfrakQbar{\yfrak_\ssQbar}

\def\seq{{\rm seq}}
\def\new{{\rm new}}

\def\calX{{\cal X}}

\def\gammaE{\gamma_{\rm\scriptscriptstyle E}}

\def\ix{{\rm i}}
\def\fx{{\rm f}}

\def\xbx{{\bar{\rm x}}}
\def\yx{{\rm y}}

\def\bbI{\mathbb{I}}

\def\qhatA{\hat q_{\rm A}}
\def\qhatF{\hat q_{\rm F}}

\def\real{{\rm real}}
\def\virt{{\rm virt}}

\def\qhat{\hat q}
\def\lstop{\ell_{\rm stop}}
\def\lstoprho{\lstop^{(\rho)}}
\def\LO{{\rm LO}}
\def\NLO{{\rm NLO}}

\def\xg{x_g}
\def\xq{x_q}
\def\Q{{\rm Q}}
\def\sQ{{\scriptstyle{\Q}}}
\def\ssQ{{\scriptscriptstyle{\Q}}}
\def\Qbar{{\bar\Q}}
\def\sQbar{{\scriptstyle{\Qbar}}}
\def\ssQbar{{\scriptscriptstyle{\Qbar}}}
\def\xQ{x_\ssQ}
\def\yQ{y_\ssQ}
\def\xQbar{x_\ssQbar}
\def\zQbar{z_\ssQbar}

\def\xe{x_e}
\def\E{{\rm E}}
\def\Ebar{{\bar\E}}
\def\ssE{{\scriptscriptstyle{\E}}}
\def\ssEbar{{\scriptscriptstyle{\Ebar}}}
\def\xE{x_\ssE}
\def\xEbar{x_\ssEbar}
\def\ye{y_e}

\def\xIR{x_{\scriptscriptstyle\rm IR}}

\def\net{{\rm net}}
\def\uqq{{\underline{q\to q}}}
\def\qq{{q\to q}}

\def\uqQ{{\underline{q\to\Q}}}
\def\uqQbar{{\underline{q\to\Qbar}}}
\def\uqg{{\underline{q\to g}}}
\def\ugQ{{\underline{g\to\Q}}}
\def\ugQbar{{\underline{g\to\Qbar}}}
\def\uij{{\underline{i\to j}}}
\def\uee{{\underline{e\to e}}}

\def\fac{{\rm fac}}
\def\form{{\rm form}}
\def\Avg{\operatorname{Avg}}

\def\seqNf{{\seq}}

\def\Ratio{\operatorname{Ratio}}

\def\IR{{\rm IR}}

\newcounter{savefootnote}
\newcommand{\astfootnote}[1]{%
  \setcounter{savefootnote}{\value{footnote}}%
  \setcounter{footnote}{0}%
  \let\oldthefootnote=\thefootnote
  \renewcommand{\thefootnote}{\fnsymbol{footnote}}%
  \footnote{#1}%
  \let\thefootnote=\oldthefootnote%
  \setcounter{footnote}{\value{savefootnote}}%
}

\begin {document}



\title{Are in-medium quark-gluon showers strongly coupled?
       Results in the large-$\Nf$ limit}

\author[a]{Peter Arnold,}
\author[b,c,a]{Omar Elgedawy,}
\author[d,e]{Shahin Iqbal}


\affiliation[a]{Department of Physics, University of Virginia,
  P.O.\ Box 400714, 
  Charlottesville, VA 22904, U.S.A.}
\affiliation[b]{
  Key Laboratory of Atomic and Subatomic Structure and Quantum Control (MOE),
  Guangdong Basic Research Center of Excellence for Structure and Fundamental
    Interactions of Matter,
  Institute of Quantum Matter,
  South China Normal University, Guangzhou 510006, China}
\affiliation[c]{
  Guangdong-Hong Kong Joint Laboratory of Quantum Matter,
  Guangdong Provincial Key Laboratory of Nuclear Science,
  Southern Nuclear Science Computing Center,
  South China Normal University, Guangzhou 510006, China}
\affiliation[d]{National Centre for Physics,
  Quaid-i-Azam University Campus,
  Islamabad, 45320 Pakistan}
\affiliation[e]{Theoretical Physics Department,
  CERN,
  CH-1211 Geneva 23, Switzerland}

\emailAdd{parnold@virginia.edu}
\emailAdd{oae2ft@virginia.edu}
\emailAdd{smi6nd@virginia.edu}

\begin {abstract}%
{%
  Inside a medium, showers originating from a very high energy particle
develop via medium-induced
splitting processes such as bremsstrahlung and pair production.
During shower development, two consecutive splittings
sometimes overlap quantum mechanically, so that
they cannot be treated independently.
Some of these effects can be absorbed into an effective value of a medium
parameter known as $\qhat$.
Previous calculations (with certain simplifying assumptions)
have found that, after adjusting the value of $\qhat$, the
leftover effect of overlapping splittings is quite
small for purely gluonic large-$\Nc$ showers but is very much larger
for large-$\Nf$ QED showers, at comparable values of $N\alpha$.
Here, by investigating the same problem for QCD \textit{with quarks} in the
large-$\Nf$ limit of many quark flavors,
we make a first study of whether the small effect in purely gluonic showers
(i) was merely an accident or (ii) is more broadly characteristic of
such overlap effects in QCD.
We also offer a qualitative explanation for the large
size of the effect in QED vs.\ QCD.
}%
\end {abstract}

\maketitle
\thispagestyle {empty}

\newpage


\section{Introduction and Preview of Results}
\label{sec:intro}

\subsection {Introduction}

When passing through matter, high energy particles lose energy by
showering, via the splitting processes of hard bremsstrahlung and pair
production.  At very high energy, the quantum mechanical duration of
each splitting process, known as the formation time, exceeds the mean
free time for collisions with the medium, leading to a significant
reduction in the splitting rate known as the Landau-Pomeranchuk-Migdal
(LPM) effect.
The LPM effect was originally worked out for QED in the 1950's
\cite{LP1,LP2,Migdal}%
\footnote{
  The papers of Landau and Pomeranchuk \cite{LP1,LP2} are also available in
  English translation \cite{LPenglish}.
}
and then later generalized to QCD in the 1990s by
Baier, Dokshitzer, Mueller, Peigne, and Schiff \cite{BDMPS1,BDMPS2,BDMPS3}
and by Zakharov \cite{Zakharov1,Zakharov2} (BDMPS-Z).

Modeling of the development of
high-energy in-medium showers typically treats each splitting
as an independent dice roll, with probabilities set by calculations
of single-splitting rates that take into account the LPM effect.
The question then arises whether consecutive splittings in a shower
can really be treated as probabilistically independent, or whether
there is any significant chance that the formation times of
splittings could overlap so that there are significant quantum interference
effects entangling one splitting with the next.
A number of years ago, several authors showed \cite{Blaizot,Iancu,Wu},
in a leading-log calculation,
that the effects of overlapping formation times in QCD showers
could become large when one of the two overlapping splittings
is parametrically softer than the other.
They also showed that those large leading logarithms could be absorbed into
a redefinition of the medium parameter $\qhat$, which parametrizes
the effectiveness with which the medium deflects high-energy particles.
A refined question arose: How large are overlapping formation
time effects that {\it cannot}\/ be absorbed into a redefinition of
$\qhat$\,?

We previously investigated this question in
refs.\ \cite{finale,finale2}
using the following thought experiment.
We focused on purely gluonic showers
(i.e.\ pure Yang-Mills theory)
in the large-$\Nc$ limit.
Consider such a shower initiated by a high-energy gluon
moving in the $z$ direction, starting at $z{=}0$.  Imagine for simplicity
that the medium is static, homogeneous, and arbitrarily large.
Let $\eps(z)$ be the distribution in $z$ of the energy that the
shower deposits
in the medium, statistically averaged over many such showers.
Define the energy stopping length $\lstop$
to be the first moment of that distribution,
$\lstop \equiv \langle z \rangle_\eps
 \equiv E_0^{-1} \int dz \> z \, \eps(z)$,
where $E_0$ is the energy of the initial gluon.
Let $\sigma$ be the width of the distribution $\eps(z)$.
Ignoring overlap effects, both $\lstop$ and $\sigma$ scale with
$\qhat$, coupling constant, and the energy $E_0$ of the initial electron as
\begin {equation}
  \sigma \sim \lstop \sim \frac{1}{\alphas} \sqrt{ \frac{E_0}{\qhat} } .
\label {eq:scale}
\end {equation}
Naively, the value of $\qhat$ then cancels in the ratio $\sigma/\lstop$.
Any effect that can be absorbed into $\qhat$ would not affect the
value of $\sigma/\lstop$, and so that ratio could be used to
test how large are overlapping formation time effects that cannot be
absorbed into $\qhat$.  (Refs.\ \cite{finale,finale2} explain how the
situation is a little more subtle than that, but we will be able to mostly
ignore those subtleties in this paper.)

To leading order in $\alphas$, refs.\ \cite{finale,finale2} found that the
\textit{relative} size of overlap effects were extremely small,
of order $O(1\%)\times\Nc\alphas$.
That was a surprising result because earlier work \cite{qedNfstop} had
made a similar calculation for large-$\Nf$ QED and found that the
corresponding overlap effects were of order $O(100\%)\times\Nf\alpha$.
(We'll present more precise numbers shortly.)

Those results raise the question of \textit{why} the QED and QCD results are
so very different for comparable values of $N\alpha$!
One possibility is that the small QCD result might
merely arise from an accidental cancellation peculiar to the special case of
purely gluonic large-$\Nc$ showers.  In this paper, we check the robustness
of the small QCD result by adding quarks to our QCD showers.
For a first look, we can tremendously simplify the calculation by
adding many flavors of quark.  We will take the large-$\Nf$ limit
(where $\Nf$ is the number of quark flavors) in addition to the
large-$\Nc$ limit.  Specifically, we study the
case $\Nf \gg \Nc \gg 1$.  The advantage of this limit is that we
will be able to adapt existing formulas for
large-$\Nf$ QED overlap rates \cite{qedNf}, and
their effect on QED showers \cite{qedNfstop,qedNfenergy},
to large-$\Nf$ QCD.

We will also present a qualitative explanation of
why the size of overlap effects is so different for QCD vs.\ QED at comparable
values of $N\alpha$.
Ultimately, the reason traces back to the fact that
colored high-energy gluons interact much more directly with a QCD medium
than high-energy neutral photons interact with a QED medium,
and that consequently pair production ($\gamma \to e\bar e)$ can have
a large qualitative effect in a
QED shower when it overlaps with the earlier LPM-suppressed
splitting ($e \to e\gamma$) that created the photon in the first place.


\subsection {Preview of Results}
\label{sec:preview}

Table \ref{tab:chi2} previews results for the overlap correction to
$\sigma/\lstop$ in various situations, contrasting large-$N$ QCD results
with large-$N$ QED results.
The first two lines show results for
the corrections to $\sigma/\lstop$ for the energy deposition distribution
$\eps(z)$ for showers initiated by either a gauge boson
(gluon vs.\ photon) or fermion (quark vs.\ electron).

The third row of the table
shows similar results for the deposition of the flavor
of the initial fermion in a fermion-initiated shower.
For both large-$\Nf$ QCD and large-$\Nf$ QED, one may follow
the progress of the initial fermion through a fermion-initiated
shower, and the
``initial flavor'' distribution is the distribution of where the
initial fermion stops and thermalizes with the medium.%
\footnote{
  Without a large-$\Nf$ approximation (more specifically $\Nf \gg \Nc$ in the
  case of QCD), identifying
  which fermion is the heir to the original parent fermion
  becomes ambiguous due to interferences in overlapping splittings.
  See the discussion of figs.\ 11a and b in ref. \cite{qedNfstop}.
  Be warned that the brief discussion of ``large-$\Nc$'' QCD in
  ref.\ \cite{qedNfstop} implicitly
  assumes that $\Nf \ll \Nc$ and does not address the
  case $\Nf \gtrsim \Nc \gg 1$.
}
For large-$\Nf$ QED, the distribution of initial-flavor deposition is the
same as the distribution of charge deposition, for slightly
non-trivial reasons that we review later.
For large-$\Nf$ QCD, it is not.

\begin {table}[tp]

\setlength{\tabcolsep}{7pt}
\begin {center}
\begin{tabular}{cccrr}
\hline
\hline
  && \multicolumn{3}{c}{overlap correction to $\sigma/\lstop$}
\\
\cline{3-5}
  deposition
  & initiating
  & \multicolumn{1}{c}{$\Nf{=}0$, large-$\Nc$}
  & \multicolumn{1}{c}{$\Nf \gg \Nc \gg 1$}
  & \multicolumn{1}{c}{large-$\Nf$}
\\[-2pt]
  distribution
  & particle
  & \multicolumn{1}{c}{QCD}
  & \multicolumn{1}{c}{QCD}
  & \multicolumn{1}{c}{QED}
\\[2pt]
\hline
  energy & $g/\gamma$
         & $1.0\%\times\Nc\alphas$
         & $-0.4\%\times\Nf\alphas$
         & $99\%\times\Nf\alpha$
\\
  energy & $q/e$
         &
         & $-0.5\%\times\Nf\alphas$
         & $113\%\times\Nf\alpha$
\\
  initial flavor & $q/e$
         &
         & $1.1\%\times\Nf\alphas$
         & $-85\%\times\Nf\alpha$
\\
\hline
  fermion number & $q/e$
         &
         & \multicolumn{1}{c}{IR-unsafe}
         & $-85\%\times\Nf\alpha$
\\
\hline
\hline
\end{tabular}
\end {center}
\caption{%
\label{tab:chi2}%
  The relative size of overlap
  corrections to the ratio $\sigma/\lstop$ of width to
  stopping distance.
  Different columns compare (i)
  purely gluonic ($\Nf{=}0$) large-$\Nc$ QCD showers \cite{finale,finale2}
  to (ii) large-$\Nf$ ($\Nf{\gg}\Nc{\gg}1$) QCD showers (this paper) to
  (iii) large-$\Nf$ QED showers \cite{qedNfenergy}.
  Different rows show results for the deposition distributions of different
  types of conserved quantities and for showers initiated by different
  types of particles (gauge bosons vs.\ fermions).
  Results in this table are given for
  the renormalization scale choice $\mu \propto (\qhat E)^{1/4}$.
  The purely gluonic ($\Nf{=}0$)
  result also requires a choice of infrared factorization
  energy scale $\Lambda_\fac$, which here is $\Lambda_\fac = E/4$.
  (The scale choices in that case are slightly different from what
  refs.\ \cite{finale,finale2} highlighted as their ``preferred''
  choices.  The result quoted here instead matches that for
  $\sigma_S$ in table 4 of ref.\ \cite{finale2})
}
\end{table}

The comparisons in table \ref{tab:chi2} (ignoring the very last line of
the table) make clear that the earlier small result
for $\qhat$-insensitive overlap effects of a purely gluonic shower is
not a peculiar accident of purely gluonic showers.
Adding many flavors of quarks does not qualitatively change the small size of
the QCD overlap effects, at least not in the
large-$\Nf$ limit we have studied.
Later in this paper, we will check that overlap corrections to
$\qhat$-insensitive ratios involving higher moments of the deposition
are also small.

The last row of table \ref{tab:chi2} is presented as a warning that
not every deposition distribution of a conserved quantity
is infrared (IR) safe.  Infrared behavior becomes an
issue for the deposition of fermion number (or electric charge)
from a quark-initiated shower.
As discussed later, this distribution is affected by
very low energy quark-antiquark pairs
produced and asymmetrically
deposited very early in the evolution of the shower.

For the purpose of this introduction,
we have not displayed more precision for the
entries in table \ref{tab:chi2} because the
exact values are sensitive to choices one makes for the renormalization
scale and, in the gluonic case ($\Nf{=}0$), also to an infrared
factorization scale that must be introduced \cite{finale,finale2}.  The
dependence of our results on these scale choices is small and does not affect
the qualitative conclusion that overlap corrections which cannot
be absorbed into $\qhat$ are very small in QCD, at least for the QCD limits
and various infrared-safe quantities that we have examined.

Though our numerical results indicate that the tiny sizes of
$\qhat$-insensitive overlap effects for QCD were not merely an
accidental cancellation peculiar to the case of
purely gluonic ($\Nf{=}0$) showers, the numbers by themselves do
not explain \textit{why} the effect is much larger in QED than QCD
for comparable values of $N\alpha$.
Our later qualitative explanation will lead to a
physically-motivated analytic approximation for the QED result that
explains most of the difference.
The steps of that explanation will rely on rough, qualitative
arguments whose credibility will be enhanced by
their agreement with features
of our numerical calculations.  For that reason,
we have chosen to organize this paper to first present
all the details behind our full, quantitative calculations,
and we will wait until the very end to present our qualitative
explanation of the difference between QED and QCD.
But readers interested mainly in qualitative takeaways may wish to skip
directly to section \ref{sec:why}.


\subsection {Outline}

In the next section, we first lay out our assumptions and approximations.
We then review the relevant BDMPS-Z in-medium splitting rates
for independent splittings and then discuss the rates needed to
account for overlapping formation time effects to first order in
$\Nf\alphas$.  In section \ref{sec:convert}, we explain how to
get formulas for those overlap effect rates in
$\Nf{\gg}\Nc{\gg}1$ QCD by adapting previous calculations of such
rates in large-$\Nf$ QED.
(The final QCD formulas are summarized in
appendix \ref{app:summary}.)
Following earlier works,
section \ref{sec:netrates} selectively integrates and repackages
all of the relevant rates into net rates for a splitting or pair of
overlapping splittings to produce one daughter of energy $xE$ from a
parent of energy $E$.  Numerical results are given for those net rates
$[d\Gamma/dx]^\net$, which are then fit to relatively simple analytic forms
for subsequent use in calculations of in-medium
shower development.
Section \ref{sec:moments}
reviews the formalism for calculating the overlap corrections
to $\sigma/\lstop$ presented in table \ref{tab:chi2}.
The formalism also provides results for similar ratios involving
higher moments of energy or initial-flavor deposition distributions.
Section \ref{sec:IRunsafe} explains why table \ref{tab:chi2}
lists the size of
overlap corrections to $\sigma/\lstop$ as ``IR-unsafe''
for the case of large-$\Nf$ QCD fermion number
deposition.

Table \ref{tab:chi2} shows that the size of overlap effects that cannot
be absorbed into $\qhat$ are very small for QCD
and so can be safely ignored.  But that will not be so useful if 
the amount that must be absorbed varies between different types of
measurement.  Section \ref{sec:universality} offers a modest
first test of whether
the size of ``corrections that must be absorbed into $\qhat$\,'' are
the same for different types of measurements.%
\footnote{
  The universality of \textit{soft} radiative corrections to $\qhat$ has already
  been established at leading-log order (LLO) by refs.\
  \cite{LMW,Blaizot,Iancu,Wu} and,
  for large $\Nc$, at NLLO by ref.\ \cite{logs2}.  However, ratios
  like $\sigma/\lstop$ also ignore significant \textit{hard} corrections that
  affect the numerator and denominator the same way, and it is also only
  in taking the ratio that one reduces extreme sensitivity to the
  choice of factorization scale for absorbing the soft radiative corrections
  into $\qhat$ \cite{finale2}.  We do not know \textit{a priori} if these
  canceling hard corrections can be absorbed into $\qhat$ in a
  process-independent way, which is why we need to check.
}

As mentioned previously, section \ref{sec:why} offers a qualitative
explanation of why the size of overlap effects presented in
table \ref{tab:chi2} are so much larger for QED than for QCD
for comparable values of $N\alpha$.
Section \ref{sec:conclusion} offers a brief conclusion.


\section{Building blocks: splitting rates}

\begin {figure}[t]
\begin {center}
  \includegraphics[scale=0.6]{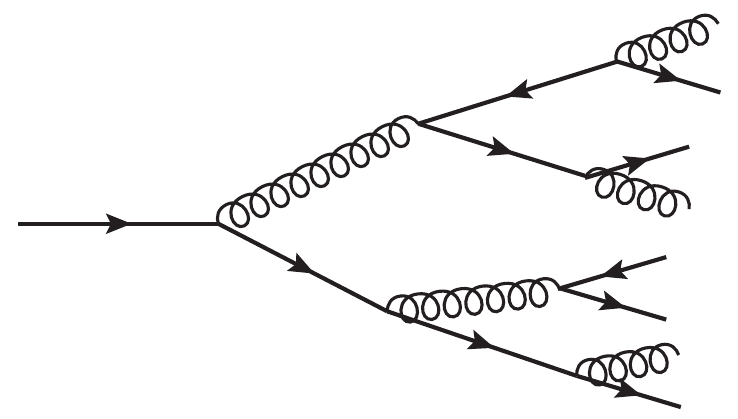}
  \caption{
     \label{fig:shower0}
     Depiction of an (in-medium) high-energy shower in large-$\Nf$ QCD.
     The interactions with the QCD medium are not explicitly shown.
     The transverse spread of the shower will be very small but has been
     exaggerated here for the sake of the drawing.
  }
\end {center}
\end {figure}

In the large-$\Nf$ limit (specifically $\Nf \gg \Nc$), QCD showers
are made up of $q{\to}qg$ and $g{\to}q\bar q$ splitting processes,
as depicted in fig.\ \ref{fig:shower0}.  The effect of
purely-gluonic splitting $g{\to}gg$ is suppressed.
In this limit, the types of relevant splitting processes are
similar to those for QED showers ($e{\to}e\gamma$ and $\gamma\to e\bar e$),
which will allow us to adapt previous large-$\Nf$ QED calculations
of overlapping splitting rates \cite{qedNf}.

\subsection{Assumptions}

In this paper, we make the same simplifying assumptions previously made
for similar analysis of purely gluonic ($\Nf{=}0$) showers in
refs.\ \cite{finale,finale2} and for large-$\Nf$ QED showers in
refs.\ \cite{qedNfstop,qedNfenergy}.
In brief summary,
we assume a static, homogeneous medium that is large enough to
contain (i) formation times in the case of splitting rate calculations
and (ii) the entire development of the shower for calculation of
overlap effects on shower characteristics like $\sigma/\lstop$.
We ignore vacuum and medium-induced masses of all high-energy particles.
We take the multiple-scattering ($\qhat$) approximation for transverse
momentum transfers from the medium.
(Given our other assumptions, this is appropriate when the daughters of
a high-energy splitting inside a quark-gluon plasma have energies large
compared to the plasma temperature $T$.%
\footnote{
  The condition for using the $\qhat$ approximation for computing
  splitting is that there be many elastic collisions with the medium during
  the formation time, which for QCD is of order
  $t_\form \sim \sqrt{\omega/\qhat}$, where
  $\omega$ is the energy of the least energetic daughter of the splitting.
  The rest is easiest to explain for
  a weakly-coupled quark-gluon plasma (which we do not assume in this paper).
  If $g \simeq g(T)$ is the QCD coupling constant at the plasma energy scale,
  then $\qhat \sim \Nc^2 g^4 T^3$ and
  the mean free time between elastic collisions is
  $\tau_{\rm el} \sim 1/\Gamma_{\rm el} \sim (\Nc g^2 T)^{-1}$.
  The condition $t_{\rm form} \gg \tau_{\rm el}$ that there are many scatterings
  during the formation time can then be reduced algebarically to
  $\omega \gg T$.
}
)
We approximate the bare value
$\qhat_{(0)}$ of $\qhat$ as constant.%
\footnote{
  In this context, the ``bare'' $\qhat_{(0)}$ represents the
  contribution from scattering of the high-energy particle with the
  medium \textit{without} any high-energy splitting.
}
We assume that the particle initiating the shower can be approximated as
on-shell.  In the case of QCD, we take the large-$\Nc$ approximation
$\Nc \gg 1$ because it simplifies color dynamics during overlapping
formation times \cite{2brem}.
We only calculate $p_\perp$-integrated rates%
\footnote{
  Especially for overlapping formation times, rates that are integrated over
  final $p_\perp$'s are enormously simpler to calculate than rates that are
  not.  See the discussion in section 4.4 of ref.\ \cite{2brem} and
  appendix F of ref.\ \cite{seq}.
}
and so only look at
$p_\perp$-insensitive observables for our thought experiments,
such as the deposition distributions
$\rho(z)$ and $\eps(z)$, for which we do not keep track of the
(parametrically small) spread of the deposition in directions transverse
to $z$.


\subsection{Leading-order rates and notation for group factors}
\label {sec:LO}

\begin {figure}[t]
\begin {center}
  \includegraphics[scale=0.6]{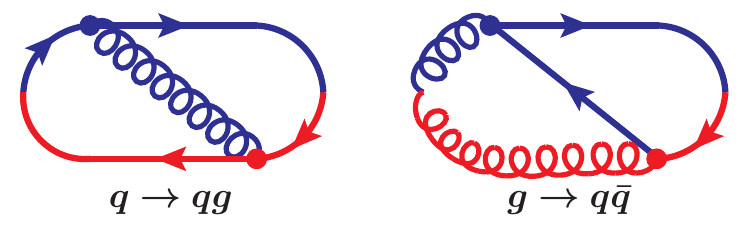}
  \caption{
     \label{fig:LOqcd}
     Time-ordered interference diagrams contributing to the rates
     of $q \to qg$ and $g \to q\bar q$.
     Time runs from left to right.
     In both cases, all lines implicitly interact with the medium.
     Only one of the two time orderings that contribute to each
     process are shown above, but both orderings can be included by
     taking $2\Re[\cdots]$.
  }
\end {center}
\end {figure}

We refer to single-splitting rates
that \textit{ignore} the possibility of
overlapping splittings as \textit{leading-order} (LO) rates,
which are the rates originally investigated by LPM and BDMPS-Z.
For us, leading order means leading order in the number of
high-energy splitting vertices and includes the effects of an
arbitrary number of interactions with the medium.
Adopting Zakharov's picture \cite{Zakharov1,Zakharov2}, we think
of rates like $q{\to}qg$ and $g{\to}q\bar q$ as time-ordered interference
diagrams, such as fig.\ \ref{fig:LOqcd}, which combine the amplitude for
splitting (blue) with the conjugate amplitude (red).
See refs.\ \cite{2brem,qcdI} for more discussion of our graphical conventions
and implementation of Zakharov's approach.

\begin {figure}[t]
\begin {center}
  \includegraphics[scale=0.6]{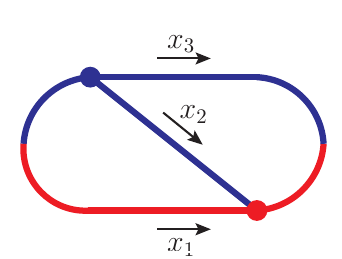}
  \caption{
     \label{fig:LOgeneric}
     A generic abstraction of LO splitting diagrams such as
     fig.\ \ref{fig:LOqcd}, here showing our labeling of the
     longitudinal momentum fractions $x_i$ of the high-energy
     particles in the amplitude and conjugate amplitude.
     Following the conventions of ref.\ \cite{2brem}, we take
     the $x$'s in the conjugate amplitude to be negative (in the
     forward time direction shown in the diagram), so
     that longitudinal momentum is conserved at the vertices
     of this interference diagram.
  }
\end {center}
\end {figure}

In order to prepare for later discussion, and for comparisons of QCD vs.\ QED,
we briefly consider a more general
(but still high-energy) situation.  Generically represent
each diagram of fig.\ \ref{fig:LOqcd} by fig.\ \ref{fig:LOgeneric}.
In Zakharov's approach, the evolution of the high-energy particles between
the two vertices in fig.\ \ref{fig:LOqcd}
is simplified
to a problem in 3-particle two-dimensional quantum mechanics in the
transverse plane, where the three particles consist of (i) the
two (blue) particles in the amplitude plus (ii) the single (red) particle
in the conjugate amplitude, with a corresponding 3-particle Hamiltonian
\begin {equation}
   \frac{\p_{1\perp}^2}{2x_1 E} + \frac{\p_{2\perp}^2}{2x_2 E}
                 + \frac{\p_{3\perp}^2}{2x_3 E}
   + V_3(\b_1,\b_2,\b_3) ,
\label {eq:Hthree}
\end {equation}
where
the kinetic terms originate from the high-energy expansion
$E_p = \sqrt{p_z^2+\p_\perp^2} \simeq p_z + \p_\perp^2/(2 p_z)$ for
each particle.  In (\ref{eq:Hthree}), $E$ is the energy of the parent, and
$(x_1,x_2,x_3)$ are the momentum fractions of the three particles as
oriented in the direction shown in the diagram of fig.\ \ref{fig:LOgeneric}.
For the first diagram of fig.\ \ref{fig:LOqcd}, for example, they would
be $(x_1,x_2,x_3) = (-1,\xg,1{-}\xg)$, where $\xg$ is momentum
fraction of the gluon.%
\footnote{
  The negative sign for $x_1$ and so the negative value of the first
  term in (\ref{eq:Hthree}) arises because particle ``1'' here represents
  the initial quark in the {\it conjugate} amplitude.
}
The 3-particle
``potential energy'' $V_3$ in (\ref{eq:Hthree})
represents the effect of correlations arising from
medium averaging the effects of the
high-energy particles' interactions with the medium.
In the $\qhat$ approximation, it is given by%
\footnote{
   For a general argument for the form (\ref{eq:Vthree}), see for example
   the argument
   in appendix A of ref.\ \cite{2brem} for eq.\ (2.21), or
   the review of earlier arguments in section III.A of ref.\ \cite{Vqhat}.
   Note that $V_3$ does not physically represent a potential energy; it merely
   plays that mathematical role in the Hamiltonian (\ref{eq:Hthree}).
   In particular, the $V_3$ of (\ref{eq:Vthree}) is imaginary-valued.
   The effective
   Hamiltonian (\ref{eq:Hthree}) need not be Hermitian because a
   medium-average of unitary time evolution is not itself unitary.
}
\begin {multline}
  V_3(\b_1,\b_2,\b_3) =
  -\frac{i}{8} \Bigl[
     (\qhat_1+\qhat_2-\qhat_3) (\b_2{-}\b_1)^2
\\
        + (\qhat_2+\qhat_3-\qhat_1) (\b_3{-}\b_2)^2
        + (\qhat_3+\qhat_1-\qhat_2) (\b_1{-}\b_3)^2
  \Bigr]
\label {eq:Vthree}
\end {multline}
in terms of the transverse
positions $(\b_1,\b_2,\b_3)$ of the three particles.
Here, $\hat q_i$ is the value of $\qhat$ associated with the type of
particle $i$.
Three-dimensional rotation invariance may be used to reduce the
two-dimensional 3-particle problem described by (\ref{eq:Hthree})
to an effective 1-particle problem in terms of a single transverse
position variable%
\footnote{
  \label{foot:B}
  An equivalent reduction was used by Zakharov \cite{Zakharov1,Zakharov2}.
  For a discussion in the language used here, see section 3 of
  ref.\ \cite{2brem}.
  The equalities in (\ref{eq:Bdef}) depend on the fact that
  $x_1{+}x_2{+}x_3 = 0$ and
  the constraint $x_1\b_1 + x_2\b_2 + x_3\b_3 = 0$.
  See, for example, section 3.1 of ref.\ \cite{2brem} for one way of
  understanding the last constraint.
}
\begin {equation}
   \B \equiv \frac{\b_1-\b_2}{x_1+x_2} = \frac{\b_2-\b_3}{x_2+x_3}
      = \frac{\b_3-\b_1}{x_3+x_1} \,.
\label {eq:Bdef}
\end {equation}
In the $\qhat$ approximation of (\ref{eq:Vthree}), the effective 1-particle
system is described by the Hamiltonian%
\footnote{
  See specifically eqs.\ (2.27) and (2.33) of ref.\ \cite{2brem}.
}
\begin {subequations}
\label {eq:calH3}
\begin {equation}
   {\cal H} =
   \frac{\P_\perp^2}{2M}
      + \tfrac12 M \, \Omega_0^2 \,  \B^2
\end {equation}
with
\begin {equation}
   M \equiv -x_1 x_2 x_3 E ,
   \qquad
   \Omega_0^2 \equiv 
   -\frac{i}{2E}
    \left(
       \frac{\hat q_1}{x_1} + \frac{\hat q_2}{x_2} + \frac{\hat q_3}{x_3}
    \right) .
\label {eq:Omega0B}
\end {equation}
\end {subequations}
Generically,
the corresponding LPM/BDMPS-Z single splitting rate (in $\qhat$ approximation)
turns out to be
\begin {equation}
  \left[ \frac{d\Gamma}{dx} \right]^\LO
  =
  \frac{\alpha}{\pi}\, P_{\scriptscriptstyle\rm DGLAP}(x)\, \Re(i\Omega_0) ,
\label{eq:dGgeneric}
\end {equation}
where $P_{\scriptscriptstyle\rm DGLAP}$ is the corresponding unregulated
Dokshitzer-Gribov-Lipatov-Alterelli-Parisi (DGLAP) vacuum splitting function.

For the splitting processes $q\to qg$ and $g\to q\bar q$,
(\ref{eq:Omega0B}) and (\ref{eq:dGgeneric}) give%
\footnote{
  The absolute values signs in (\ref{eq:LOrate0}) are unnecessary for
  $0 < \xq < 1$.  They have been included here for an obscure reason, which is
  to avoid any possible confusion about the applicability of formulas
  if considering the front-end transformations
  used to relate some diagrams for overlapping splitting effects, as
  described in refs.\ \cite{qedNf} and \cite{qcd}.  Those transformations
  can replace $\xq$ by a negative value.
}
\begin {subequations}
\label {eq:LOrate0}
\begin {align}
  \left[ \frac{d\Gamma}{d\xq} \right]^\LO_{q\to qg}
  &= \frac{\alphas}{2\pi} \, P_{q\to q}(\xq)
     \sqrt{ \frac{1}{E} \left|
       \frac{\qhatF}{\xq} + \frac{\qhatA}{1{-}\xq} - \qhatF
     \right| }
    ,
\\
  \left[ \frac{d\Gamma}{d\xq} \right]^\LO_{g\to q\bar q}
  &= \frac{\Nf \alphas}{2\pi} \, P_{g\to q}(\xq)
     \sqrt{ \frac{1}{E} \left|
       \frac{\qhatF}{\xq} + \frac{\qhatF}{1{-}\xq} - \qhatA
     \right| }
    ,
\end {align}
\end {subequations}
where $\qhatF$ and $\qhatA$ are respectively the values of $\qhat$ for
quarks (fundamental color representation) and gluons (adjoint color
representation), and
\begin {equation}
  P_{q\to q}(x) = \CF\frac{1+x^2}{1-x} ,
  \qquad
  P_{g\to q}(x) = \tF[ x^2+(1{-}x)^2 ].
  \label {eq:DGLAP_Ps}
\end {equation}
$C_R$ and $t_R$ are the quadratic Casimir and trace normalization
defined in terms of the color generators $T_R^a$ of representations $R$
by
\begin {equation}
  T_R^a T_R^a = C_R \openone , \qquad \tr(T_R^a T_R^b) = t_R \delta^{ab} .
\end {equation}

In the large-$\Nc$ limit, which we take to simplify the calculation of overlap
effects, $\qhatA = 2\qhatF$, which is an example of Casimir scaling of
$\qhat$ since $\CA = 2\CF$ in that same limit.
At the moment, however, we are only discussing leading-order splitting
rates (no overlaps yet), and so we may be a little more general but
will still assume Casimir scaling of $\qhat$.
(Casimir scaling is automatic in the limit of weakly-coupled quark gluon
plasmas.  Beyond weak coupling, ref.\ \cite{color}
suggests that Casimir scaling is, more generally,
a necessary consistency condition for any $\qhat$ approximation,
but there are caveats.%
\footnote{
\label{foot:Casimir}
  See in particular section 3.3 of ref.\ \cite{color}.
  The argument there was specific to the SU($N$) generalizations
  of SU(3) color representations
  $R = \bm 8$, $\bm{10}$, and $\bm{27}$.
  We presume that similar arguments could be constructed for other
  color representations, but (as far as we know) this has not yet been done.
  Ref.\ \cite{color} implicitly assumed that the $\qhat_i$ were
  constant, and so the argument is
  restricted to what in this paper we
  have called our bare $\qhat_{(0)}$ (which we have taken to be constant).
  Additionally, the argument assumed that there is no difference between
  the values of $\qhat$ for (i) an amplitude (blue) line correlated with a
  conjugate amplitude (red) line and
  (ii) an amplitude (blue) line correlated with another amplitude (blue)
  line.  Once one includes radiative corrections, it is known \cite{logs2}
  that these are no longer exactly equivalent.
}%
)
With Casimir scaling, (\ref{eq:LOrate0}) may be written solely
in terms
of $\qhatF$ as
\begin {subequations}
\label {eq:LOrate}
\begin {align}
  \left[ \frac{d\Gamma}{d\xq} \right]^\LO_{q\to qg}
  &= \frac{\alphas}{2\pi} \, P_{q\to q}(\xq)
     \sqrt{ \frac{\qhatF}{E} \left|
       \frac{1}{\xq} + \frac{\CA}{\CF(1{-}\xq)} - 1
     \right| }
    ,
\label {eq:LOrateq}
\\
  \left[ \frac{d\Gamma}{d\xq} \right]^\LO_{g\to q\bar q}
  &= \frac{\Nf \alphas}{2\pi} \, P_{g\to q}(\xq)
     \sqrt{ \frac{\qhatF}{E} \left|
       \frac{1}{\xq} + \frac{1}{1{-}\xq} - \frac{\CA}{\CF}
     \right| }
    .
\label {eq:LOpair}
\end {align}
\end {subequations}
There is a factor of $\Nf$ in the pair production rate (\ref{eq:LOpair})
because the produced pair can have any flavor.

For QCD, the group constants are
\begin {align}
  \CA&=\Nc, &
  \CF&= \tfrac{\Nc^2-1}{2\Nc}, &
  \tF&=\tfrac12 &
  &\mbox{(QCD)} ,
\\
\intertext{but, for the results obtained in this paper,
           we use the large-$\Nc$ limit}
  \CA&=\Nc, &
  \CF&=\tfrac{\Nc}2 , &
  \tF&=\tfrac12 &
  &\mbox{($\Nc{\gg}1$ QCD)} ,
\label {eq:largeNcCoeffs}
\\
\intertext{and so $\CA/\CF = 2$ in (\ref{eq:LOrate}).
           One may also recover the leading-order rates for QED by using}
  \CA&=0, &
  \CF&=1, &
  \tF&=1 &
  &\mbox{(QED)}
\end {align}
and replacing $\alphas$ by $\alpha$, but the difference between
large-$\Nf$ QCD and QED formulas
for \textit{overlap} calculations will be slightly more
involved (see section \ref{sec:convert}).


\subsection{NLO Diagrams}

\begin {figure}[t]
\begin {center}
  \includegraphics[scale=0.55]{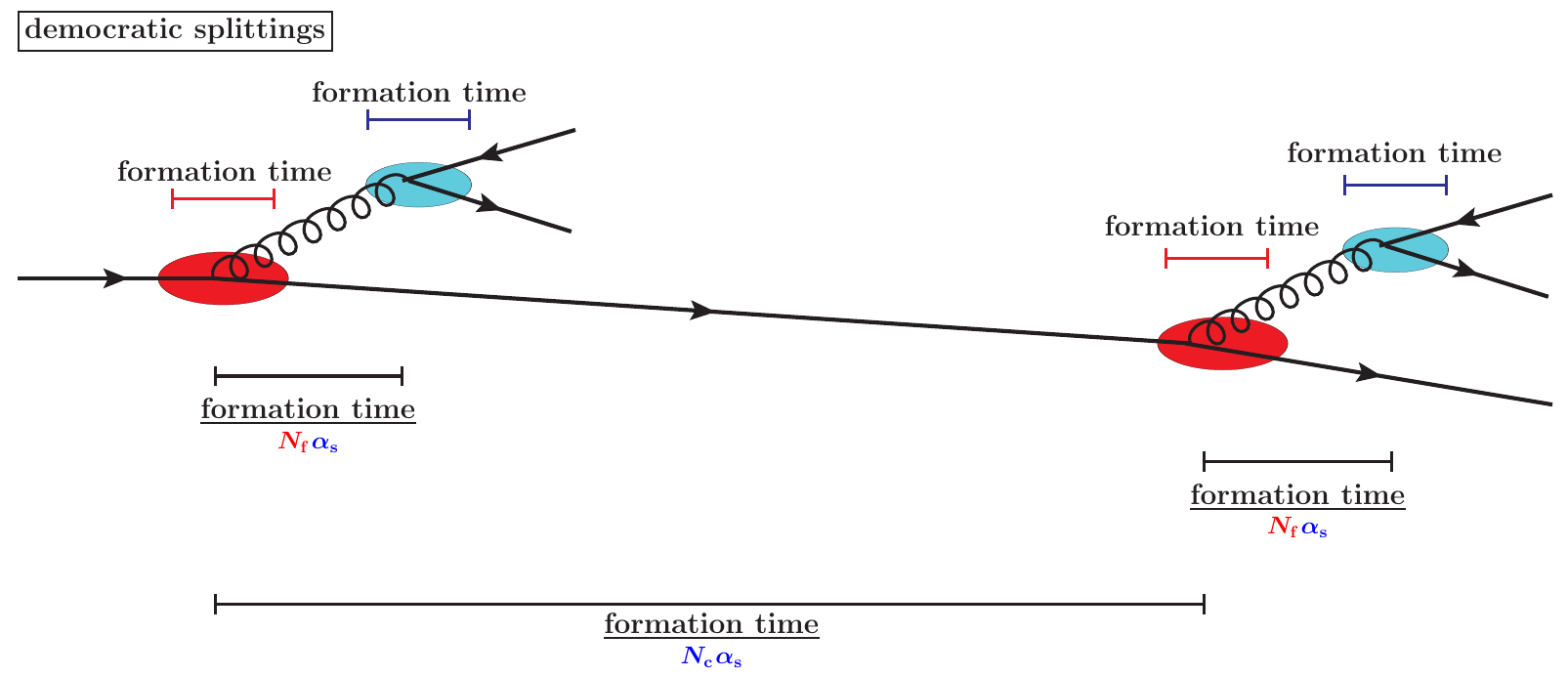}
  \caption{
     \label{fig:typicalNf_qcd}
     Parametric summary of the relative size of typical formation lengths
     and distances between \textit{democratic} splitting for large-$\Nf$
     QCD when $\Nc\alphas \ll \Nf\alphas \ll 1$.
     Large-$\Nf$ QED looks the same \cite{qedNf}
     for democratic splittings, with
     $\alphas$ replaced by $\alpha$, and $\Nc$ replaced by 1. 
  }
\end {center}
\end {figure}

For $\Nf{\gg}\Nc{\gg}1$ QCD,
the relevant diagrams and many details of the calculation of
overlap effects will closely parallel the discussion of large-$\Nf$
QED in refs.\ \cite{qedNfstop,qedNfenergy,qedNf}.
For a start, fig.\ \ref{fig:typicalNf_qcd} depicts a section of a shower
and the parametric size of typical distance scales
for the case of democratic splittings.  Democratic means
that neither daughter of a splitting is soft compared to its parent.
(We'll later have reason in section \ref{sec:why} to instead discuss the case
where one of the gluons is soft.)
For democratic splittings, formation times are of order%
\footnote{
  In terms of formulas, the parametric
  scale of the formation time is the scale
  of $1/|\Omega_0|$ from (\ref{eq:Omega0B}), here specialized to
  democratic splittings.
}
\begin {equation}
   t_\form \sim \sqrt{ \frac{E}{\qhat} }
   \qquad \mbox{(democratic splittings)}
\label {eq:tformdem}
\end {equation}
and are all of the same order for the first few (the most important)
splittings in the shower.
When the LPM effect is relevant (high energies and so formation times
large compared to mean-free times), the probability of democratic
bremsstrahlung $q{\to}qg$
is parametrically of order $\CF\alphas$ per formation time, and so
the typical distance between successive bremsstrahlungs is of
order the formation time divided by $\CF\alphas$, as depicted in
the figure (except that there we've used the fact that $\CF\sim\Nc$).
In contrast, pair production (\ref{eq:LOpair})
is enhanced by a factor of $\Nf$ and
so is much more rapid than bremsstrahlung in the large-$\Nf$
limit.  In detail, the probability of pair production is
of order $\Nf\tF\alphas \sim \Nf\alphas$ per formation time.
In this paper, we take the large-$\Nf$ limit of QCD to mean
\begin {equation}
  \Nf \gg \Nc \gg 1 .
\label {eq:Nlimit}
\end {equation}
We then have a parametric hierarchy of scales in
fig.\ \ref{fig:typicalNf_qcd}, and the possibility of (i) two brem\-sstrahlungs
overlapping each other
is suppressed (by a factor of $\Nc/\Nf \ll 1$) compared to
the possibility that (ii) a bremsstrahlung overlaps a following
pair production.  In computing overlap effects, we therefore
only need to compute the latter case in the limit (\ref{eq:Nlimit})
[at least as far as real (i.e.\ non-virtual) double splitting is concerned].

We also assume that
\begin {equation}
   \Nf\alphas(\mu) \ll 1 ,
\end {equation}
where $\alphas(\mu)$ is the value of $\alphas$ appropriate to
high-energy splitting vertices, meaning that
the renormalization scale $\mu$
has been chosen to be of order the typical transverse
momenta of the daughters of typical splittings, which is
$\mu \sim (\qhat E)^{1/4}$ for democratic splittings.%
\footnote{
  See, for instance, the discussion in refs.\ \cite{finale, finale2}.
}
There is then a double hierarchy of scales
\begin {equation}
  \mbox{formation time}
  \ll \frac{\mbox{formation time}}{\Nf\alphas}
  \ll \frac{\mbox{formation time}}{\Nc\alphas}
\end {equation}
in fig.\ \ref{fig:typicalNf_qcd}.

\begin {figure}[tp]
\begin {center}
  \includegraphics[scale=0.6]{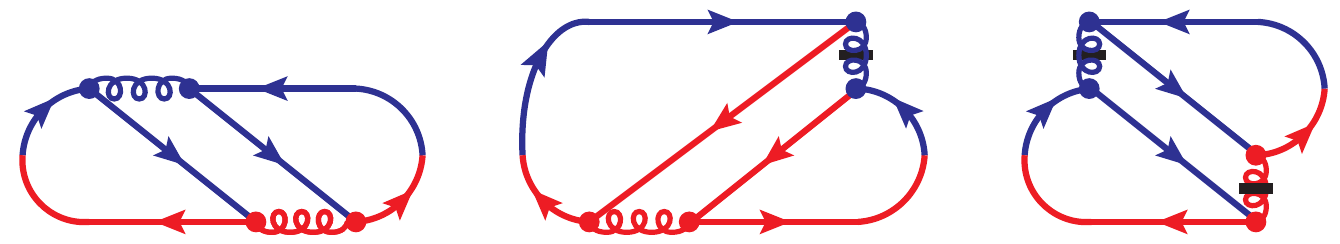}
  \caption{
     \label{fig:realExamples}
     A few examples of lightcone-time ordered interference diagrams for
     $q \to qq \bar q$ in $\Nf{\gg}\Nc{\gg}1$ QCD.
     Here, the gluons are transverse polarized unless they are
     crossed by a black bar.  The bar-crossed gluon lines indicate the
     exchange of a longitudinally-polarized gluon in light-cone gauge,
     which is instantaneous in lightcone time.
  }
\end {center}
\end {figure}

\begin {figure}[tp]
\begin {center}
  \includegraphics[scale=0.6]{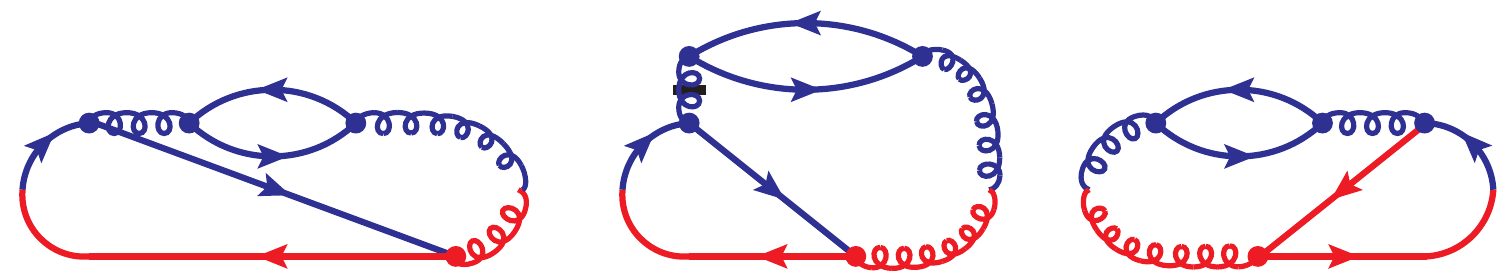}
  \caption{
     \label{fig:virtExamples}
     A few examples of
     lightcone-time-ordered interference diagrams for the virtual
     correction to single splitting
     $q \to qg$ or $g \to q\bar q$ in
     $\Nf{\gg}\Nc{\gg}1$ QCD.
  }
\end {center}
\end {figure}

With that preparation,
fig.\ \ref{fig:realExamples} shows some examples of interference diagrams
contributing to overlapping $q\to qg\to qq\bar q$.
We refer to overlap corrections
as next-to-leading-order (NLO) effects because these diagrams are
suppressed by $O\bigl(\Nf\alphas(\mu)\bigr)$ compared to leading-order ones.
At the same order, there are also virtual corrections to
single splitting, some examples of which are shown in
fig.\ \ref{fig:virtExamples}.  The full set of NLO
diagrams for large-$\Nf$ QED may be found in ref.\ \cite{qedNf}
(also reproduced in our fig.\ \ref{fig:qeddiags} here in
appendix \ref{app:summary}).
Replacing photon lines
with gluon lines gives the full set for
$\Nf{\gg}\Nc{\gg}1$ QCD.

\subsection{Notation for Rates}
\label{sec:RateNotation}

\begin {figure}[t]
\begin {center}
  \includegraphics[scale=0.70]{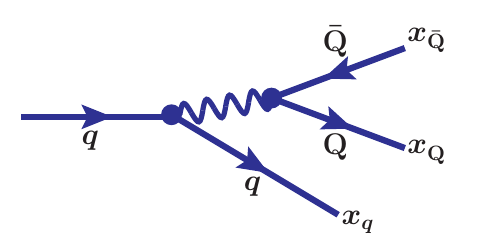}
  \caption{
     \label{fig:qQnotation}
     Our notation (\ref{eq:qQQ})
     for distinguishing pair-produced quarks from the
     original quark in $q \to qg \to qq\bar q$
     in the large-$\Nf$ limit.  The $x$'s are the energy fractions of
     the original electron, and $\xQbar = 1 - \xq - \xQ$.
  }
\end {center}
\end {figure}

The original motivation of taking the large-$\Nf$ limit was simply to reduce
the number of interference diagrams that had to be calculated.
However, following ref.\ \cite{qedNfenergy},
the large-$\Nf$ limit also somewhat simplifies
the analysis of shower development by distinguishing all the daughters of
bremsstrahlung overlapping with pair production, labeled as in
fig.\ \ref{fig:qQnotation}.  The two daughter quarks are distinguishable
particles in the large-$\Nf$ limit
because the probability that the flavor of the pair-produced
quark is the same as that of the initial quark scales like $1/\Nf$.
When convenient, we will emphasize this distinguishability by
using the symbol $\Q$ for pair-produced quarks and so refer to
overlapping bremsstrahlung followed by pair production as
\begin {equation}
   q \to qg \to q\Q\Qbar .
\label {eq:qQQ}
\end {equation}
We will also often refer to leading-order pair production as
$g \to \Q\Qbar$ and the corresponding one-loop virtual correction
(such as the last diagram of fig.\ \ref{fig:virtExamples}) as
the interference of
\begin {equation}
   g \to \Q'\Qbar' \to g \to \Q\Qbar
\end {equation}
with leading-order $g \to \Q\Qbar$.

With this notation, the initial building blocks
that we need (closely following the
treatment \cite{qedNfenergy} of large-$\Nf$ QED)
will be referred to as
\begin {subequations}
\label {eq:rates}
\begin {align}
   \mbox{$1{\to}2$ rates:} \quad &
   \left[ \frac{d\Gamma}{d\xq} \right]_{q\to qg} =
   \left[ \frac{d\Gamma}{d\xq} \right]^{\rm LO}_{q\to qg} +
   \left[ \Delta \frac{d\Gamma}{d\xq} \right]_{q\to qg}^{\rm NLO} ,
\label {eq:rateqqg}
\\ &
   \left[ \frac{d\Gamma}{d\xQ} \right]_{g\to \Q\Qbar} =
   \left[ \frac{d\Gamma}{d\xQ} \right]^{\rm LO}_{g\to \Q\Qbar} +
   \left[ \Delta \frac{d\Gamma}{d\xQ} \right]_{g \to \Q\Qbar}^{\rm NLO} ,
\label {eq:rategQQ}
\\
   \mbox{effective $1{\to}3$ rate:} \quad &
   \left[ \Delta \frac{d\Gamma}{d\xq\,d\xQ} \right]_{q \to q\Q\Qbar} .
\label {eq:1to3rate}
\end {align}
\end {subequations}
The $1{\to}3$
rate (\ref{eq:1to3rate}) represents the overlap \textit{correction} to
treating real double splitting (\ref{eq:qQQ})
as two, consecutive, independent
LO splittings.%
\footnote{
  See the discussion in section 1.1 of ref.\ \cite{seq}.
}
The NLO pieces of (\ref{eq:rateqqg}) and (\ref{eq:rategQQ})
represent virtual corrections to the leading-order rates.
In terms of the above building blocks, overlap effects of two
consecutive splittings can be accounted for by
\textit{classical} probability analysis of a shower developing
via $1{\to}2$ and $1{\to}3$ splittings described by (\ref{eq:rates}).


\section{\boldmath Converting large-$\Nf$ QED rates to
         $\Nf{\gg}\Nc{\gg}1$ QCD rates}
\label {sec:convert}

\begin {figure}[t]
\begin {center}
  \includegraphics[scale=0.8]{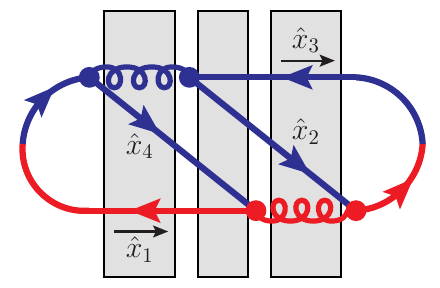}
  \caption{
     \label{fig:qcdseqovlap}
     The first time-ordered interference
     diagram of fig.\ \ref{fig:realExamples}, here showing how we label
     various energy fractions as
     $(\hat x_1,\hat x_2,\hat x_3,\hat x_4)=(-1,\xQ,\xQbar,\xq)$ following
     the convention of ref.\ \cite{qedNf} inherited from the conventions
     of refs.\ \cite{2brem,seq}.
     The medium-averaged evolution in the leftmost and rightmost shaded areas
     are described
     by 3-body quantum mechanics with non-Hermitian Hamiltonians,
     similar to the discussion of
     leading-order rates in section \ref{sec:LO}.  The middle
     shaded area corresponds to a 4-body quantum mechanics problem.
  }
\end {center}
\end {figure}

We now discuss how large-$\Nf$ QCD
formulas for evaluating the building blocks (\ref{eq:rates})
may be obtained by adapting known large-$\Nf$ QED results \cite{qedNf}.
There are two adaptations to make.
\begin {itemize}
\item
  We need to include overall SU($\Nc$) color factors associated with
  high-energy splitting vertices in NLO contributions like
  figs.\ \ref{fig:realExamples} and \ref{fig:virtExamples}.
\item
  In section \ref{sec:LO}, we reviewed
  the 3-body ``potential'' $V_3(\b_1,\b_2,\b_3)$
  representing medium-averaged correlations between particle interactions
  in the analysis of leading-order LPM/\linebreak[1]BDMPS-Z rates.
  NLO corrections involve times when
  corresponding 4-body potentials are needed, such as the middle
  shaded area of fig.\ \ref{fig:qcdseqovlap}.  We need to change
  the 4-body $\bar ee\bar ee$ potential appropriate for large-$\Nf$
  QED \cite{qedNf} to a 4-body $\bar qq\bar qq$ potential for
  $\Nf{\gg}\Nc{\gg}1$ QCD.
\end {itemize}

We will discuss the differences more thoroughly below.
In addition to identifying the isolated differences below, we
provide a complete summary in appendix \ref{app:summary} of the
formulas for the basic rates (\ref{eq:rates}), in a form that
covers both large-$\Nf$ QCD and large-$\Nf$ QED.


\subsection {Overall group factors}

\subsubsection{Basics}

The reason that we have taken the large-$\Nc$ limit in this and previous
work on overlap effects in QCD is because the large-$\Nc$ limit
removes the need to keep track of the evolution of the
color state of the high-energy particles during 4-body evolution.%
\footnote{
  In the context of the formalism used for our overlap calculations, this
  use of the large-$\Nc$ approximation goes back to ref.\ \cite{2brem}.
  In the same context, discussion of how to go beyond the large-$\Nc$
  limit may be found in refs.\ \cite{color,1overN}.
  The same issue arises in the somewhat different problem of
  calculating $p_\perp$ dependence for \textit{leading-order} splitting
  rates.
  See for example, refs.\ \cite{NSZ6j,Zakharov6j,Konrad1}
  and the more recent ref.\ \cite{KonradNc3}.
}
Because of this simplification, the overall color factors of diagrams
decouple from the evolution of color in the medium, and one can read
them off by briefly pretending that interference diagrams such as
fig.\ \ref{fig:qcdseqovlap} were vacuum diagrams.
Summing over final-state colors and averaging over initial-state ones,
the overall color factors for fig.\ \ref{fig:qcdseqovlap} are then
\begin {equation}
  \frac{\tr(\TF^a\TF^b) \tr(\TF^a\TF^b)}{\dF}
  = \frac{\tF^2\dA}{\dF}
  = \tF\CF
  \qquad \mbox{[NLO $q\to qg$ or $q\Q\Qbar$]} ,
\label {eq:overallq}
\end {equation}
where $d_R$ is the dimension of color representation $R$
and where the last equality follows from the (general-$\Nc$) relation that
\begin {equation}
  \tF \dA = \tr(\TF^a \TF^a) = \CF \dF .
\label {eq:Crelation}
\end {equation}
Nothing about (\ref{eq:overallq}) depends on time ordering or whether gluons
are transverse or longitudinally polarized, and so the same factor applies
to all our $\Nf{\gg}\Nc{\gg}1$ quark-initiated NLO diagrams.
For $\Nf{\gg}\Nc{\gg}1$ gluon-initiated NLO diagrams, such as the last diagram
of fig.\ \ref{fig:virtExamples}, the only difference is that we
average over $\dA$ initial colors instead of $\dF$ initial colors,
leading to an overall factor
\begin {equation}
  \frac{\tr(\TF^a\TF^b) \tr(\TF^a\TF^b)}{\dA}
  = \tF^2
  \qquad \mbox{[NLO $g\to \Q\Qbar$]} .
\label {eq:overallg}
\end {equation}

For future reference, note that in both cases (\ref{eq:overallq})
and (\ref{eq:overallg}),
the \textit{ratio} of NLO rates to LO rates (\ref{eq:LOrate})%
\footnote{
  The overall color factors for the LO rates are the $\CF$ or $\tF$
  in (\ref{eq:DGLAP_Ps}).
}
will have an overall factor of $\tF\alphas$ in $\Nf{\gg}\Nc{\gg}1$ QCD
formulas in place of the factor of $\alpha$ in large-$\Nf$ QED formulas.

Later, we will present numerical results in units of the overall
color factors described above just for the sake of eliminating this
rather simple difference between large-$\Nf$ QED and $\Nf{\gg}\Nc{\gg}1$ QCD
when comparing numbers.  Then the only difference between the
numbers will come from the difference in how the high-energy particles
interact with the medium, captured by section \ref{sec:potentials} below.


\subsubsection{Implication for certain relations among diagrams}
\label {sec:frontend}

\begin {figure}[t]
\begin {center}
  \includegraphics[scale=0.6]{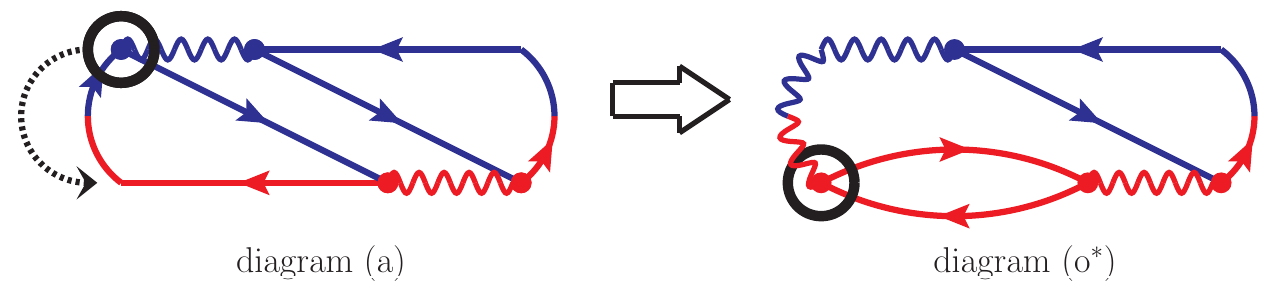}
  \caption{
     \label{fig:frontend}
     Graphical example of a ``front-end'' transformation
     taken from fig.\ 25 of ref. \cite{qedNf}.
     The diagram names (a) and (o) correspond to those of
     ref.\ \cite{qedNf} and fig.\ \ref{fig:qeddiags}
     except that (o$^*$) here means the
     complex conjugate of diagram (o) there.
  }
\end {center}
\end {figure}

Ref.\ \cite{qedNf} reduced the work of calculating all interference
diagrams for large-$\Nf$ QED by showing how some diagrams were
related to each other through what were called front-end and back-end
transformations.  Diagrammatically, a front-end transformation
corresponds to taking the earliest-time vertex in an
interference diagram and sliding it around the front of the diagram
so that it moves from the amplitude to the conjugate amplitude or
vice versa.  Fig.\ \ref{fig:frontend} shows an example which
converts a diagram contributing to real double splitting
$e \to e\gamma \to e\E\Ebar$ into a diagram contributing to
NLO single splitting $\gamma \to \E\Ebar$.
Ref.\ \cite{qedNf} found that there was a relation
among the corresponding contributions to rates,%
\footnote{
  Taking $2\Re[\cdots]$ of our eq.\ (\ref{eq:frontendqed}) here corresponds
  to the first case covered by eq.\ (4.3) of ref.\ \cite{qedNf}.
  Our $\xE$ here is called $\ye$ there.
}
\begin {equation}
   \left[ \frac{d\Gamma}{d\xE} \right]_{\rm (o^*)}
   = - \Nf \int_0^1 d\xe \>
   \left\{
     \left[ \frac{d\Gamma}{d\xe\,d\xE} \right]_{\rm (a)}
     \mbox{with}~
     (\xe,\xE,E) \to
     \Bigl(\frac{-\xe}{1-\xe}\,,\,\frac{\xE}{1-\xe}\,,\,(1{-}\xe)E\Bigr)
   \right\}.
\label {eq:frontendqed}
\end {equation}
In the case of QCD, we have to modify this formula to account for the
fact that the rate for $q \to qg \to q\Q\Qbar$ is averaged over
the $\dF{=}\Nc$ colors of the initial quark, whereas the rate for
$g \to \Q\Qbar$ should be averaged over the $\dA$ colors of the
initial gluon.  So the front-end relation in this case needs
a corresponding factor of $\dF/\dA$ on the right-hand side:
\begin {multline}
   \left[ \frac{d\Gamma}{d\xQ} \right]_{\rm (o^*)}
   =
\\
 - \frac{\dF}{\dA} \, \Nf\int_0^1 d\xq \>
   \left\{
     \left[ \frac{d\Gamma}{d\xq\,d\xQ} \right]_{\rm (a)}
     \mbox{with}~
     (\xq,\xQ,E) \to
     \Bigl(\frac{-\xq}{1-\xq}\,,\,\frac{\xQ}{1-\xq}\,,\,(1{-}\xq)E\Bigr)
   \right\}.
\label {eq:frontend}
\end {multline}
Note that (\ref{eq:Crelation}) means that the new factor may alternatively
be written as
\begin {equation}
  \frac{\dF}{\dA} = \frac{\tF}{\CF} \,,
\end {equation}
which (as it must be) is just the ratio of the previously-discussed
overall color factor for NLO $g \to \Q\Qbar$ (\ref{eq:overallg}) 
to that for $q \to q\Q\Qbar$ (\ref{eq:overallq}).

\subsection {Potentials}
\label {sec:potentials}

\begin {figure}[t]
\begin {center}
  \includegraphics[scale=0.5]{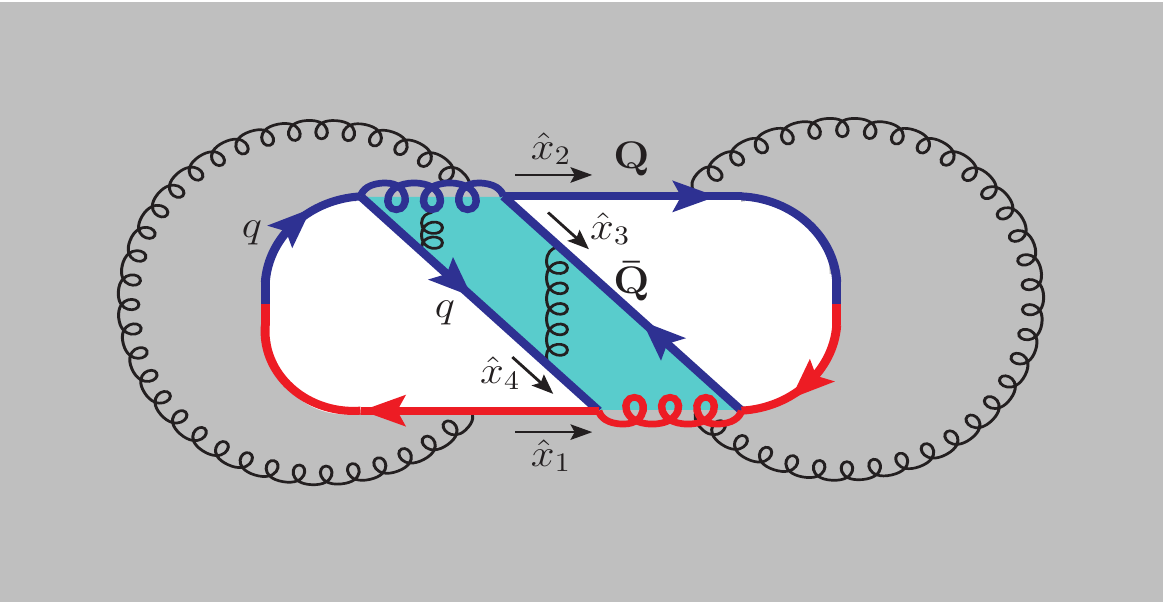}
  \caption{
     \label{fig:largeNc0shaded}
     A large-$\Nc$ planar drawing of (i)
     fig.\ \ref{fig:qcdseqovlap} supplemented
     by (ii) a few
     simple examples of (2-point) medium-averaged correlations of high-energy
     particle interactions with the medium (represented by thin
     black gluon lines).
     The shaded region (both colors) is the section of the plane where
     large-$\Nc$ gluon lines are drawn, with the quark loops
     defining the region's
     boundary.
     Note that
     we had to interchange where
     the $\Q$ and $\Qbar$ lines of fig.\ \ref{fig:qcdseqovlap} are drawn in
     order that the region have a consistently oriented boundary.
  }
\end {center}
\end {figure}

In order to understand how to write down the 4-body ``potential''
$V_4(\b_1,\b_2,\b_3,\b_4)$ appropriate to
medium-averaged evolution in the middle
shaded area of fig.\ \ref{fig:qcdseqovlap}, we need to understand
what correlations are allowed between high-energy particles in
the large-$\Nc$ limit.
The general diagrammatic rules for large-$\Nc$ QCD are that
(i) Feynman diagrams should be planar and (ii) fermion lines should define
an oriented boundary to the planar region in which the gluon lines are
drawn \cite{tHooft}.
A corresponding version of fig.\ \ref{fig:qcdseqovlap} is depicted by
the thick colored lines in fig.\ \ref{fig:largeNc0shaded}.
Unlike the previous figure, fig.\ \ref{fig:largeNc0shaded} also explicitly shows
a few simple examples of medium-averaged correlations between
interactions of the high-energy particles with the medium.
Those examples of correlations are represented by the thin black gluon lines.
We've used 2-point correlations as an example that's easy to draw,
but higher-point correlations are possible too, since we do not assume that
the medium is itself weakly-coupled.
In this graph, the two fermion loops define holes
cut out of the plane.

\begin {figure}[t]
\begin {center}
  \begin{picture}(350,200)(0,0)
    \put(5,5){\includegraphics[scale=0.8]{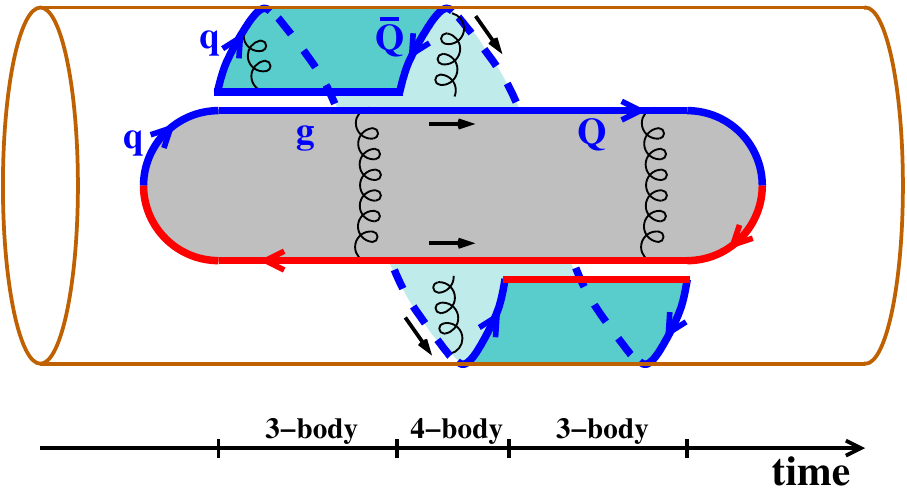}}
    \put(174,105){\large$\hat x_1$}
    \put(174,132){\large$\hat x_2$}
    \put(198,179){\large$\hat x_3$}
    \put(152,58){\large$\hat x_4$}
%
  \end{picture}
  \caption{
     \label{fig:largeNc}
     A re-drawing of fig.\ \ref{fig:largeNc0shaded}
     on a time-ordered cylinder.
     The dashed lines indicate lines that are
     wrapping around the back of the cylinder.
     All of the thick
     lines correspond
     to the fundamental (quark) color representation flowing in the
     direction of the arrows.
     The color of large-$\Nc$ gluons is represented by
     a pair of fundamental color lines flowing in opposite directions,
     which we have drawn explicitly for the high-energy gluons.
     (One may draw the correlations using double-line representation as well,
     but we didn't find that helpful for this discussion.)
     The thick lines all interact with the medium, and the correlations
     (black) arising from
     medium-averaging of those interactions must (in the large-$\Nc$ limit)
     each be routable entirely through one of the
     two disjoint shaded regions above, without
     crossing any other lines.
     The simple examples of correlations shown here correspond to those
     in fig.\ \ref{fig:largeNc0shaded}, and the third correlation
     (which is between particles 3 and 4) runs around the back
     of the cylinder, through the cyan region.
  }
\end {center}
\end {figure}

Correlations should be
effectively instantaneous compared to the scale of the diagram because
high-energy formation times are large compared to the correlation length of
the medium, and this hierarchy is what makes possible the treatment of
each shaded area of fig.\ \ref{fig:qcdseqovlap} as a type of non-relativistic
quantum mechanics problem.
But the planar drawing of the correlators in fig.\ \ref{fig:largeNc0shaded}
obscures the approximately instantaneous nature of the correlations.
Following refs.\ \cite{2brem,seq},%
\footnote{
  See in particular section 4.3 of ref.\ \cite{2brem} and
  section 2.2.1 of ref.\ \cite{seq}.
}
a better way to visualize a
large-$\Nc$ planar diagram when time ordering is important
is to draw the diagram on a cylinder instead of on a plane, with time ordered
along the length of the cylinder.  Any Feynman diagram that can be
drawn in the plane without crossing lines can also be drawn on the
surface of a cylinder without crossing lines, and vice versa,
but the cylinder allows us to simultaneously visualize the time ordering.
Fig.\ \ref{fig:largeNc} shows a version of fig.\ \ref{fig:largeNc0shaded}
drawn on a time-ordered cylinder.
Here, we also find it useful to use 't Hooft double line notation
for the high-energy gluon lines.
In the large-$\Nc$ limit all
correlations generated by
the medium must lie completely within one of the two shaded areas
shown (which is the same shading as fig.\ \ref{fig:largeNc0shaded} but is
unrelated to the shading of fig.\ \ref{fig:qcdseqovlap}).

Using this large-$\Nc$ picture, we now revisit 3-body evolution and implement
4-body evolution.


\subsubsection{Initial 3-body time evolution}

Consider first the initial region of 3-body time
evolution in
fig.\ \ref{fig:largeNc}.  We will refer to the thick lines as
fundamental ($\rm F$) or anti-fundamental ($\rm\bar F$)
color lines depending on how they
are oriented in the diagram relative to the forward time direction.
In the large-$\Nc$ limit, correlations
(whether 2-point or higher) between interactions of the high-energy particles
are only allowed
(i) across the gray region between the red $\rm\bar F$ line ($\hat x_1$) and
the blue $\rm F$-line component of the gluon, or
(ii) across the cyan region between the daughter-quark blue $\rm F$ line
($x_q{=}\hat x_4$) and the blue $\rm\bar F$-line component of the gluon.
The corresponding potential should therefore decomposes into
the sum of 2-body $\rm F\bar F$ potentials for (i) and (ii):%
\footnote{
  The factor of $\frac14$ in
  $V_{\rm F\bar F}(\Delta b) = -\frac{i\qhatF}{4} (\Delta b)^2$ is the
  same factor of 4 that relates the diffusion constant $\kappa$ in
  the $\p_\perp$ diffusion equation
  $\partial_t\rho(\p_\perp,t) = \kappa \nabla_{\p_\perp}^2 \rho(\p_\perp,t)$
  to the original definition of $\qhat$ in terms of the time dependence of
  the second moment of the diffusion equation's solution,
  $\langle \p_\perp^2 \rangle = \qhat t$.
  See, for example, section 2.2 of ref.\ \cite{BDMPS3}.
}
\begin {equation}
   V_{3,\ix}(\b_1,\b_g,\b_q)
   = V_{\rm F\bar F}(\b_g{-}\b_1) + V_{\rm F\bar F}(\b_q{-}\b_g)
   = -\frac{i\qhatF}{4} (\b_g{-}\b_1)^2 -\frac{i\qhatF}{4} (\b_g{-}\b_q)^2 .
\label {eq:V3i}
\end {equation}
This is the same as the more general 3-particle potential
(\ref{eq:Vthree}) with particles $(1,2,3)$ there referring to
$(1,g,q)$ here and using the large-$\Nc$ relation $\qhatA = 2\qhatF$.
As reviewed earlier, the 3-body quantum mechanics
problem can be reduced to an effective 1-body problem.
The mass and frequency (\ref{eq:Omega0B}) of the latter correspond
here to
\begin {equation}
   M_\ix = \xq(1{-}\xq)E ,
   \qquad
   \Omega_\ix
   = \sqrt{ -\frac{i\qhatF}{2E}
       \left( -1 + \frac{1}{\xq} + \frac{2}{\xg} \right) }
   \qquad
   \mbox{($\Nc{\gg}1$ QCD)} .
\label {eq:MLambdai0}
\end {equation}
The above formula for $\Omega_\ix$ is an example
of a low-level QCD modification that needs to be made to
the large-$\Nf$ QED formulas of ref.\ \cite{qedNf} for overlap corrections.
The convention of ref.\ \cite{qedNf} is to write final formulas in terms
of the momentum fractions
\begin {equation}
   (\hat x_1,\hat x_2,\hat x_3,\hat x_4) = (-1,\xQ,\xQbar,\xq)
\label {eq:x1234}
\end {equation}
of the four particles appearing in the 4-body evolution section of
figures like figs.\ \ref{fig:qcdseqovlap} and \ref{fig:largeNc}.
So we rewrite (\ref{eq:MLambdai0}) as
\begin {equation}
   M_\ix = \hat x_1 \hat x_4 (\hat x_1{+}\hat x_4) E ,
   \qquad
   \Omega_\ix
   = \sqrt{ -\frac{i\qhatF}{2E}
       \left( \frac{1}{\hat x_1} + \frac{1}{\hat x_4}
                    - \frac{2}{\hat x_1+\hat x_4} \right) }
   \qquad
   \mbox{($\Nc{\gg}1$ QCD)}.
\label {eq:MLambdai}
\end {equation}


\subsubsection{4-body time evolution}

Now turn to the region of 4-body time evolution in fig.\ \ref{fig:largeNc}.
Here again the potential splits into a sum of two $\rm F\bar F$ potentials:
one for correlations through the gray region of
the red parent-quark line ($\hat x_1$) with $\Q$ and the other
through the cyan region of $q$ with $\Qbar$.
The large-$\Nc$ 4-body potential is then
\begin {equation}
   V_4(\b_1,\b_\Q,\b_\Qbar,\b_q)
   = V_{\rm F\bar F}(\b_\Q{-}\b_1) + V_{\rm F\bar F}(\b_q{-}\b_\Qbar)
   = -\frac{i\qhatF}{4} (\b_\Q{-}\b_1)^2 -\frac{i\qhatF}{4} (\b_q{-}\b_\Qbar)^2
\end {equation}
or, using the particle numbering scheme of (\ref{eq:x1234}) throughout,
\begin {equation}
   V_4(\b_1,\b_2,\b_3,\b_4)
   = -\frac{i\qhatF}{4} \left[
        (\b_2{-}\b_1)^2 + (\b_4{-}\b_3)^2
     \right] .
\label {eq:Vfour0}
\end {equation}
Ref.\ \cite{2brem} discuss how 3-dimensional rotational symmetry can be
used to generally reduce such 4-body problems to effective 2-body problems
in terms of any two of the set of variables%
\footnote{
  The normalization $(x_i+x_j)^{-1}$ in (\ref{eq:Cdef}) is just a convenience
  that makes the $\C$'s natural generalizations of the variable $\B$
  discussed earlier for 3-body evolution (\ref{eq:Bdef}).
  Putting that convenience
  aside, we could have said more simply
  that the 4-body problem could be reduced to an effective 2-body problem
  in terms of any two of the variables $\b_i{-}\b_j$.
}
\begin {equation}
  \C_{ij} \equiv \frac{\b_i{-}\b_j}{x_i+x_j} .
\label {eq:Cdef}
\end {equation}
Focusing just on the potential for the moment, and using the fact that
$x_1+x_2+x_3+x_4=0$, the potential (\ref{eq:Vfour0}) may be written as
\begin {equation}
   V_4 = -\frac{i\qhatF}{4}(x_3+x_4)^2 (C_{12}^2 + C_{34}^2) .
\label {eq:V4C12C34}
\end {equation}
The kinetic terms appropriate to this problem do not depend on the
details of how the particles interact with the medium; they are the same
as those for the pure gluon case \cite{2brem} or the QED case \cite{qedNf}.
In terms of the variables $(\C_{12},\C_{34})$ and their conjugate momenta
$(\P_{12},\P_{34})$, the effective Hamiltonian is
\begin {equation}
  {\cal H}_4
    = \frac{P_{12}^2}{2x_1 x_2(x_1{+}x_2)E} + \frac{P_{34}^2}{2x_3 x_4(x_3{+}x_4)E}
      + V_4(\C_{12},\C_{34}) .
\label {eq:H4two}
\end {equation}
For our particular
potential (\ref{eq:V4C12C34}), the $\C_{12}$ and $\C_{34}$ degrees of
freedom completely decouple.
In order to draw on results for overlap corrections from previous work,
it will be useful to adopt the same notation as ref.\ \cite{2brem}
and rewrite this harmonic oscillator problem as
\begin {equation}
  {\cal H}_4
    = \frac12 \begin{pmatrix} \P_{34} \\ \P_{12} \end{pmatrix}^{\!\top}
          \!{\mathfrak M}^{-1}
        \begin{pmatrix} \P_{34} \\ \P_{12} \end{pmatrix}
        +
        \frac12 \begin{pmatrix} \C_{34} \\ \C_{12} \end{pmatrix}^{\!\top}
          \!{\mathfrak K}
        \begin{pmatrix} \C_{34} \\ \C_{12} \end{pmatrix}
\end {equation}
where
\begin {equation}
   {\mathfrak M}
   =
   \begin{pmatrix}
      x_3 x_4 (x_3+x_4) & \\ & x_1 x_2 (x_1+x_2)
   \end {pmatrix} E
   =
   \begin{pmatrix}
      x_3 x_4 & \\ & -x_1 x_2
   \end {pmatrix} (x_3+x_4) E
\label {eq:frakM}
\end {equation}
is the analog of a mass matrix and
\begin {equation}
   {\mathfrak K}
   =
   -\frac{i\qhatF}{2}(x_3+x_4)^2
   \begin{pmatrix}
      1 & \\ & ~~1
   \end {pmatrix}
\end {equation}
is the analog of a spring constant matrix, taken here from (\ref{eq:V4C12C34}).
Because $\C_{34}$ and $\C_{12}$ decouple here, and the two transverse
spatial directions decouple, the normal modes are
simply
\begin {equation}
   \begin{pmatrix} C_{34}^+ \\ C_{12}^+ \end{pmatrix} =
   \frac{1}{\sqrt{x_3 x_4 (x_3{+}x_4)E}} \begin{pmatrix} 1 \\ 0 \end{pmatrix}
     \qquad \mbox{and} \qquad
   \begin{pmatrix} C_{34}^- \\ C_{12}^- \end{pmatrix} =
   \frac{1}{\sqrt{x_1 x_2 (x_1{+}x_2)E}} \begin{pmatrix} 0 \\ 1 \end{pmatrix} ,
\label {eq:NM0}
\end {equation}
for each transverse direction.
Again following ref.\ \cite{2brem} for compatibility, we have normalized our
normal modes so that
\begin {equation}
   \begin{pmatrix} \C^j_{34} \\ \C^j_{12} \end{pmatrix}^{\!\top}
   \!{\mathfrak M}
   \begin{pmatrix} \C^{j'}_{34} \\ \C^{j'}_{12} \end{pmatrix}
   = \delta^{jj'} .
\label {eq:NMnorm}
\end {equation}
The corresponding normal mode frequencies are
\begin {equation}
  \Omega_+ =
    \sqrt{
       -\frac{i\qhatF}{2E} \left( \frac{1}{x_3} + \frac{1}{x_4} \right)
    } ,
  \qquad
  \Omega_- =
    \sqrt{
       -\frac{i\qhatF}{2E} \left( \frac{1}{x_1} + \frac{1}{x_2} \right)
    } .
\label {eq:Omegapm}
\end {equation}

Earlier work such as \cite{qedNf,2brem,seq}, that we want to adapt to
the case of large-$\Nf$ QCD, package all the normal modes into
columns of a matrix.  We will call the corresponding matrix
\begin {equation}
   a_{34,12} \equiv
   \begin{pmatrix} C_{34}^+ & C_{34}^- \\ C_{12}^+ & C_{12}^- \end{pmatrix}
   =
   \begin{pmatrix}
     \frac{1}{\sqrt{x_3 x_4 (x_3{+}x_4)E}} & 0 \\
     0 & \frac{1}{\sqrt{x_1 x_2 (x_1{+}x_2)E}}
   \end{pmatrix} ,
\label{eq:a3412}
\end {equation}
where here we use the subscript on $a_{34,12}$ to indicate that we have
specified the normal modes in terms of the values of $(\C_{34},\C_{12})$.

Though the choice of variables $(\C_{34},\C_{12})$ diagonalized the
Hamiltonian in this particular case, that is a quirk of large-$\Nf$ QCD.
Previous work on diagrams analogous to fig.\ \ref{fig:qcdseqovlap}
for large-$\Nf$ QED \cite{qedNf} and purely gluonic splitting \cite{seq}
instead used the basis $(\C_{41},\C_{23})$, which is convenient because
$\C_{23}=0$ at the start of the 4-body
evolution because of the vertex that brings particles 2 and 3
together there, and $\C_{41}=0$ at the end of 4-body evolution because of the
vertex that brings particles 4 and 1 together.
The two sets of variables are related by%
\footnote{
  This relation comes from eq.\ (5.31) of ref.\ \cite{2brem}.
  Similar to our earlier footnote \ref{foot:B}, the relation may
  be derived using the fact that $x_1{+}x_2{+}x_3{+}x_4=0$
  and the constraint
  $x_1\b_1 + x_2\b_2 + x_3\b_3 + x_4\b_4 = 0$.
}
\begin {equation}
   \begin{pmatrix} C_{41} \\ C_{23} \end{pmatrix}
   =
   {\cal S}
   \begin{pmatrix} C_{34} \\ C_{12} \end{pmatrix} .
\label {eq:changebasis}
\end {equation}
with transformation matrix
\begin {subequations}
\label {eq:ay}
\begin {equation}
   {\cal S}
   =
   \frac{1}{(x_1{+}x_4)}
   \begin{pmatrix}
       -x_3 & -x_2 \\
        \phantom{-}x_4 &  \phantom{-}x_1
   \end {pmatrix} .
\label {eq:S}
\end {equation}
For the new choice of variables, the matrix of normal modes, which we
might call $a_{41,23}$ here but instead call $a_y$ in
earlier work \cite{2brem,seq,qedNf}, becomes
\begin {equation}
   a_y \equiv a_{41,23} \equiv
   \begin{pmatrix} C_{41}^+ & C_{41}^- \\ C_{23}^+ & C_{23}^- \end{pmatrix}
   = {\cal S} \, a_{34,12} .
\end {equation}
\end {subequations}

Equations (\ref{eq:Omegapm}), (\ref{eq:a3412}), and (\ref{eq:ay}) for
$\Omega_\pm$ and $a_y$ give the specific formulas needed to update the
large-$\Nf$ 4-body QED evolution in ref.\ \cite{qedNf} to
large-$\Nf$ QCD.

There is a related change to cover.
In large-$\Nf$ QED, one of the 4-body
evolution frequencies $\Omega_\pm$ vanished.  Ref.\ \cite{qedNf} labeled
that frequency $\Omega_-$, and so $\Omega_- = 0$.  In general,
the 4-body frequencies appear in formulas for overlap effects
in the combinations $\Omega_\pm \cot(\Omega_\pm\,\Delta t)$ and
$\Omega_\pm \csc(\Omega_\pm\,\Delta t)$ (see, for instance,
appendix A of ref.\ \cite{qcd}).  But in the large-$\Nf$ QED case \cite{qedNf},
the $\Omega_-$ combinations were replaced by their
$\Omega_-{=}0$ limits,
\begin {equation}
  \Omega_- \cot(\Omega_-\,\Delta t) \longrightarrow \frac{1}{\Delta t}
  \qquad \mbox{and} \qquad
  \Omega_- \csc(\Omega_-\,\Delta t) \longrightarrow \frac{1}{\Delta t}
  \qquad
  \mbox{(large-$\Nf$ QED)}.
\label{eq:OmmReplace}
\end {equation}
In large-$\Nf$ QCD, neither $\Omega_+$ nor $\Omega_-$
vanish (\ref{eq:Omegapm}), and so we must undo the replacement
(\ref{eq:OmmReplace}) to convert large-$\Nf$ QED formulas to
large-$\Nf$ QCD.


\subsubsection{Final 3-body time evolution}

The final region of 3-body evolution in fig.\ \ref{fig:largeNc} is
similar to the first region except that the particles involved are different.
Eqs. (\ref{eq:V3i}) and (\ref{eq:MLambdai}) become
\begin {equation}
   V_{3,\fx}(\b_\sQbar,\b_g,\b_\sQ)
   = V_{\rm F\bar F}(\b_g{-}\b_\sQbar) + V_{\rm F\bar F}(\b_\sQ{-}\b_g)
   = -\frac{i\qhatF}{4} (\b_g{-}\b_\sQbar)^2 -\frac{i\qhatF}{4} (\b_g{-}\b_\sQ)^2
   ,
\end {equation}
where $b_g$ now refers to the transverse position of the
\textit{conjugate}-amplitude
(red) gluon in fig.\ \ref{fig:largeNc},
and%
\footnote{
  The superscript ``seq'' stands for ``sequential diagrams,'' which was
  a necessary distinction to draw in early work \cite{2brem,seq} on
  overlapping $g\to gg \to ggg$ where $M$ for the final 3-body evolution
  was different for
  different classes of diagrams.
  We keep the superscript here for consistency of
  notation.
}
\begin {equation}
   M_\fx^\seq = \hat x_2 \hat x_3 (\hat x_2{+}\hat x_3) E ,
   \qquad
   \Omega_\ix^\seq
   = \sqrt{ -\frac{i\qhatF}{2E}
       \left( \frac{1}{\hat x_2} + \frac{1}{\hat x_3}
                    - \frac{2}{\hat x_2+\hat x_3} \right) }
   \qquad
   \mbox{($\Nc{\gg}1$ QCD)}.
\label {eq:MLambdaf}
\end {equation}


\section {Net rates}
\label {sec:netrates}

\subsection{Definition}

Shower development can be conveniently packaged in terms of ``net'' rates
for a splitting or pair of overlapping splittings to produce one daughter
of energy $xE$ (plus any other daughters) from a parent of energy $E$
\cite{qcd,finale,finale2,qedNfenergy}.
Here, we exactly follow ref.\ \cite{qedNfenergy} to define our net rates
$[d\Gamma/dx]_\uij^\net$ for parents of type $i$ producing a daughter
of type $j$,
where underlining of subscripts like $\uqq$ indicates that we are using
the large-$\Nf$ limit to distinguish the pair-produced quark ($\Q$) from
the other quark daughter ($q$) in overlapping splitting rates.
The leading-order contributions are simply
\begin {subequations}
\label {eq:LOrates}
\begin {align}
   \left[ \frac{d\Gamma}{dx} \right]^\LO_{\uqq}
   \equiv& \left[ \frac{d\Gamma}{d\xq} \right]^\LO_{q\to qg}
         &\quad\mbox{with $\xq=x$,$\phantom{1-{}}$}
\label{eq:LOqq}
\\
   \left[ \frac{d\Gamma}{dx} \right]^\LO_{\uqg}
   \equiv& \left[ \frac{d\Gamma}{d\xq} \right]^\LO_{q\to qg}
         &\quad\mbox{with $\xq=1-x$,}
\\
   \left[ \frac{d\Gamma}{dx} \right]^\LO_{\ugQ}
   \equiv& \left[ \frac{d\Gamma}{d\xQ} \right]^\LO_{g\to\Q\Qbar}
         &\quad\mbox{with $\xQ=x$,$\phantom{1-{}}$}
\end {align}
\end {subequations}
and the NLO contributions are
\begin {subequations}
\label {eq:NLOrates}
\begin {align}
   \left[ \frac{d\Gamma}{dx} \right]^\NLO_{\uqq}
  &\equiv
   \left[ \Delta \frac{d\Gamma}{d\xq} \right]_{q \to qg}^{\rm NLO}
   +
   \int_0^{1-\xq} d\xQ \>
   \left[ \Delta \frac{d\Gamma}{d\xq\,d\xQ} \right]_{q \to q\Q\Qbar}
  && \mbox{with $\xq=x$,}
\label {eq:NLOqq}
\\[15pt]
   \left[ \frac{d\Gamma}{dx} \right]^\NLO_{\uqQ}
  &\equiv
   \int_0^{1-\xQ} d\xq \>
   \left[ \Delta \frac{d\Gamma}{d\xq\,d\xQ} \right]_{q \to q\Q\Qbar}
  && \mbox{with $\xQ=x$,}
\label{eq:NLOeE}
\\[15pt]
   \left[ \frac{d\Gamma}{dx} \right]^\NLO_{\uqQbar}
  &\equiv
   \int_0^{1-\xQbar} d\xq \>
   \biggl(
   \left[ \Delta \frac{d\Gamma}{d\xq\,d\xQ} \right]_{q \to q\Q\Qbar}
   \biggr)_{\!\!\xQ = 1-\xq-\xQbar}
  && \mbox{with $\xQbar=x$,}
\\[15pt]
   \left[ \frac{d\Gamma}{dx} \right]^\NLO_{\uqg}
  &\equiv
   \left[ \Delta \frac{d\Gamma}{d\xq} \right]_{q \to qg}^{\rm NLO}
  && \mbox{with $\xq=1-x$,}
\\[15pt]
   \left[ \frac{d\Gamma}{dx} \right]^\NLO_{\ugQ}
  &\equiv
   \left[ \Delta \frac{d\Gamma}{d\xQ} \right]_{g \to\Q\Qbar}^{\rm NLO}
  && \mbox{with $\xQ=x$.}
\end {align}
\end {subequations}
\vspace{4pt}


\subsection{Numerics and fits for net rates}

\subsubsection{Basic net rates}

The NLO net rates are computed using (\ref{eq:NLOrates}) and the
formulas of appendix \ref{app:summary}.  Since the latter
formulas are complicated, and the numerical integrals
involved are computationally expensive, we then approximate the
results for the net rates
by reasonably accurate fits to relatively simple analytic
functions of $x$.  Those fit functions may
then be used for numerically efficient calculations of shower
development.

Following ref.\ \cite{qedNfenergy}, we find it easier to first
transform the NLO net rates $[ d\Gamma/dx ]^\NLO_\uij$ into
smoother functions $f_\uij(x)$ before fitting them.
We find it easiest to factor out the power-law behavior of the
rates as $x{\to}0$ and $x{\to}1$ and also to remove some of
the logarithmic dependence.  Specifically, we define
the $f_\uij(x)$ by
\begin {equation}
  \left[ \frac{d\Gamma}{dx} \right]^\NLO_\uij =
  L_\uij(x,\mu) + f_\uij(x) \, R_\uij(x) ,
\label{eq:fdef}
\end {equation}
where
\begin {subequations}
\begin {align}
  R_\uqq(\xq) &\equiv
    \xq^{-1/2} (1{-}\xq)^{-3/2} \,
    \frac{\Nf\tF\CF\alphas^2}{2\pi} \sqrt{ \frac{\qhatF}{E} } \,,
\label{eq:Rqq}
\\
  R_\uqQ(\xQ) &\equiv
    \xQ^{-3/2} (1{-}\xQ)^{+1/2} \,
    \frac{\Nf\tF\CF\alphas^2}{2\pi} \sqrt{ \frac{\qhatF}{E} } \,,
\label{eq:RqQ}
\\
  R_\uqQbar(\xQbar) &\equiv R_\uqQ(\xQbar) ,
\\
  R_\uqg(\xg) &\equiv R_\uqq(1{-}\xg) ,
\\
  R_\ugQ(\xQ) &\equiv
    \xQ^{-1/2} (1{-}\xQ)^{-1/2} \,
    \frac{\Nf^2\tF^2\alphas^2}{2\pi} \sqrt{ \frac{\qhatF}{E} }
\end {align}
\end {subequations}
and
\begin {subequations}
\label {eq:L}
\begin {align}
  & L_\uqq(\xq,\mu) \equiv
    - \frac{\beta_0\alphas}{2}
    \left[ \frac{d\Gamma}{d\xq} \right]_\LO^{q\to qg}
    \ln\left( \frac{ \mu^2 }{ \sqrt{\frac{(1{-}\xq)\qhatF E}{\xq}} } \right) ,
\label{eq:Lqq}
\\
  & L_\uqQ \equiv L_\uqQbar \equiv 0 ,
\\
  & L_\uqg(\xg,\mu) \equiv L_\uqq(1{-}\xg,\mu) ,
\\
  & L_\ugQ(\xQ,\mu) \equiv
    - \frac{\beta_0\alphas}{2}
    \left[ \frac{d\Gamma}{d\xQ} \right]_\LO^{g\to\Q\Qbar}
    \ln\left( \frac{ \mu^2 }{ \sqrt{\qhatF E} } \right) .
\label {eq:LPH}
\end {align}
\end {subequations}
Above, $\beta_0$ is the coefficient of the 1-loop renormalization group
$\beta$ function for $\alphas$ in the large-$\Nf$ limit,
\begin {equation}
   \beta_0 = \frac{2\tF\Nf}{3\pi} \,.
\label {eq:beta0}
\end {equation}

As in the previous QED calculations of ref.\ \cite{qedNfenergy},
we shall see shortly that the $L$'s above do not capture all of the
logarithmic dependence of the net rates on $x$.%
\footnote{
  Specifically, see the additional logarithmic dependence in
  eq.\ (\ref{eq:fqg}).
}
Our particular choice of $x$ dependence (or lack of it) inside the
logarithms of (\ref{eq:Lqq}) and (\ref{eq:LPH}) is just a matter of
convention for our definition (\ref{eq:fdef}) for $f_\uij(x)$.
Readers need not ponder too deeply the logic of our choices for
what logarithmic $x$ dependence to separate from the $f$'s into
the $L$'s; the choices do not affect final results except
to the extent that it makes it a little easier to find good fit
functions to the numerical data points for the $f$'s.
Historically, our particular choice of $L$'s
came from a combination of guesses made early in our work combined with
some choices to help make $f_\uij$'s smooth enough to more easily
find good fits.

\begin {figure}[t]
\begin {center}
  \includegraphics[scale=0.5]{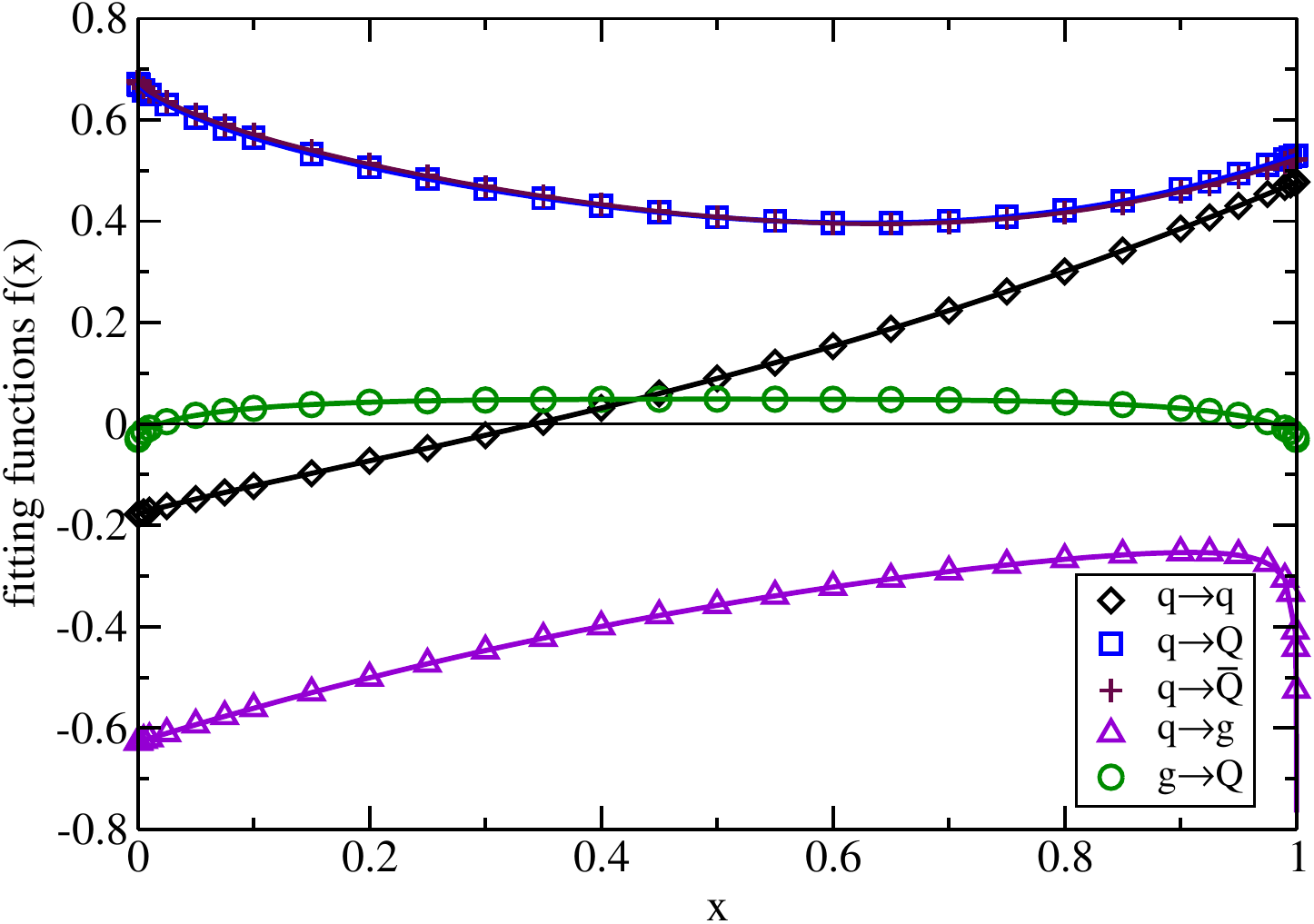}
  \caption{
     \label{fig:dGnet}
     Plots of numerically-computed data points (table \ref{tab:dGnet})
     and fits (\ref{eq:fits}) for
     the functions $f_\uij(x)$ defined by (\ref{eq:fdef}).
     The $q{\to}\Q$ and $q{\to}\Qbar$ points are almost but
     (importantly) not quite on top
     of each other.
  }
\end {center}
\end {figure}

With these definitions, table \ref{tab:dGnet} and figure \ref{fig:dGnet}
present our numerical results for the functions $f_\uij(x)$,
from which our original numerical results for NLO net rates
can be recovered using (\ref{eq:fdef}).

\begin {table}[tp]

\setlength{\tabcolsep}{7pt}
\begin{center}
\begin{tabular}{lrrrrr}
\hline
\hline
  \multicolumn{1}{c}{$x$} &
  \multicolumn{1}{c}{$f_\uqq$} &
  \multicolumn{1}{c}{$f_\uqQ$} &
  \multicolumn{1}{c}{$f_\uqQbar$} &
  \multicolumn{1}{c}{$f_\uqg$} &
  \multicolumn{1}{c}{$f_\ugQ$} \\[12pt]
\hline
  0.0001 & -0.1786 & 0.6710 & 0.6759 & -0.6274 & -0.0304 \\
  0.0005 & -0.1777 & 0.6694 & 0.6743 & -0.6272 & -0.0272 \\
  0.001 & -0.1773 & 0.6678 & 0.6727 & -0.6269 & -0.0249 \\
  0.005 & -0.1744 & 0.6586 & 0.6637 & -0.6241 & -0.0151 \\
  0.01 & -0.1711 & 0.6500 & 0.6552 & -0.6206 & -0.0082 \\
  0.025 & -0.1620 & 0.6301 & 0.6356 & -0.6103 & 0.0045 \\
  0.05 & -0.1482 & 0.6044 & 0.6104 & -0.5934 & 0.0171 \\
  0.075 & -0.1351 & 0.5832 & 0.5895 & -0.5770 & 0.0250 \\
  0.1 & -0.1223 & 0.5647 & 0.5712 & -0.5609 & 0.0308 \\
  0.15 & -0.0975 & 0.5331 & 0.5396 & -0.5301 & 0.0384 \\
  0.2 & -0.0728 & 0.5065 & 0.5127 & -0.5009 & 0.0430 \\
  0.25 & -0.0478 & 0.4835 & 0.4891 & -0.4732 & 0.0457 \\
  0.3 & -0.0222 & 0.4633 & 0.4681 & -0.4471 & 0.0473 \\
  0.35 & 0.0041 & 0.4457 & 0.4496 & -0.4225 & 0.0482 \\
  0.4 & 0.0315 & 0.4305 & 0.4334 & -0.3995 & 0.0487 \\
  0.45 & 0.0600 & 0.4178 & 0.4197 & -0.3780 & 0.0490 \\
  0.5 & 0.0898 & 0.4078 & 0.4087 & -0.3578 & 0.0490 \\
  0.55 & 0.1210 & 0.4007 & 0.4006 & -0.3391 & 0.0490 \\
  0.6 & 0.1536 & 0.3969 & 0.3959 & -0.3217 & 0.0487 \\
  0.65 & 0.1879 & 0.3967 & 0.3947 & -0.3057 & 0.0482 \\
  0.7 & 0.2238 & 0.4005 & 0.3977 & -0.2912 & 0.0473 \\
  0.75 & 0.2614 & 0.4087 & 0.4050 & -0.2782 & 0.0457 \\
  0.8 & 0.3008 & 0.4217 & 0.4173 & -0.2671 & 0.0430 \\
  0.85 & 0.3421 & 0.4399 & 0.4348 & -0.2583 & 0.0384 \\
  0.9 & 0.3853 & 0.4638 & 0.4580 & -0.2536 & 0.0308 \\
  0.925 & 0.4076 & 0.4780 & 0.4719 & -0.2541 & 0.0250 \\
  0.95 & 0.4304 & 0.4937 & 0.4873 & -0.2589 & 0.0171 \\
  0.975 & 0.4537 & 0.5110 & 0.5044 & -0.2748 & 0.0045 \\
  0.99 & 0.4680 & 0.5223 & 0.5154 & -0.3057 & -0.0082 \\
  0.995 & 0.4728 & 0.5261 & 0.5193 & -0.3338 & -0.0151 \\
  0.999 & 0.4766 & 0.5293 & 0.5224 & -0.4081 & -0.0249 \\
  0.9995 & 0.4771 & 0.5296 & 0.5228 & -0.4424 & -0.0272 \\
  0.9999 & 0.4775 & 0.5299 & 0.5230 & -0.5249 & -0.0304 \\
\hline
\hline
\end{tabular}
\end{center}
\caption{
   \label{tab:dGnet}
   Results for the functions $f_\uij(x)$ extracted from numerical
   computation of NLO net rates (\ref{eq:NLOrates}) using the
   explicit formulas of appendix \ref{app:summary}.
   Numerical results for the rates were translated into numerical
   results for $f_\uij$ using the definition (\ref{eq:fdef}) of
   the $f_\uij$.
}
\end{table}

In large-$\Nf$ QED, we found in ref.\ \cite{qedNfenergy} that
$f_{\underline{e\to\E}}(x) = f_{\underline{e\to\Ebar}}(x)$.
Table \ref{tab:dGnet} shows that in large-$\Nf$ QCD, there is instead
a small difference between $f_\uqQ(x)$ and $f_\uqQbar(x)$.
This difference, and its persistence in the limit of small $x$,
turns out to be the source of the feature in the last row
of table \ref{tab:chi2} that characteristics of the fermion-number
deposition distribution are infrared unsafe for large-$\Nf$ QCD.
But we save that discussion for later.

Our fits to the numerical data of table \ref{tab:dGnet} are
\begin {subequations}
\label {eq:fits}
\begin {multline}
  f_\uqq(x) =
    0.55622 - 1.17795\,x + 0.49222\,x^2 - 0.61229\,x^3
\\
    + 0.02989\,x^{1/2} + 0.02248\,x^{3/2} + 1.16702\,x^{5/2}
\\
    - 0.00178\,(1{-}x)^{1/2} - 0.16099\,(1{-}x)^{3/2} - 0.57232\,(1{-}x)^{5/2}
  ,
\label {eq:fqq}
\end {multline}
\begin {multline}
  f_\uqQ(x) =
    0.99693 - 1.93966\,x - 1.67427\,x^2 + 0.73141\,x^3
\\
    - 0.11846\,x^{1/2} + 2.41297\,x^{3/2} + 0.12113\,x^{5/2}
\\
    - 0.00132\,(1{-}x)^{1/2} - 0.14081\,(1{-}x)^{3/2} - 0.18232\,(1{-}x)^{5/2}
  ,
\label {eq:fqQ}
\end {multline}
\begin {multline}
  f_\uqQbar(x) =
    0.99862 - 1.91570\,x - 1.99018\,x^2 + 0.58225\,x^3
\\
    - 0.11797\,x^{1/2} + 2.44638\,x^{3/2} + 0.51974\,x^{5/2}
\\
    - 0.00119\,(1{-}x)^{1/2} - 0.13183\,(1{-}x)^{3/2} - 0.18823\,(1{-}x)^{5/2}
  ,
\label {eq:fqQbar}
\end {multline}
\begin {multline}
  f_\uqg(x) =
    \tfrac{1}{6\pi} \ln (1{-}x)
    - 4.00177 + 8.58463\,x - 4.00386\,x^2 + 1.63579\,x^3
\\
    - 0.00727\,x^{1/2} - 0.50996\,x^{3/2} - 1.73122\,x^{5/2}
\\
    - 0.24648\,(1{-}x)^{1/2} + 1.19122\,(1{-}x)^{3/2} + 2.42961\,(1{-}x)^{5/2}
  ,
\label {eq:fqg}
\end {multline}
\begin {multline}
  f_\ugQ(x) =
    - 0.03304
    + 0.26349\,\bigl(x(1{-}x)\bigr)^{1/2}
    - 0.11723\,x(1{-}x)
\\
    - 0.18078\,\bigl(x(1{-}x)\bigr)^{3/2}
    + 0.03594\,\bigl(x(1{-}x)\bigr)^{2}
  .
\label {eq:fgQ}
\end {multline}
\end {subequations}
The fits (\ref{eq:fits}) match the original numerical data points of
table \ref{tab:dGnet} to better than 0.0003 absolute error.

The $\ln(1{-}x)$ term in $f_\uqg(x)$ is the same as it was for large-$\Nf$ QED
\cite{qedNfenergy}.  However, large-$\Nf$ QED had a $\ln(1{-}x)$ term
in $f_{\underline{e\to e}}(x)$ and a $\ln x$ term in
$f_{\underline{e\to\E}}(x) = f_{\underline{e\to\Ebar}}(x)$ that do not appear
in large-$\Nf$ QCD.  The logarithms or lack of them
are discussed in appendix \ref{app:logs} and section \ref{sec:why}.


\subsubsection{Decomposition into real and virtual parts}

Some readers may be interested in how the NLO net rate
$[d\Gamma/dx]^\NLO_\uqq$ splits into real and virtual
processes.
Following ref.\ \cite{qedNfenergy}, we split (\ref{eq:NLOqq})
into the sum of (i) a real double splitting
($q \to qg \to q\Q\Qbar$) piece,
\begin {equation}
  \int_0^{1-\xq} d\xQ \>
    \left[ \Delta \frac{d\Gamma}{d\xq\,d\xQ} \right]_{q \to q\Q\Qbar}
    = 
    f_\qq^\real(x) \, R_\uqq(x)
\label {eq:feeRealDef}
\end {equation}
and (ii) the virtual correction to single splitting $q \to qg$,
\begin {equation}
  \left[ \Delta \frac{d\Gamma}{d\xq} \right]_{q \to qg}^{\rm NLO}
    =
    L_\uqq(x,\mu) + f_\qq^\virt(x) \, R_\uqq(x) .
\end {equation}
Then
\begin {subequations}
\label {eq:fdecomposeAll}
\begin {equation}
   f_\uqq(x) = f_\qq^\virt(x) + f_\qq^\real(x)
\label {eq:fdecompose0}
\end {equation}
with the pieces related to other $f_\uij$ by
\begin {equation}
   f_\qq^\virt(x) = f_\uqg(1{-}x)
   \quad \mbox{and} \quad
   f_\qq^\real(x) = f_\uqq(x) - f_\uqg(1{-}x) .
\label {eq:fdecompose}
\end {equation}
\end {subequations}
From (\ref{eq:fqq}) and (\ref{eq:fqg}), the corresponding fits are
\begin {subequations}
\label {eq:fitrealvirt}
\begin {multline}
  f_\qq^\real(x) =
    - \tfrac{1}{6\pi} \ln x
    - 1.65857 + 4.30633\,x - 0.41129\,x^2 + 1.02350\,x^3
\\
    + 0.27637\,x^{1/2} - 1.16874\,x^{3/2} - 1.26259\,x^{5/2}
\\
    + 0.00549\,(1{-}x)^{1/2} + 0.34897\,(1{-}x)^{3/2} + 1.15890\,(1{-}x)^{5/2}
  ,
\label {eq:fqqReal}
\end {multline}
\begin {multline}
  f_\qq^\virt(x) =
    \tfrac{1}{6\pi} \ln x
    + 2.21479 - 5.48428\,x + 0.90351\,x^2 - 1.63579\,x^3
\\
    - 0.24648 \,x^{1/2} + 1.19122\,x^{3/2} + 2.42961\,x^{5/2}
\\
    - 0.00727\,(1{-}x)^{1/2} - 0.50996\,(1{-}x)^{3/2} - 1.73122\,(1{-}x)^{5/2}
  .
\label {eq:fqqVirt}
\end {multline}
\end {subequations}


\subsection{Choice of renormalization scale}
\label{sec:mu}

For democratic splittings, the natural choice of renormalization scale
is given parametrically by the typical scale of relative transverse momenta in
a leading-order splitting process, which is
$\mu \sim (\qhat E)^{1/4}$ \cite{qedNf,finale,finale2}
in the case of an infinite medium considered here.
Following ref.\ \cite{qedNfenergy}, our canonical choice of renormalization
scale (and the one used to preview results in table \ref{tab:chi2})
is therefore
\begin {equation}
   \mu \propto (\qhatF E)^{1/4} ,
\label {eq:muE}
\end {equation}
where $E$ is the energy of the parent to the splitting.
Our results for $\qhat$-insensitive quantities like $\sigma/\lstop$
turn out not to
depend on the exact choice of the proportionality constant in
(\ref{eq:muE}), as long as it is truly a constant.

However, as discussed in ref.\ \cite{qedNfenergy}, it is less clear
what the best choice of renormalization scale should be when one of
the daughters is soft.  One might consider the total transverse momentum
kick during a typical formation time $t_\form$, which would be
\begin {equation}
   \mu \sim \Delta p_\perp \sim \sqrt{\qhat t_\form} .
\label {eq:mux0}
\end {equation}
For leading-order splittings,
the formation time scale is determined by the
frequency scale (\ref{eq:Omega0B}) as $t_{\rm form} \sim 1/|\Omega_0|$.
In the particular case of QCD, that's parametrically equivalent to
\begin {equation}
   t_{\rm form}
   \sim \min\left( \sqrt{\frac{xE}{\qhatF}} ,
                   \sqrt{\frac{(1{-}x)E}{\qhatF}} \right)
   \sim \sqrt{\frac{x(1{-}x)E}{\qhatF}}
\label {eq:tformx}
\end {equation}
for both $q \to qg$ and $g \to\Q\Qbar$, where $x$ is the momentum
fraction of either daughter.%
\footnote{
  Eq.\ (\ref{eq:tformx}) is also parametrically the formation time for
  $g\to gg$, but that's irrelevant for $\Nf \gg \Nc$.
}
So (\ref{eq:mux0}) suggests the choice
\begin {equation}
  \mu \propto \bigl( x(1{-}x)\qhatF E \bigr)^{1/4} .
\label {eq:mux}
\end {equation}

A different plausible choice for an $x$-dependent renormalization scale
would be to choose the typical invariant mass of the two daughters
of the splitting, which is equivalent to $2|\vec p_{\rm COM}|^2$ where
$\pm\vec p_{\rm COM}$ are the 3-momenta of the daughters in their
center-of-momentum frame.  Ref.\ \cite{qedNfenergy} showed that this choice
is parametrically
\begin {equation}
   \mu^2 \sim |\vec p_{\rm COM}|^2 \sim \frac{E}{t_\form} .
\end {equation}
For QCD, this suggests the choice
\begin {equation}
   \mu \propto
   \left( \frac{\qhatF E}{x(1{-}x)} \right)^{1/4} .
\label {eq:mualt}
\end {equation}

It's useful to use these alternatives to get an idea of how sensitive our
results are to such choices.  Table \ref{tab:chi2more} expands on the
$\Nf{\gg}\Nc{\gg}1$ QCD results of table \ref{tab:chi2} to compare
results for all three choices of $\mu$ introduced above.
Because our canonical choice (\ref{eq:muE}) of renormalization scale $\mu$
happens to be the geometric mean of the other two, and rates
(\ref{eq:fdef}) are only logarithmically dependent on the explicit value
of $\mu$, the results listed for $\mu \propto [x(1{-}x)\qhatF E]^{1/4}$
and $\mu \propto [\qhatF E/x(1{-}x)]^{1/4}$ symmetrically bracket
(within round-off error) the results for our canonical choice
$\mu \propto (\qhatF E)^{1/4}$.  Following ref. \cite{qedNfenergy},
we therefore need only show one of the two alternatives in the future
in order to convey the sensitivity of our results to these choices of
renormalization scale.

\begin {table}[tp]

\setlength{\tabcolsep}{7pt}
\begin {center}
\begin{tabular}{rccc}
\hline
\hline
  & \multicolumn{3}{c}{overlap correction to $\sigma/\lstop$}
\\
\cline{2-4}
  & $\mu\propto\bigl(\qhat E/x(1{-}x)\bigr)^{1/4}$
  & $\mu\propto(\qhat E)^{1/4}$
  & $\mu\propto\bigl(x(1{-}x)\qhat E\bigr)^{1/4\strut}$
\\
\hline
  energy, $g$
               & $-0.70\%\times\Nf\alphas$
               & $-0.40\%\times\Nf\alphas$
               & $-0.11\%\times\Nf\alphas$
\\
  energy, $q$
               & $-1.03\%\times\Nf\alphas$
               & $-0.53\%\times\Nf\alphas$
               & $-0.03\%\times\Nf\alphas$
\\
  initial flavor, $q$
               & $+0.72\%\times\Nf\alphas$
               & $+1.05\%\times\Nf\alphas$
               & $+1.38\%\times\Nf\alphas$
\\
\hline
\hline
\end{tabular}
\end {center}
\caption{%
\label{tab:chi2more}%
  The $\Nf{\gg}\Nc{\gg}1$ QCD column of table \ref{tab:chi2} but here
  expanded to show the dependence on different plausible choices of
  renormalization scale.
  We have included an extra significant digit
  in our results to make clear that the values in the first and third
  columns of numbers symmetrically bracket those in the second column
  (within round-off error).
}
\end{table}

Though the ambiguity of how best to choose the
renormalization scale introduces a significant {\it relative} uncertainty
in the exact percentage size of overlap effects, as shown in table
\ref{tab:chi2more}, the percentages remain very small for any reasonable size
of $\Nf\alphas$, and so our qualitative
conclusion that $\qhat$-insensitive overlap effects are small is unaffected.


\section{Numerical results for moments of shape functions}
\label {sec:moments}

\subsection {Initial-flavor deposition}
\label {sec:flavor}

Consider a shower initiated by a high-energy quark with energy $E_0$.
As explained in the sections \ref{sec:preview} and \ref{sec:RateNotation},
in the large-$\Nf$ limit
we may unambiguously follow the fate of that initial quark through the
shower and so construct a statistically-averaged distribution,
call it $\rho(z)$,
for where that particular quark
is eventually deposited in the medium.
In the language of conserved quantities, this is equivalent to studying
initial-flavor deposition since the chance that
a pair-produced quark has the same flavor as the initial quark vanishes
in the large-$\Nf$ limit.  So we may view $\rho(z)$ as the
deposition distribution of the particular quark flavor of
the quark that initiated the shower.

In large-$\Nf$ QED, following the heir of the original electron also gave
the charge deposition distribution.
That's because the net rates
$[d\Gamma/dx]_{\underline{e\to\E}}$ and $[d\Gamma/dx]_{\underline{e\to\Ebar}}$
happen to be equal in large-$\Nf$ QED, and
so the $\E$ and $\Ebar$ in overlapping $e \to e\gamma \to e\E\Ebar$ are
produced with the same distribution of energy and so
subsequently deposit charge in the exact same way (statistically)
except for sign.  Their contributions to total charge deposition therefore
exactly cancel.  Only the original electron, tracked by
$[d\Gamma/dx]_{\underline{e \to e}}$, contributed to charge deposition.
The situation in large-$\Nf$ QCD is different because
$[d\Gamma/dx]_{\underline{q\to\Q}}$ and $[d\Gamma/dx]_{\underline{q\to\Qbar}}$ are
\textit{not} equal.

For initial-flavor deposition, however, we need follow \textit{only}
the heir of the initial quark through the shower for
large-$\Nf$ QCD as well as large-$\Nf$ QED, and so we only
need the rate $[d\Gamma/dx]_\uqq$.
We can then carry over the
formalism that was used for large-$\Nf$ QED charge deposition in
refs.\ \cite{qedNfstop,qedNfenergy}.  A simple recursion relation
was found for moments of the charge distribution, which we simply quote here
after translating electrons and charge to quarks and initial flavor:
\begin {subequations}
\label {eq:znMasterRho}
\begin {equation}
   \langle z^n \rangle_\rho =
   \frac{ n \langle z^{\,n-1} \rangle_\rho }
        { \Avg_\uqq[ 1{-}x^{n/2} ] }
   \,,
\label {eq:znrho}
\end {equation}
where
\begin {equation}
   \Avg_{i\to j}[g(x)] \equiv
   \int_0^1 dx \> \biggl[\frac{d\Gamma}{dx}(E_0,x)\biggr]^\net_{i\to j} g(x) .
\end {equation}
\end {subequations}
We then expand the relevant net rate as
\begin {equation}
  \biggl[\frac{d\Gamma}{dx}(E_0,x)\biggr]^\net_\uqq
  =
  \biggl[\frac{d\Gamma}{dx}(E_0,x)\biggr]^\LO_\uqq +
  \biggl[\frac{d\Gamma}{dx}(E_0,x)\biggr]^\NLO_\uqq ,
\label {eq:rateexpand}
\end {equation}
similarly expand moments of the initial deposition distribution
that we want to calculate as
\begin {equation}
   \langle z^n \rangle_\rho \simeq
   \langle z^n \rangle^\LO_\rho + \delta \langle z^n \rangle_\rho \,,
\label {eq:znexpand}
\end {equation}
and then expand the recursion relation (\ref{eq:znrho}) into recursion
relations for $\langle z^n\rangle_\rho^\LO$ and its NLO correction
$\delta\langle z^n \rangle_\rho$.  Details may be found in
ref.\ \cite{qedNfenergy}.%
\footnote{
  See in particular section 5.3 of ref.\ \cite{qedNfenergy}.
}

Computing leading-order moments and their NLO corrections due to overlapping
formation times then simply involves using the recursion relation,
using the LO rate (\ref{eq:LOrateq}) and its NLO correction from
(\ref{eq:fdef}) and (\ref{eq:fqq}), and numerically calculating
some integrals.  Our results for $\Nf{\gg}\Nc{\gg}1$ QCD are
given in table \ref{tab:momentsrho} in units of $\ell_0^n$ where
\begin {equation}
   \ell_0 \equiv \frac{1}{\CF\alphas} \sqrt{ \frac{E_0}{\qhatF} } \,.
\label {eq:ell0}
\end {equation}

\begin {table}[t]

\setlength{\tabcolsep}{7pt}
\begin{center}
\begin{tabular}{lccc}
\hline\hline
  \multicolumn{1}{c}{$z^n$}
  & \multicolumn{1}{c}{$\langle z^n\rangle_\rho^\LO$}
  & \multicolumn{2}{c}{$\delta\langle z^n \rangle_\rho$}
 \\
\cline{3-4}
  & &
  \multicolumn{1}{c}{$\mu = (\qhatF E)^{1/4}$}
  &
  \multicolumn{1}{c}{
    $\mu = \bigl(x(1{-}x)\qhatF E\bigr)^{1/4\strut}$
  }
  \\[6pt]
\cline{2-4}
  & \multicolumn{3}{c}{in units of $\ell_0^{\kern1pt n}$} \\
\hline
$\langle z\rangle$   & 2.0402 & -0.04523\,$\tF\Nf\alphas$
    & -0.3665\,$\tF\Nf\alphas$ \\
$\langle z^2\rangle$ & 4.9333 & -0.1864\,$\tF\Nf\alphas$
    & -1.7299\,$\tF\Nf\alphas$ \\
$\langle z^3\rangle$ & 13.377 & -0.6591\,$\tF\Nf\alphas$
    & -6.9428\,$\tF\Nf\alphas$ \\
$\langle z^4\rangle$ & 39.559 & -2.2706\,$\tF\Nf\alphas$
    & -27.153\,$\tF\Nf\alphas$ \\
\hline\hline
\end{tabular}
\end{center}
\caption{
   \label{tab:momentsrho}
   Expansions (\ref{eq:znexpand}) of moments $\langle z^n \rangle_\rho$
   of the initial-flavor deposition distribution $\rho(z)$
   for renormalization scale choices (\ref{eq:muE}) and (\ref{eq:mux}).
   The unit $\ell_0$ is defined by (\ref{eq:ell0}), and
   $\tF = \frac12$ as in (\ref{eq:largeNcCoeffs}).
}
\end{table}

However, our interest is in $\qhat$-insensitive quantities like
$\sigma/\lstop$.  Following refs.\ \cite{finale,finale2,qedNfenergy},
we can more generally consider what we call the shape
$S_\rho(Z)$ of the initial-flavor deposition distribution $\rho(z)$,
defined by rescaling $\rho(z)$ to units where the stopping length
and the area under the curve are both one.  That is
\begin {equation}
   S_\rho(Z) \equiv \lstoprho \, \rho(Z\lstoprho)
\end {equation}
where $Z$ is position in units of
$\lstoprho \equiv \langle z \rangle_\rho$.
The shape and all of its moments are $\qhat$-insensitive and so
good tests of the size of overlap corrections that cannot
be absorbed into $\qhat$.
The numerical values in table \ref{tab:momentsrho} can be used
to compute the moments of $S_\rho$ expanded to first order in
NLO corrections.
Results are tabulated in table \ref{tab:shaperho} for moments
$\langle Z^n \rangle$, reduced moments
$\mu_{n,S} \equiv \bigl\langle (Z-\langle Z\rangle)^n \bigr\rangle$
and cumulants $k_{n,S}$ of the shape $S_\rho$.
However, for the sake of comparing apples to apples, we have
followed refs.\ \cite{finale2,qedNfenergy} by first converting all of
these quantities into corresponding lengths:
$\langle Z^n\rangle^{1/n}$, $\mu_{n,S}^{1/n}$, and $k_{n,S}^{1/n}$.
For each such quantity $Q$, the table gives the LO value $Q_\LO$,
the NLO correction $\delta Q$ when expanded to first order,
and the relative size of overlap corrections
\begin {equation}
   \chi\alphas \equiv \frac{\delta Q}{Q_\LO} \,.
\end {equation}
The row for $\mu_2^{1/2}$, which is the width $\sigma_S$ of the shape
function, is the ratio $\sigma/\lstop$ previewed in the introduction,
where the ``overlap correction'' quoted there for
$\Nf{\gg}\Nc{\gg}1$ QCD corresponds to the value of
$\chi\alphas$ here for $\mu \propto (\qhat E)^{1/4}$.

\begin {table}[t]
\begin {center}

\setlength{\tabcolsep}{4pt}
\begin{tabular}{lcrrcrr}
\hline\hline
  \multicolumn{1}{l}{quantity $Q$}
  & \multicolumn{1}{c}{$Q_\rho^\LO$}
  & \multicolumn{1}{c}{$\delta Q_\rho$}
  & \multicolumn{1}{c}{$\chi\alphas$}
  & 
  & \multicolumn{1}{c}{$\delta Q_\rho$}
  & \multicolumn{1}{c}{$\chi\alphas$}
 \\
\cline{3-4}\cline{6-7}
  &&
  \multicolumn{2}{c}{$\mu \propto (\qhat E)^{1/4}$}
  &&
  \multicolumn{2}{c}{
    $\mu \propto \bigl(x(1{-}x)\qhat E\bigr)^{1/4\strut}$
  }
  \\[6pt]
\hline
$\langle Z \rangle$
   & $1$ \\
$\langle Z^2 \rangle^{1/2}$
   & $1.0887$
   & $0.0036\,\tF\Nf\alphas$
   & $0.0033\,\tF\Nf\alphas$
   &
   & $0.0047\,\tF\Nf\alphas$
   & $0.0043\,\tF\Nf\alphas$ \\
$\langle Z^3 \rangle^{1/3}$
   & $1.1636$
   & $0.0067\,\tF\Nf\alphas$
   & $0.0057\,\tF\Nf\alphas$
   &
   & $0.0077\,\tF\Nf\alphas$
   & $0.0066\,\tF\Nf\alphas$ \\
$\langle Z^4 \rangle^{1/4}$
   & $1.2293$
   & $0.0096\,\tF\Nf\alphas$
   & $0.0078\,\tF\Nf\alphas$
   &
   & $0.0099\,\tF\Nf\alphas$
   & $0.0080\,\tF\Nf\alphas$ \\[1pt]
\hline
$\mu_{2,S}^{1/2} {=} k_{2,{\rm S}}^{1/2} {=} \sigma_S$
   & $0.4304$
   & $0.0090\,\tF\Nf\alphas$
   & $0.0210\,\tF\Nf\alphas$
   &
   & $0.0119\,\tF\Nf\alphas$
   & $0.0276\,\tF\Nf\alphas$ \\
$\mu_{3,S}^{1/3} {=} k_{3,{\rm S}}^{1/3}$
   & $0.2691$
   & $0.0177\,\tF\Nf\alphas$
   & $0.0658\,\tF\Nf\alphas$
   &
   & $0.0032\,\tF\Nf\alphas$
   & $0.0118\,\tF\Nf\alphas$ \\
$\mu_{4,S}^{1/4}$
   & $0.5536$
   & $0.0139\,\tF\Nf\alphas$
   & $0.0251\,\tF\Nf\alphas$
   &
   & $0.0138\,\tF\Nf\alphas$
   & $0.0250\,\tF\Nf\alphas$ \\[2pt]
$(-k_{4,S})^{1/4}$
   & $0.3086$
   & $-0.0066\,\tF\Nf\alphas$
   & $-0.0214\,\tF\Nf\alphas$
   &
   & $0.0169\,\tF\Nf\alphas$
   & $0.0547\,\tF\Nf\alphas$ \\[2pt]
\hline\hline
\end{tabular}
\end{center}
\caption{
   \label{tab:shaperho}
   Expansions involving moments $\langle Z^n \rangle$, reduced moments
   $\mu_{n,S}$, and cumulants $k_{n,S}$ of the initial-flavor deposition
   shape function $S_\rho(Z)$,
   for renormalization scale choices (\ref{eq:muE}) and (\ref{eq:mux}).
   There are no NLO entries for $\langle Z \rangle$ because
   $\langle Z \rangle = 1$ and $\langle Z \rangle_\LO = 1$
   by definition of $Z \equiv z/\langle z\rangle$.
   The last row gives $(-k_{4,S})^{1/4}$ instead of $k_{4,S}^{1/4}$ because
   the leading-order 4th cumulant $k_{4,S}^\LO$
   turns out to be negative in this case.
}
\end{table}

We put the factor of $\CF$ into the definition (\ref{eq:ell0}),
and wrote the $\delta Q$ entries of the tables in
terms of $\tF \Nf\alphas$ instead of just $\Nf\alphas$, so that the
simple overall color factors are made explicit.
This means that the only difference
between the explicit numbers in the table here and those in the corresponding
table of ref.\ \cite{qedNfenergy} for large-$\Nf$ QED is the difference
in 3-body and 4-body frequencies $\Omega$ and normal modes, arising
from the different nature of interactions with the medium and their
correlations for
$\Nf{\gg}\Nc{\gg}1$ QCD vs.\ large-$\Nf$ QED.%
\footnote{
  There is also another way to put this.
  Imagine that the quarks all had the same
  electric charge, were splitting via gluon
  bremsstrahlung and quark pair production, but were getting their kicks from
  a QED medium instead of a QCD medium.
  The results for $\chi\alphas$ would be identical to
  the large-$\Nf$ QED tables of ref.\ \cite{qedNfenergy}
  except for replacing
  $\Nf\alpha$ there by $\tF\Nf\alphas$ and $\ell_0$ there by the
  definition (\ref{eq:ell0}) of $\ell_0$ here.
}
That difference arises because
(i) gluons have color and (ii) the correlations
of medium interactions were forced to be ``planar'' in large-$\Nc$ QCD,
in the sense of figs.\ \ref{fig:largeNc0shaded} and \ref{fig:largeNc}.

The conclusion to take away from the $\chi\alphas$ columns of
table \ref{tab:shaperho} is that the overlap corrections
reported for $\sigma/\lstop$ in table \ref{tab:chi2} are not merely isolated
examples of very small overlap corrections for a $\qhat$-insensitive quantity
in $\Nf{\gg}\Nc{\gg}1$ QCD.
All of the moments in table \ref{tab:shaperho} have small overlap corrections
for reasonable values of $\Nf\alphas$.

We content ourselves with having tested the size of overlap
corrections to various moments of $S_\rho(Z)$.  At the cost of
significantly more
numerical effort, one could more generally compute the overlap corrections
to the full functional form of $S_\rho(Z)$, as was done
by refs.\ \cite{finale,finale2}
for the energy deposition distribution in
$\Nf{=}0$, large-$\Nc$ QCD.
For now, we think that checking the moments
is an adequate test.


\subsection {Energy deposition}

Ref.\ \cite{qedNfenergy}
derived recursion relations for moments of
large-$\Nf$ energy deposition distributions,%
\footnote{
  Specifically, see section 6.3 of ref.\ \cite{qedNfenergy}.
}
which we also simply quote
here after translating electrons and photons to quarks and gluons.
Let $\eps_i(z)$ be the energy deposition distribution for a shower initiated
by a particle of type $i=q$ or $g$, and let $\langle z^n \rangle_{\eps,i}$
represent the moments of that distribution.  Then the
large-$\Nf$ recursion relations are
\begin {subequations}
\label {eq:znRecursion}
\begin {equation}
  \langle z^n \rangle_{\eps,q}
  =
  \frac{n M_{(n),gg}}{\det M_{(n)}}
  \,
  \langle z^{n-1} \rangle_{\eps,q}
\label {eq:zneRecursion}
\end {equation}
for quark-initiated showers
and a dependent result
\begin {equation}
  \langle z^n \rangle_{\eps,g}
  =
  - \frac{M_{(n),gq}}{M_{(n),gg}}
  \,
  \langle z^n \rangle_{\eps,q}
\label {eq:zngRecursion}
\end {equation}
\end {subequations}
for moments of gluon-initiated showers.
Above, the constants $M_{(n)}$ are defined by
\begin {align}
  M_{(n)} &\equiv
  \begin{pmatrix}
    M_{(n),qq} & M_{(n),qg} \\ M_{(n),gq} & M_{(n),gg}
  \end{pmatrix}
\nonumber\\
  &=
  \Avg_{\,q\to q\,{\rm or}\,\bar q}
    \begin{pmatrix}
       x{-}x^{1+\frac{n}{2}} & 0 \\ 0 & 0
    \end{pmatrix}
  +
  \Avg_\uqg
    \begin{pmatrix}
       x & - x^{1+\frac{n}{2}} \\ 0 & 0
    \end{pmatrix}
  +
  \Avg_{\,g\to q\,{\rm or}\,\bar q}
    \begin{pmatrix}
       0 & 0 \\ - x^{1+\frac{n}{2}} & x
    \end{pmatrix}
   ,
\label {eq:M}
\end {align}
where
\begin {subequations}
\begin {align}
   \left[ \frac{d\Gamma}{dx} \right]^\net_{q\to q\,{\rm or}\,\bar q}
   &\equiv
   \left[ \frac{d\Gamma}{dx} \right]^\net_\uqq
   +
   \left[ \frac{d\Gamma}{dx} \right]^\net_\uqQ
   +
   \left[ \frac{d\Gamma}{dx} \right]^\net_\uqQbar \,,
\label {eq:dGneteepm}
\\
   \left[ \frac{d\Gamma}{dx} \right]^\net_{g\to q\,{\rm or}\,\bar q}
   &\equiv
   \left[ \frac{d\Gamma}{dx} \right]^\net_\ugQ
   +
   \left[ \frac{d\Gamma}{dx} \right]^\net_\ugQbar
\label {eq:dGnetgam}
\end {align}
\end {subequations}
are defined as the sum of net rates to produce \textit{any} type of quark or
anti-quark daughter from the specified type of parent.
One then expands the net rates $[ d\Gamma/dx ]^\net_\uij$ and moments
$\langle z^n \rangle_{\eps,i}$ as LO+NLO, similar to
(\ref{eq:rateexpand}) and (\ref{eq:znexpand}), and then correspondingly
expands the recursion relations (\ref{eq:znRecursion}).

Our results for moments of $\eps_i(z)$ are given in table \ref{tab:momentseps}.
Using those, we then extract the various types of moments,
shown in table \ref{tab:shapeeps},
of the corresponding $\qhat$-insensitive shape functions
\begin {equation}
   S_{\eps,i}(Z) \equiv \frac{\lstop^{(\eps,i)}}{E_0} \, \eps_i(Z\lstop^{(\eps,i)}) ,
\end {equation}
where $\lstop^{(\eps,i)} \equiv \langle z \rangle_{\eps,i}$
and $E_0$ is the energy of the particle that initiated the shower.
Once again,
all of the moments in table \ref{tab:shapeeps} have small overlap corrections
for reasonable values of $\Nf\alphas$.

\begin {table}[t]

\setlength{\tabcolsep}{7pt}
\begin{center}
\begin{tabular}{lcrc}
\hline\hline
  \multicolumn{1}{c}{$z^n$}
  & \multicolumn{1}{c}{$\langle z^n\rangle_{\eps,i}^\LO$}
  & \multicolumn{2}{c}{$\delta\langle z^n \rangle_{\eps,i}$}
 \\
\cline{3-4}
  & &
  \multicolumn{1}{c}{$\mu = (\qhatF E)^{1/4}$}
  &
  \multicolumn{1}{c}{$\mu = \bigl(x(1{-}x)\qhatF E\bigr)^{1/4\strut}$}
 \\[6pt]
\cline{2-4}
  & \multicolumn{3}{c}{in units of $\ell_0^{\kern1pt n}$} \\
\hline
\multicolumn{4}{l}{\textbf{quark initiated ($i=q$):}}\\
$\langle z\rangle$   & 1.3703 & 0.0029\,$\tF\Nf\alphas$
                                         & -0.2232\,$\tF\Nf\alphas$ \\
$\langle z^2\rangle$ & 2.4734 & -0.0021\,$\tF\Nf\alphas$
                                         & -0.8066\,$\tF\Nf\alphas$ \\
$\langle z^3\rangle$ & 5.3271 & -0.0201\,$\tF\Nf\alphas$
                                         & -2.6085\,$\tF\Nf\alphas$ \\
$\langle z^4\rangle$ & 13.033 & -0.0676\,$\tF\Nf\alphas$
                                         & -8.5177\,$\tF\Nf\alphas$ \\
\hline
\multicolumn{4}{l}{\textbf{gluon initiated ($i=g$):}}\\
$\langle z\rangle$   & 1.2118 & -0.0041\,$\tF\Nf\alphas$
                                         & -0.1931\,$\tF\Nf\alphas$ \\
$\langle z^2\rangle$ & 2.0059 & -0.0224\,$\tF\Nf\alphas$
                                         & -0.6415\,$\tF\Nf\alphas$ \\
$\langle z^3\rangle$ & 4.0377 & -0.0726\,$\tF\Nf\alphas$
                                         & -1.9424\,$\tF\Nf\alphas$ \\
$\langle z^4\rangle$ & 9.3382 & -0.2115\,$\tF\Nf\alphas$
                                         & -6.0032\,$\tF\Nf\alphas$ \\
\hline\hline
\end{tabular}
\end{center}
\caption{
   \label{tab:momentseps}
   Like table \ref{tab:momentsrho} but showing moments $\langle z^n\rangle$
   for energy deposition
   instead of quark-initiated
   initial-flavor deposition.
   The unit $\ell_0$ is defined by (\ref{eq:ell0}),
   and $\tF=\tfrac12$ as in (\ref{eq:largeNcCoeffs}).
}
\end{table}

\begin {table}[t]
\begin {center}

\setlength{\tabcolsep}{3.2pt}
\begin{tabular}{lcrrcrr}
\hline\hline
  \multicolumn{1}{l}{quantity $Q$}
  & \multicolumn{1}{c}{$Q_{\eps,i}^\LO$}
  & \multicolumn{1}{c}{$\delta Q_{\eps,i}$}
  & \multicolumn{1}{c}{$\chi\alphas$}
  & 
  & \multicolumn{1}{c}{$\delta Q_{\eps,i}$}
  & \multicolumn{1}{c}{$\chi\alphas$}
 \\
\cline{3-4}\cline{6-7}
  &&
  \multicolumn{2}{c}{$\mu \propto (\qhat E)^{1/4}$}
  &&
  \multicolumn{2}{c}{$\mu \propto \bigl(x(1{-}x)\qhat E\bigr)^{1/4\strut}$}
  \\[6pt]
\hline
\multicolumn{4}{l}{\textbf{quark initiated ($i=q$):}}\\
$\langle Z \rangle$
   & $1$ \\
$\langle Z^2 \rangle^{1/2}$
   & $1.1477$
   & $-0.0029\,\tF\Nf\alphas$
   & $-0.0026\,\tF\Nf\alphas$
   &
   & $-0.0002\,\tF\Nf\alphas$
   & $-0.0002\,\tF\Nf\alphas$ \\
$\langle Z^3 \rangle^{1/3}$
   & $1.2746$
   & $-0.0043\,\tF\Nf\alphas$
   & $-0.0034\,\tF\Nf\alphas$
   &
   & $-0.0004\,\tF\Nf\alphas$
   & $-0.0003\,\tF\Nf\alphas$ \\
$\langle Z^4 \rangle^{1/4}$
   & $1.3866$
   & $-0.0048\,\tF\Nf\alphas$
   & $-0.0034\,\tF\Nf\alphas$
   &
   & $-0.0007\,\tF\Nf\alphas$
   & $-0.0005\,\tF\Nf\alphas$ \\[1pt]
\hline
$\mu_{2,S}^{1/2} {=} k_{2,{\rm S}}^{1/2} {=} \sigma_S$
   & $0.5633$
   & $-0.0060\,\tF\Nf\alphas$
   & $-0.0106\,\tF\Nf\alphas$
   &
   & $-0.0004\,\tF\Nf\alphas$
   & $-0.0007\,\tF\Nf\alphas$ \\
$\mu_{3,S}^{1/3} {=} k_{3,{\rm S}}^{1/3}$
   & $0.4913$
   & $-0.0011\,\tF\Nf\alphas$
   & $-0.0023\,\tF\Nf\alphas$
   &
   & $-0.0010\,\tF\Nf\alphas$
   & $-0.0021\,\tF\Nf\alphas$ \\
$\mu_{4,S}^{1/4}$
   & $0.7513$
   & $-0.0041\,\tF\Nf\alphas$
   & $-0.0055\,\tF\Nf\alphas$
   &
   & $-0.0010\,\tF\Nf\alphas$
   & $-0.0013\,\tF\Nf\alphas$ \\[2pt]
$k_{4,S}^{1/4}$
   & 0.3584$$
   & $0.0320\,\tF\Nf\alphas$
   & $0.0892\,\tF\Nf\alphas$
   &
   & $-0.0048\,\tF\Nf\alphas$
   & $-0.0135\,\tF\Nf\alphas$ \\[2pt]
\hline\hline
\multicolumn{4}{l}{\textbf{gluon initiated ($i=g$):}}\\
$\langle Z \rangle$
   & $1$ \\
$\langle Z^2 \rangle^{1/2}$
   & $1.1687$
   & $-0.0025\,\tF\Nf\alphas$
   & $-0.0022\,\tF\Nf\alphas$
   &
   & $-0.0007\,\tF\Nf\alphas$
   & $-0.0006\,\tF\Nf\alphas$ \\
$\langle Z^3 \rangle^{1/3}$
   & $1.3140$
   & $-0.0034\,\tF\Nf\alphas$
   & $-0.0026\,\tF\Nf\alphas$
   &
   & $-0.0013\,\tF\Nf\alphas$
   & $-0.0010\,\tF\Nf\alphas$ \\
$\langle Z^4 \rangle^{1/4}$
   & $1.4425$
   & $-0.0033\,\tF\Nf\alphas$
   & $-0.0023\,\tF\Nf\alphas$
   &
   & $-0.0020\,\tF\Nf\alphas$
   & $-0.0014\,\tF\Nf\alphas$ \\[1pt]
\hline
$\mu_{2,S}^{1/2} {=} k_{2,{\rm S}}^{1/2} {=} \sigma_S$
   & $0.6049$
   & $-0.0049\,\tF\Nf\alphas$
   & $-0.0081\,\tF\Nf\alphas$
   &
   & $-0.0013\,\tF\Nf\alphas$
   & $-0.0021\,\tF\Nf\alphas$ \\
$\mu_{3,S}^{1/3} {=} k_{3,{\rm S}}^{1/3}$
   & $0.5550$
   & $0.0001\,\tF\Nf\alphas$
   & $0.0003\,\tF\Nf\alphas$
   &
   & $-0.0024\,\tF\Nf\alphas$
   & $-0.0043\,\tF\Nf\alphas$ \\
$\mu_{4,S}^{1/4}$
   & $0.8192$
   & $-0.0019\,\tF\Nf\alphas$
   & $-0.0023\,\tF\Nf\alphas$
   &
   & $-0.0025\,\tF\Nf\alphas$
   & $-0.0031\,\tF\Nf\alphas$ \\[2pt]
$k_{4,S}^{1/4}$
   & $0.4697$
   & $0.0213\,\tF\Nf\alphas$
   & $0.0454\,\tF\Nf\alphas$
   &
   & $-0.0052\,\tF\Nf\alphas$
   & $-0.0110\,\tF\Nf\alphas$ \\[2pt]
\hline\hline
\end{tabular}
\end{center}
\caption{
   \label{tab:shapeeps}
   Expansions of moments $\langle Z^n \rangle$, reduced moments $\mu_{n,S}$,
   and cumulants $k_{n,S}$ of the energy deposition shape functions
   $S_{\eps,q}(Z)$ and $S_{\eps,g}(Z)$ for quark-initiated and
   gluon-initiated showers, respectively. 
   Like table \ref{tab:shaperho} but for energy deposition
   instead of quark-initiated
   initial-flavor deposition.
   (Unlike table \ref{tab:shaperho}, the leading-order 4th cumulant
   $k_{4,S}^\LO$ is positive here.)
}
\end{table}


\section {IR sensitivity of QCD fermion number deposition}
\label {sec:IRunsafe}

We will now explore what goes wrong if we try to use the same formalism
to explore quark-number stopping instead of initial-flavor or
energy stopping.  The same issue would also arise for electric charge
stopping in QCD.  (Indeed, if all quarks had the same electric charge,
then electric charge would simply be directly proportional to quark number.)

In the context of the infinite-size medium case we are studying,
the phrase ``infrared senstive'' in the following will mean
sensitive to radiation energies $\omega$ that are
parametrically small compared to the energy $E_0$ of the
shower.  More specifically, in the case of showering in a quark-gluon
plasma, it will mean sensitivity to radiation of energy
$\omega \sim T \ll E_0$,%
\footnote{
  More precisely, infrared sensitivity will refer to radiation with
  $\omega \sim [1+(\Nf/\Nc)]\, T \ll E_0$.
  See (\ref{eq:xIRwrapper}) and footnotes
  \ref{foot:xIR1} and \ref{foot:xIR2}.
}
which is so soft that those splittings cannot be treated in the $\qhat$
approximation.


\subsection {Recursion equation for moments}

In large-$\Nf$ QED, the contribution to electron-number
(or charge) deposition from pair-produced electrons and positrons
canceled.
Because
$[d\Gamma/dx]_\uqQ \not= [d\Gamma/dx]_\uqQbar$ in large-$\Nf$ QCD,
the analysis of quark-number stopping cannot just follow the
fate of the initiating quark using $[d\Gamma/dx]_\uqq$
but must also account for how
pair-produced quarks and anti-quarks deposit \textit{their}
quark number.
The derivation needed is similar but slightly different from
those made in ref.\ \cite{qedNfenergy} that resulted in
the recursion relations quoted in our (\ref{eq:znrho}) and
(\ref{eq:znRecursion}).  So we will take time here to explain
the differences.

Let $\phi_q(E_0,z)$ represent the quark-number deposition distribution
from a shower initiated by a quark with energy $E_0$.
By charge conjugation, the corresponding
distribution for an anti-quark initiated shower would be
\begin {equation}
  \phi_{\bar q}(E_0,z) = - \phi_q(E_0,z) .
\label {eq:phiqbar}
\end {equation}
Also by charge conjugation, $\phi_g(E_0,z) = 0$.
Following the logic of refs.\ \cite{qedNfstop,qedNfenergy},%
\footnote{
  See in particular sections 5.1--5.3 and 6.1 of ref.\ \cite{qedNfenergy}.
}
our starting equation is
\begin {multline}
  \phi_q(E,z+\Delta z)
  =
  [1 - \Gamma_q(E)\,\Delta z] \, \phi_q(E,z)
  + \int_0^1 dx \> \left[\frac{d\Gamma}{dx}(E,x)\right]^\net_\uqq
     \Delta z \, \phi_q(x E,z)
\\
  + \int_0^1 dx \> \left[\frac{d\Gamma}{dx}(E,x)\right]^\net_\uqQ
     \Delta z \, \phi_q(x E,z)
  + \int_0^1 dx \> \left[\frac{d\Gamma}{dx}(E,x)\right]^\net_\uqQbar
     \Delta z \, \phi_{\bar q}(x E,z)
\end {multline}
in the limit of small $\Delta z$.
The first term accounts for the chance that the quark does not
split in the first $\Delta z$ of distance, and the remaining terms sum
up the contributions from each daughter if the original quark did
split in the first $\Delta z$ of distance.
Rearranging terms, taking the limit $\Delta z \to 0$, and using
(\ref{eq:phiqbar}) gives an integro-differential equation for
$\phi_q$,
\begin {align}
  \frac{\partial \phi_q(E,z)}{\partial z}
  =&~
  - \Gamma_q(E) \, \phi_q(E,z)
  + \int_0^1 dx \> \left[\frac{d\Gamma}{dx}(E,x)\right]^\net_\uqq \,
     \phi_q(x E,z)
\nonumber\\ &
  + \int_0^1 dx \> \left[\frac{d\Gamma}{dx}(E,x)\right]^\net_\uqQ \,
     \phi_q(x E,z)
  - \int_0^1 dx \> \left[\frac{d\Gamma}{dx}(E,x)\right]^\net_\uqQbar \,
     \phi_q(x E,z) .
\label {eq:start1}
\end {align}
Now rewrite the total quark splitting rate (including NLO corrections)
as%
\footnote{
  The validity of eq.\ (\ref{eq:Gammaq}) depends subtly on the
  fact that all the daughters of our LO+NLO quark splitting processes
  $q{\to}qg$ and $q{\to}q\Q\Qbar$ are distinguishable in the large-$\Nf$
  limit.  Otherwise the relations would be complicated by
  identical-particle final state factors such as those appearing in the
  analysis of $g{\to}gg$ and $g{\to}ggg$ in refs.\ \cite{finale,finale2}.
}
\begin {equation}
  \Gamma_q = \int_0^1 dx \> \left[ \frac{d\Gamma}{dx} \right]^\net_\uqq
\label{eq:Gammaq}
\end {equation}
to put (\ref{eq:start1}) into the form
\begin {multline}
  \frac{\partial \phi_q(E,z)}{\partial z}
  =
  \int_0^1 dx \> \left[\frac{d\Gamma}{dx}(E,x)\right]^\net_\uqq \,
     \bigl[ \phi_q(x E,z) - \phi_q(E,z) \bigr]
\\
  + \int_0^1 dx \>
     \left(
        \left[\frac{d\Gamma}{dx}(E,x)\right]^\net_\uqQ -
        \left[\frac{d\Gamma}{dx}(E,x)\right]^\net_\uqQbar
     \right)
     \phi_q(x E,z) .
\end {multline}
Following arguments similar to ref.\ \cite{qedNfenergy}, we may use
the fact that the LO and NLO rates scale with energy as
$\sqrt{\qhat/E}$ to introduce rescaled variables $d\tilde\Gamma$,
$\tilde z$, and $\tilde\rho$ by%
\footnote{
  This is the same rescaling as eq.\ (5.4) of ref.\ \cite{qedNfenergy}.
}
\begin {equation}
    \left[ \frac{d\Gamma}{dx}(E,x) \right]_\uij^\net
    = E^{-1/2} \left[ \frac{d\tilde\Gamma}{dx}(x) \right]_\uij^\net ,
  \qquad
  z = E^{1/2} \tilde z ,
  \qquad
  \phi_q(E,z) = E^{-1/2} \, \tilde\phi_q(\tilde z)
\label {eq:tilde}
\end {equation}
to get
\begin {multline}
  \frac{\partial \tilde\phi_q(\tilde z)}{\partial \tilde z}
  =
  \int_0^1 dx \> \left[\frac{d\tilde\Gamma}{dx}(x)\right]^\net_\uqq \,
     \bigl[ x^{-1/2} \tilde\phi_q(x^{-1/2}\tilde z) - \tilde\phi_q(\tilde z) \bigr]
\\
  + \int_0^1 dx \>
     \left(
        \left[\frac{d\tilde\Gamma}{dx}(x)\right]^\net_\uqQ -
        \left[\frac{d\tilde\Gamma}{dx}(x)\right]^\net_\uqQbar
     \right)
     x^{-1/2} \tilde\phi_q(x^{-1/2}\tilde z) .
\end {multline}
Then we may convert the scaled variables back to the original
unscaled variables to get a simple equation for the desired
quark-number deposition distribution $\phi_q(z) \equiv \phi_q(E_0,z)$
of the shower:
\begin {multline}
  \frac{\partial\phi_q(z)}{\partial z}
  =
  \int_0^1 dx \> \left[\frac{d\Gamma}{dx}(E_0,x)\right]^\net_\uqq \,
     \bigl[ x^{-1/2} \phi_q(x^{-1/2} z) - \phi_q(z) \bigr]
\\
  + \int_0^1 dx \>
     \left(
        \left[\frac{d\Gamma}{dx}(E_0,x)\right]^\net_\uqQ -
        \left[\frac{d\Gamma}{dx}(E_0,x)\right]^\net_\uqQbar
     \right)
     x^{-1/2} \phi_q(x^{-1/2} z) .
\label{eq:start2}
\end {multline}
To get a recursion relation for the moments, multiply
both sides of (\ref{eq:start2}) by $z^n$ and integrate over $z$ to get
\begin {multline}
  -n \langle z^{n-1} \rangle_{\phi}
  =
  \int_0^1 dx \> \left[\frac{d\Gamma}{dx}(E_0,x)\right]^\net_\uqq \,
     \bigl[ x^{n/2} \langle z^n \rangle_{\phi}
            - \langle z^n \rangle_{\phi} \bigr]
\\
  + \int_0^1 dx \>
     \left(
        \left[\frac{d\Gamma}{dx}(E_0,x)\right]^\net_\uqQ -
        \left[\frac{d\Gamma}{dx}(E_0,x)\right]^\net_\uqQbar
     \right)
     x^{n/2} \langle z^n \rangle_{\phi} \,,
\end {multline}
which may be reorganized algebraically into
\begin {subequations}
\begin {equation}
  \langle z^n \rangle_{\phi} =
  \frac{ n \langle z^{\,n-1} \rangle_{\phi} }{ {\cal M}_n }
\label {eq:znphi}
\end {equation}
where the constants ${\cal M}_n$ are given by
\begin {equation}
   {\cal M}_n =
         \Avg_\uqq[ 1{-}x^{n/2} ]
         + \bigl\{ \Avg_\uqQbar[ x^{n/2} ] - \Avg_\uqQ[ x^{n/2} ] \bigr\}
   .
\label {eq:Mn}
\end {equation}
\end {subequations}
Note that if $[d\Gamma/dx]^\net_\uqQ$ and $[d\Gamma/dx]^\net_\uqQbar$ were
equal, this would reduce to exactly the same recursion
relation (\ref{eq:znrho}) as for the initial-flavor distribution $\rho(z)$.


\subsection{IR logarithm}
\label {sec:IRlog}

\subsubsection{Divergence in \boldmath$\qhat$ approximation}

Both the LO and NLO contributions to the
net rate $[d\Gamma/d\xq]^\net_\uqq$ have power-law dependence
$x^{-1/2}$ as $x{\to}0$ and $(1{-}x)^{-3/2}$ as $x{\to}1$.
[See eqs.\ (\ref{eq:LOrateq}) and (\ref{eq:LOqq}) for the LO rate and
eqs.\ (\ref{eq:fdef}), (\ref{eq:Rqq}), (\ref{eq:Lqq}) and (\ref{eq:fqq})
for the NLO correction.]  This behavior is mild enough that the integral
over $x$ in the first term of (\ref{eq:Mn}) converges, which is why we
had no issues with our numerical evaluations of moments of 
initial-flavor deposition in section \ref{sec:flavor}.
Parametrically, the size of the first term of (\ref{eq:Mn}) is
\begin {equation}
    \Avg_\uqq[ 1{-}x^{n/2} ] \sim \frac{1}{\ell_0} ,
\label {eq:1stterm}
\end {equation}
where $\ell_0$ is the parametric scale for the length of the shower
given by (\ref{eq:ell0}).

But now look at the remaining terms of (\ref{eq:Mn}).
The $q{\to}\Q$ and $q{\to}\Qbar$ net rates do not have any LO contribution,
and so these rates are given by
$[d\Gamma/dx]^\NLO_\uqQ$ and $[d\Gamma/dx]^\NLO_\uqQbar$.
Everything is okay
for $x{\to}1$ because
neither rate diverges
in that limit.  [See eq.\ (\ref{eq:RqQ}) in particular.]
But there is a problem for $x{\to}0$.
In that limit, eqs.\ (\ref{eq:fqQ}) and (\ref{eq:fqQbar}) give
\begin {equation}
  f_\uqQ(x) \to 0.6725,
  \qquad
  f_\uqQbar(x) \to 0.6774,
\end {equation}
and so the difference
$f_\uqQbar(x) - f_\uqQ(x) \to 0.0049$ is small but non-zero as $x{\to}0$.
Eqs.\ (\ref{eq:fdef}) and (\ref{eq:RqQ}) then give that the corresponding
difference in net rates scales as
\begin {equation}
  \left[ \frac{d\Gamma}{dx} \right]^\NLO_\uqQbar
  -
  \left[ \frac{d\Gamma}{dx} \right]^\NLO_\uqQ
  \sim
  x^{-3/2} .
\end {equation}
That means that the
\begin {equation}
  \bigl\{ \Avg_\uqQbar[ x^{n/2} ] - \Avg_\uqQ[ x^{n/2} ] \bigr\}
  =
  \int_0^1 dx \>
     \left(
        \left[\frac{d\Gamma}{dx}\right]^\NLO_\uqQbar -
        \left[\frac{d\Gamma}{dx}\right]^\NLO_\uqQ
     \right)
     x^{n/2}
\label {eq:newpiece}
\end {equation}
contribution to (\ref{eq:Mn}) is convergent for $n>1$ but logarithmically
IR divergent for $n=1$.  Parametrically,
\begin {equation}
  \bigl\{ \Avg_\uqQbar[ x^{n/2} ] - \Avg_\uqQ[ x^{n/2} ] \bigr\}
  \sim
  \frac{1}{\ell_0} \times
  \begin{cases}
     \Nf\alphas \times (\mbox{IR log divergence}), & n{=}1 ; \\
     \Nf\alphas, & n{>}1 ,
  \end {cases}
\label {eq:philog0}
\end {equation}
where the factor of $\Nf\alphas$ appears because the effect only arises
from {\it overlapping} $q \to qg \to q\Q\Qbar$, which is NLO.%
\footnote{
   If $q\to qg$ is instead followed by
   \textit{non}-overlapping $g\to\Q\Qbar$,
   then \textit{that} contribution of pair-produced quarks to quark-number
   deposition will average to zero because, as explained
   earlier, $\phi_g(E,z) = 0$ by charge conjugation.
}
Compare to (\ref{eq:1stterm}).


\subsubsection{IR scale}

The logarithmic IR divergence is a consequence of the non-canceling
$x^{-3/2}$ small-$x$ behavior of the
net rates for $q{\to}\Q$ and $q{\to}\Qbar$.
This $x^{-3/2}$ behavior also has the consequence that the total rate
for overlapping $q\to qg\to q\Q\Qbar$, which may be written as
\begin {equation}
   \Gamma_{q\to q\Q\Qbar} =
   \int_0^1 d\xQ \left[\frac{d\Gamma}{d\xQ}\right]^\net_\uqQ \,,
\label {eq:GqQQbar}
\end {equation}
is \textit{power-law} infrared divergent.
That's a common situation in the context of the LPM effect for QCD
in the $\qhat$ approximation.  As an example, consider the total
LO rate for bremsstrahlung,
\begin {equation}
   \Gamma_{q\to qg}^\LO =
   \int_0^1 dx_g \left[\frac{d\Gamma}{dx_g}\right]^\LO_{q\to qg} \,,
\end {equation}
The integrand behaves like $x_g^{-3/2} = (1{-}\xq)^{-3/2}$, and so
integration gives a power-law soft-gluon divergence for the total
LO rate as opposed to the more familiar logarithmic infrared
divergences in the absence of the LPM effect.
That power-law divergence is closely related
to the one for (\ref{eq:GqQQbar})
since a soft gluon in $q \to qg \to q\Q\Qbar$ produces a soft $\Q$ and
soft $\Qbar$.  The $x_g^{-3/2}$ behavior of $q \to qg$ arises
because in QCD the formation time (\ref{eq:tformx}) for a bremsstrahlung
gluon shrinks as the gluon becomes softer, and so there is less LPM
suppression of softer gluon bremsstrahlung.  This trend continues until
the gluon is so soft that the formation time becomes less than the
mean free time $\tau_0$ between collisions with the medium, at which
point the multiple-scattering ($\qhat$) approximation used throughout this
paper breaks down, and the $x_g^{-3/2}$ softens.
Parametrically, this takes place at $x_g$ of order
what we will call%
\footnote{
\label{foot:xIR1}
  The IR cut-off on $x$ was called $\delta$ in ref.\ \cite{qcd}.
  The parametric estimate (\ref{eq:xIR}) arises because the validity
  of the multiple scattering ($\qhat$) approximation requires that
  the formation time $t_\form \sim \sqrt{\omega/\qhat}$ be large compared
  to the time $\tau_0$ between collisions with the medium, which requires
  $\omega \gg \omega_\IR$ with $\omega_\IR \sim \qhat \tau_0^2$.
  That corresponds to energy fraction $x_\IR \equiv \omega_\IR/E$ in
  (\ref{eq:xIR}).
}
\begin {subequations}
\label {eq:xIRwrapper}
\begin {equation}
   \xIR \sim \frac{\qhat \tau_0^2}{E} \,,
\label {eq:xIR}
\end {equation}
which in the specific case of a quark-gluon plasma would be%
\footnote{
\label{foot:xIR2}
  In the parametric equation (\ref{eq:xIR2}), the factor $1+(\Nf/\Nc)$ means
  $O(1) + O(\Nf/\Nc)$, which is just $O(\Nf/\Nc)$ in the
  $\Nf{\gg}\Nc$ limit considered in this paper.
  Eq.\ (\ref{eq:xIR2}) can be understood by first rewriting (\ref{eq:xIR})
  as $\omega_\IR \sim \qhat/\Gamma_0^2$, where $\Gamma_0$ is the rate of
  collisions with the medium.
  The rate $\qhat$ for $(\Delta p_\perp)^2$ to grow with time is given
  by $\qhat \sim \Gamma_0 \mD^2$ up to logs, where the Debye mass
  $\mD$ is the
  typical transverse momentum kick per collision.  So
  $\omega_{\rm IR} \sim \mD^2/\Gamma_0$ (up to logs).
  Now consider a weakly-coupled plasma for the sake of keeping track of
  factors of coupling $g(T)$ in the plasma.  Then
  $\Gamma_0 \sim g^4 n/\mD^2$ and so $\omega_{\rm IR} \sim \mD^4/g^4 n$,
  where $n$ is the density of partons.
  Finally,
  $\mD^2 \sim (\Nc+\Nf) g^2 T^2$ and $n \sim (\Nc^2 + \Nf\Nc)T^3$, which
  gives (\ref{eq:xIR2}) for $x_{\rm IR} \equiv \omega_{\rm IR}/E$.
  Medium-induced masses of on-shell gluons (which we have ignored)
  also become important at roughly the same scale,
  and they also cut off divergences in splitting rates.
  For strongly-coupled plasmas, just ignore the factors of $g$ in the
  preceding parametric estimates.
}
\begin {equation}
   \xIR \sim \left(1 + \frac{\Nf}{\Nc}\right) \frac{T}{E} \,.
\label {eq:xIR2}
\end {equation}
\end {subequations}
The softening of the IR divergence for $x \ll \xIR$ will cut off
the logarithmic divergence of (\ref{eq:newpiece}) for $n=1$, resolving
(\ref{eq:philog0}) as
\begin {equation}
  \bigl\{ \Avg_\uqQbar[ x^{n/2} ] - \Avg_\uqQ[ x^{n/2} ] \bigr\}
  \sim
  \frac{1}{\ell_0} \times
  \begin{cases}
     \Nf\alphas \ln(1/\xIR), & n{=}1 ; \\
     \Nf\alphas, & n{>}1 .
  \end {cases}
\label {eq:philog}
\end {equation}

Infrared sensitivity means
that, with our $\qhat$-based NLO rates, we cannot compute the NLO
correction to quark-number deposition to better than leading-log order,
which is why we summarized the case of quark-number deposition as
IR-unsafe in table \ref{tab:chi2}.
But also, for large enough energy, $\Nf\alphas\ln(1/\xIR)$ may no
longer be small even for small $\Nf\alphas$.
In that case, the NLO effect (\ref{eq:philog}) will no longer
be small compared to the LO result (\ref{eq:1stterm}) for
${\cal M}_1$.
Then one may wonder about the uncalculated NNLO correction and so
forth, which is another complication of
IR-unsafe quantities.

Ultimately, the difference between quark-number deposition and
energy deposition is that softer particles carry less energy and
so contribute less to determining where the shower's energy stops.
In contrast, softer quarks carry just as much quark number as harder ones,
and so the details of how even arbitrarily soft quarks stop
may significantly affect where (on average)
overall quark number is deposited.
We give a qualitative picture of why this happens in
appendix \ref{app:phi}.
The reason that
QCD initial-flavor deposition avoided this problem is special
to the large-$\Nf$ limit because pair-produced quarks
(soft or otherwise) do not carry the flavor of the initial quark.


\subsection{\boldmath What is responsible for
            $[d\Gamma/dx]^\net_{q\to\Q} \not= [d\Gamma/dx]^\net_{q\to\Qbar}$
            as $x \to 0$?}
\label {sec:WhatQQbar}

The first thing to note about the relationship between
$[d\Gamma/dx]^\net_{q\to\Q}$ and $[d\Gamma/dx]^\net_{q\to\Qbar}$ is that
there is no simple reason that they should be equal.
Charge conjugation symmetry, for example, will interchange $\Q$ and $\Qbar$,
but it equates $q\to q\Q\Qbar$ to $\bar q\to\bar q\Qbar\Q$ and not to
$q\to q\Qbar\Q$.  If there were no ``interactions'' between $\Q\Qbar$
and $q$ in our calculation, then there would be an
independent charge conjugation symmetry for $\Q$'s that only swapped
$\Q{\leftrightarrow}\Qbar$ and left $q$ alone.
However, in our overlap calculations, the evolution of the $q$'s and
$\Q$'s or $\Qbar$'s is linked via the medium-averaged correlations of their
interactions with the medium, and the correlations between $Q$ and $q$
will differ from those between $\Qbar$ and $q$ because of the
difference in color states.  (This difference is especially acute in
large-$\Nc$ QCD.)
So, charge conjugation symmetry does not directly provide a reason to
think that $[d\Gamma/dx]^\net_{q\to\Q}$ and $[d\Gamma/dx]^\net_{q\to\Qbar}$ should
be equal.  Numerically, we find that they are not.

\begin {figure}[tp]
\begin {center}
  \includegraphics[scale=0.6]{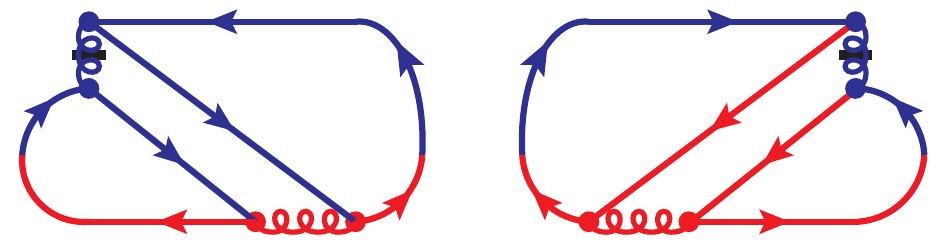}
  \caption{
     \label{fig:phidiags}
     The complete set of diagrams responsible for the IR sensitivity of
     quark number deposition in $\Nf{\gg}\Nc{\gg}1$ QCD.
  }
\end {center}
\end {figure}

Despite similar absence of a simple charge conjugation argument,
we found in ref.\ \cite{qedNfenergy} that
the $e\to e\E\Ebar$ rate
\begin {equation}
   \left[ \Delta\, \frac{d\Gamma}{d\xe\,d\xE} \right]_{e\to e\E\Ebar}
\end {equation}
in QED \textit{was} for some reason symmetric under the exchange of the values
\begin {equation}
   \xE \leftrightarrow \xEbar = 1{-}\xe{-}\xE ,
\label{eq:swapEEbar}
\end {equation}
resulting in
$[d\Gamma/dx]_{\underline{e\to\E}} = [d\Gamma/dx]_{\underline{e\to\Ebar}}$.
We were unable to articulate a simple symmetry or other high-level
explanation expressible
in a few sentences, but we did give a more involved,
lower-level analysis of how invariance under (\ref{eq:swapEEbar})
arose in the calculation of the $e\to e\E\Ebar$ rate.%
\footnote{
  See Appendix A of ref.\ \cite{qedNfenergy}.
}
For the $\Nf{\gg}\Nc{\gg}1$ QCD case, we may
similarly examine the behavior of
the unintegrated rate $[\Delta\,d\Gamma/d\xq\,d\xQ]_{q\to q\Q\Qbar}$
under
\begin {equation}
   \xQ \leftrightarrow \xQbar = 1{-}\xq{-}\xQ .
\label {eq:swapQQbar}
\end {equation}
There is no invariance, else the net rates would have been equal.
But the lack of IR-safety for quark number deposition depends specifically
on divergences in the limits where $\xQ$ or $\xQbar$ are small.
And when $\xQ$ is small, the IR divergence of the net rate
$[d\Gamma/dx]^\net_\uqQ$ comes from the limit of soft gluon emission
in overlapping $q \to qg \to q\Q\Qbar$, in which case $\xQbar$ is also
comparably small.
In that limit where
\begin {equation}
   \xQ \sim \xQbar \to 0 ,
\label {eq:softlim}
\end {equation}
we find numerically that each interference diagram that contributes
to $[\Delta\,d\Gamma/d\xq\,d\xQ]_{q\to q\Q\Qbar}$ is individually
either symmetric or anti-symmetric under (\ref{eq:swapQQbar}).
If they were all symmetric, then the dominant small-$x$ behavior of
the rates $[d\Gamma/dx]^\net_{q\to\Q}$ and $[d\Gamma/dx]^\net_{q\to\Qbar}$
would have been equal, and so quark-number deposition would have been infrared
safe.  The IR sensitivity of quark-number deposition therefore
arises exclusively from the diagrams that are anti-symmetric in this
soft limit.  Interestingly, those diagrams have a simple
characterization: they are the
$q \to q\Q\Qbar$ diagrams with a single instantaneous longitudinal
gluon exchange.  The complete set of such diagrams (excluding one
that vanishes) is shown in fig.\ \ref{fig:phidiags}.

Unfortunately, we do not know any simple, short, high-level explanation
of why these particular diagrams are anti-symmetric in the
soft limit.
We instead provide a lower-level explanation in appendix \ref{app:QQbar}.
There also exist diagrams similar to
fig.\ \ref{fig:phidiags} in large-$\Nf$ QED,
but in the QED case \textit{all} diagrams are symmetric under
(\ref{eq:swapEEbar}).


\section{Process independence}
\label{sec:universality}

Our previous results for charge and energy stopping lengths can be
used for a first exploration of an important issue, which is whether
the effects that must be absorbed into the value of $\qhat$ are
process independent.

Recall that $\eps_\LO(E_0,z;\qhat)$ refers
to a leading-order ($\LO$) calculation of energy deposition,
based solely on
BDMPS-Z splitting rates (expressed in terms of $\qhat$) that completely
ignore the possibility of overlapping formation times.
In our problem, $\qhatA$ and $\qhatF$ are related by $\qhatA = 2\qhatF$
in the large-$\Nc$ limit and more generally by Casimir scaling;%
\footnote{
  See footnote \ref{foot:Casimir} for caveats.
}
so pick one (e.g.\ $\qhatF$) to be the ``$\qhat\,$'' of this discussion.
Imagine (as a thought experiment!)\ that we
(i) could somehow measure the {\it actual}\/ energy
deposition $\eps(z)$ of an in-medium gluon-initiated shower
in an infinite medium and extract its
first moment $\lstop = \langle z \rangle_\eps$, and then (ii) equate
that measurement of $\lstop$ to the formula for the
first moment of $\eps_\LO(E_0,\qhat,z)$
to extract a ``measured'' effective $\qhat$.
The results from table \ref{tab:chi2} (combined with similar results
presented in this paper and in ref.\ \cite{finale2} for
$\qhat$-insensitive ratios involving higher moments)
indicate that, with \textit{that} $\qhat$ extracted from
measurement, everything \textit{else} about the
overlap-ignoring distribution $\eps_\LO(E_0,\qhat,z)$ should
also be in good agreement with measurement.

But what if we had instead extracted the value of $\qhat$ from
the first moment of the energy deposition of a \textit{quark}-initiated
shower instead of from a gluon-initiated shower?
Would we get almost the same value of effective $\qhat$,
or would we need a very
different value in order to accurately describe energy deposition in the
quark-initiated case?

A program of ignoring overlap effects by extracting the value of $\qhat$
from one measurement, and then using it to predict other measurements,
will hold together only if there is little sensitivity (besides
the known sensitivity to energy scale) to what type of measurement one
uses (within reason) to extract $\qhat$.
We can turn this around to say it another way.  A necessary condition
for success would be that
\textit{overlap corrections} to a LO calculation
of the ratio
\begin {equation}
   \frac{
           \lstop(q, \mbox{energy};\qhat,E_0)
        }{
           \lstop(g, \mbox{energy};\qhat,E_0)
        }
\label {eq:processratio}
\end {equation}
must be small.  Above, $\lstop(i, \mbox{energy})$ means the
first moment of the energy deposition $\eps_i(z)$
of a shower initiated by a particle
of type $i$.
The numerator and denominator of the ratio (\ref{eq:processratio}) should
be calculated using the same value of $\qhat$, in which case the ratio
itself is $\qhat$-insensitive, just like the $\sigma/\lstop$ ratios of
table \ref{tab:chi2}.

Numerical results for the size of overlap corrections to ratios like
(\ref{eq:processratio}) are given in table
\ref{tab:lstopratios} for $\Nf{\gg}\Nc{\gg}1$ QCD.
The corrections are small for any reasonable
value of $\Nf\alphas$, which means that extracting an effective value of
$\qhat$ from one type of measurement would make LO analysis work very
well for all types of (IR-safe) measurements considered here.

\begin {table}[t]

\setlength{\tabcolsep}{7pt}
\begin {center}
\begin{tabular}{rcc}
\hline
\hline
  \multicolumn{3}{c}{$\Nf{\gg}\Nc{\gg}1$ QCD}
\\
\hline
  \multicolumn{1}{c}{ratio} & \multicolumn{2}{c}{overlap correction}
\\
\cline{2-3}
  & $\mu\propto(\qhat E)^{1/4}$
  & $\mu\propto\bigl(x(1{-}x)\qhat E\bigr)^{1/4\strut}$
\\
\hline
  $\lstop(g,\mbox{energy})/\lstop(q,\mbox{energy})$
    & $-0.3\%\times\Nf\alphas$
    & $\phantom{-}0.2\%\times\Nf\alphas$
\\
  $\lstop(q,\mbox{init flavor})/\lstop(q,\mbox{energy})$
    & $-1.2\%\times\Nf\alphas$
    & $-0.8\%\times\Nf\alphas$
\\
\hline
\hline
\end{tabular}
\end {center}
\caption{%
\label{tab:lstopratios}%
  The relative size of overlap
  corrections to ratios of different types of stopping distances
  in large-$\Nf$ QCD.
}
\end{table}

Small overlap corrections to the $\lstop$
ratio (\ref{eq:processratio}) would not be
surprising or interesting if the overlap corrections to the
individual stopping distances $\lstop$ were already small to begin with.
For the purely gluonic showers ($\Nf{=}0$) studied in
refs.\ \cite{finale,finale2}, NLO corrections to the energy stopping
distance were $O(100\%)\times\Nc\alphas$ and were extremely
sensitive to the exact choice of IR factorization scale $\Lambda_\fac$.
The fact that NLO corrections to $\qhat$-insensitive ratios like
$\sigma/\lstop$ were found to be extremely small
was therefore very significant.
The ratios of table \ref{tab:lstopratios} can only
be examined once we add quarks to the theory, which in this paper
has only been done in the large-$\Nf$ limit.  In that case,
the corrections to stopping distances are more moderate, as shown
in table \ref{tab:chilstop}.
Judging only from the $\mu = (\qhatF E)^{1/4}$ column, the NLO
corrections to individual stopping distances might appear very small.  But this
is an accident of that choice of $\mu$, as can be seen from
the much larger NLO corrections in the
$\mu = \bigl(x(1{-}x) \qhatF E\bigr)^{1/4}$ column, which are roughly $8\%$.
So, the much reduced sensitivity to renormalization scale exhibited by
the $\lstop$ ratios in
table \ref{tab:lstopratios}, and the uniformly small NLO corrections
there, is still a significant effect.

\begin {table}[t]

\setlength{\tabcolsep}{7pt}
\begin{center}
\begin{tabular}{rrc}
\hline\hline
 & \multicolumn{2}{c}{$\delta\langle z \rangle/\langle z\rangle^\LO$} \\
\cline{2-3}
  &
  \multicolumn{1}{c}{$\mu = (\qhatF E)^{1/4}$}
  &
  \multicolumn{1}{c}{$\mu = \bigl(x(1{-}x)\qhatF E\bigr)^{1/4\strut}$}
 \\[6pt]
\hline
  energy, $g$ & $-0.2\%\times\Nf\alphas$ & $-8.0\%\times\Nf\alphas$ \\
  energy, $q$ & $ 0.1\%\times\Nf\alphas$ & $-8.1\%\times\Nf\alphas$ \\
  initial-flavor, $q$ & $-0.1\%\times\Nf\alphas$ & $-9.0\%\times\Nf\alphas$ \\
\hline
\hline
\end{tabular}
\end{center}
\caption{
   \label{tab:chilstop}
   Relative size of overlap corrections to $\lstop \equiv \langle z \rangle$,
   given by the ratio of $\delta\langle z \rangle$ and $\langle z \rangle^\LO$
   from tables \ref{tab:momentsrho} and \ref{tab:momentseps}.
}
\end{table}

Also, it's not just a matter of whether one includes some sort of
$x$ dependence in $\mu$.
Unlike $\qhat$-insensitive quantities like $\sigma/\lstop$,
and $\lstop$ ratios (\ref{eq:processratio}),
the NLO corrections to $\lstop$ itself are sensitive to the overall scale
of $\mu$.  For example, we find that
$\mu = \tfrac12 (\qhat E)^{1/4}$ gives NLO corrections to
$\lstop$ that are comparable to those for
$\mu = \bigl(x(1{-}x)E\bigr)^{1/4}$ in table \ref{tab:chilstop}
while still giving exactly the same
small numbers for corrections to the $\lstop$ ratios 
in the $\mu \propto (\qhat E)^{1/4}$ column of table \ref{tab:lstopratios}.

We should note that the
second ratio of table \ref{tab:lstopratios}
is probably not a sensible test
of process independence outside of the large-$\Nf$ approximation.
Remember that we have found fermion-number deposition to be IR unsafe
in large-$\Nf$ QCD,
and our argument for IR-safety of initial-flavor deposition depended on the
large-$\Nf$ approximation.  But the ratio (\ref{eq:processratio}) of
energy stopping distances should always be a good test, even outside of
the large-$\Nf$ limit.


\section {\boldmath Why is QED overlap $\gg$ QCD overlap
          for comparable $N\alpha$?}
\label {sec:why}

\begin {figure}[t]
\begin {center}
  \includegraphics[scale=0.5]{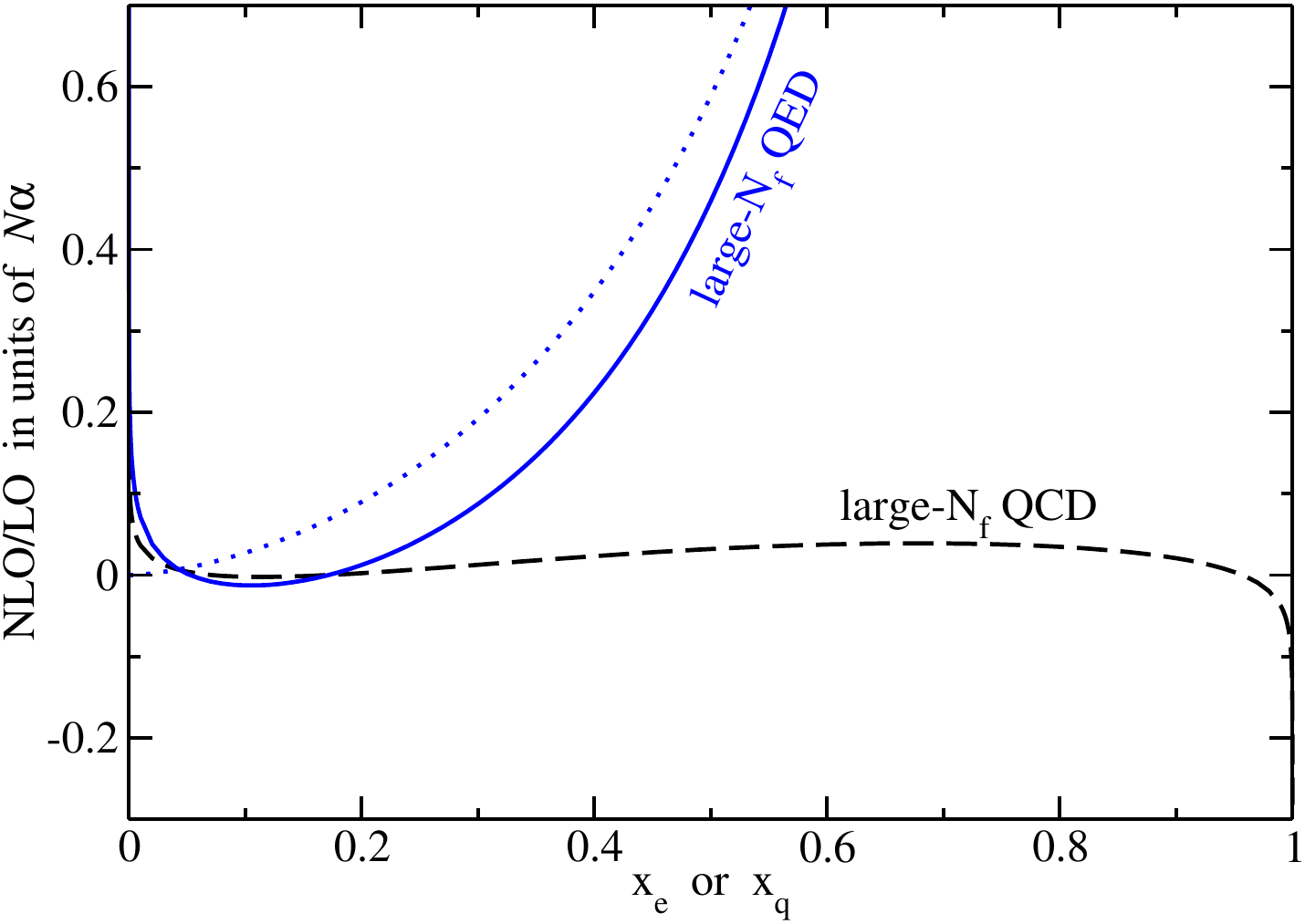}
  \caption{
     \label{fig:ratio}
     The ratio (\ref{eq:Ratio}) of NLO to LO contributions to the net rates
     for $e{\to}e$ (solid blue curve) and
     $q{\to}q$ (dashed black curve), in units of $\Nf\alpha$ for
     $\mu = (\qhatF E)^{1/4}$.  The dotted blue curve will be explained
     later in the main text.
  }
\end {center}
\end {figure}

In this section, we propose a qualitative explanation for why
the relative size of QED overlap
effects are so much larger than for QCD overlap effects for comparable values
of $N\alpha$.  For the sake of simplicity, we will focus our attention on
initial-flavor stopping, which involves only the $e{\to}e$ net rate in QED
versus the $q{\to}q$ net rate in QCD.
The size of overlap effects that cannot be absorbed into $\qhat$
depends on the $x$ dependence of the ratio of
the NLO to LO contributions to that rate.
If the ratio
\begin {equation}
   \Ratio(x) \equiv \frac{[d\Gamma/dx]_\uee^\NLO}{[d\Gamma/dx]_\uee^\LO}
\label {eq:Ratio}
\end {equation}
and/or its QCD analog were independent of $x$, the NLO contribution
$[d\Gamma/dx]_\uee^\NLO \propto \sqrt{\qhat}$
to the rate could be completely absorbed
into the leading-order rate $[d\Gamma/dx]_\uee^\LO \propto \sqrt{\qhat}$
by replacing $\qhat$ by $\qhat_{\rm eff} = (1{+}\Ratio)^2 \qhat$ in
the leading-order rate.
The actual variation with $x$ is shown in fig.\ \ref{fig:ratio}
for QED vs.\ QCD.
The QED curve varies drastically more than the QCD curve; so much so that
we can't even fit much of the QED curve into the figure while showing
variation of
the QCD curve. For reasons to be discussed, the QED ratio has a power-law
divergence in the soft photon limit ($1{-}\xe \to 0$), whereas the
QCD ratio has only a mild logarithmic divergence.%
\footnote{
  The logarithmic divergence of the QCD ratio as
  $\xq{\to}1$ comes solely from the logarithm $L_\uqq(\xq,\mu)$
  of (\ref{eq:Lqq}), and in detail depends on the $\xq$ dependence of
  one's choice of renormalization scale $\mu$.
  (See various alternatives discussed in section \ref{sec:mu}.)
}

To understand the different behaviors in fig.\ \ref{fig:ratio}
of the NLO/LO ratio for $e{\to}e$ vs.\ $q{\to}q$ in the
soft photon or gluon limit,
we need to (i) review that limit for the LO bremsstrahlung rate
and (ii) understand it for the overlap (NLO) correction.


\subsection {Soft bremsstrahlung limit: LO review}

The LPM effect depends on the near-collinearity of high-energy bremsstrahlung
in the high-energy limit. (The same is true of pair production, but
the ensuing discussion here will focus on bremsstrahlung.)
The near-collinearity allows for phase coherence in single bremsstrahlung
of (i) emission at some time (call it $t{=}0$) with
(ii) emission at a time $t_\form \gg \lambda_{\rm rad}$ later,
where $\lambda_{\rm rad} \sim 1/\omega_{\rm rad}$
is the wavelength of the bremsstrahlung radiation.
There are two competing effects when considering how the formation
time behaves for soft bremsstrahlung.  First, softer radiation means
larger $\lambda_{\rm rad}$, which by itself would make coherence over
larger times easier and so might be expected to increase $t_{\rm form}$.
On the other hand, longer $t_{\rm form}$ allows more opportunity for
high-energy particles to pick up more $p_\perp$ kick from the medium
over the duration of the bremsstrahlung process and so be driven
to be less collinear, which would work toward reducing rather than
increasing the time over which the process is coherent.
The actual behavior of $t_{\rm form}$ in the soft limit is a balance
between these effects.  For QED,%
\footnote{
  In terms of more-general formulas, the scale of the QED bremsstrahlung
  formation time is
  $t_\form \sim 1/|\Omega_0|$ with $\Omega_0$ given by (\ref{eq:Omega0B}) with
  $(x_1,x_2,x_3) = (-1,\xe,x_\gamma)$ and
  $(\qhat_1,\qhat_2,\qhat_3) = (\qhat,\qhat,0)$, since the
  high-energy photon does
  not interact directly with the medium.
  Eq.\ (\ref{eq:tformsoftqed}) corresponds to the $x_\gamma \ll 1$ limit.
}
\begin {equation}
   t_{\rm form}(x_\gamma) \sim \sqrt{ \frac{E^2}{\omega_{\rm rad} \qhat} }
   \sim \sqrt{ \frac{E}{x_\gamma \qhat} }
   \qquad
   (x_\gamma \ll 1) ,
\label {eq:tformsoftqed}
\end {equation}
where $E$ is the energy of the parent.  Overall, the QED formation time
for bremsstrahlung grows, and so LPM suppression becomes even more
significant, in the soft limit.

The situation is very different in QCD because gluons, unlike photons,
scatter easily from the medium.  The softer the bremsstrahlung gluon,
the more a small kick from the medium will change its direction, and
so the less collinear the bremsstrahlung process will become.
So the effect of scattering from the medium toward reducing formation
times is much stronger in the soft limit of QCD bremsstrahlung than
QED bremsstrahlung.  The result is%
\footnote{
   Eq.\ (\ref{eq:tformsoftqcd}) is just the soft limit of
   the formula (\ref{eq:tformx}) presented previously.
}
\begin {equation}
   t_{\rm form}(x_g) \sim \sqrt{ \frac{\omega_{\rm rad}}{\qhat} }
   \sim \sqrt{ \frac{x_g E}{\qhat} }
   \qquad
   (x_g \ll 1) .
\label {eq:tformsoftqcd}
\end {equation}
In QCD, the formation time for bremsstrahlung shrinks, so that there
is \textit{less} LPM suppression, in the soft limit (until
the formation time becomes as small as the mean free path for
scattering from the medium, at which point the $\qhat$ approximation
breaks down).
This difference between
soft-gluon vs.\ soft-photon bremsstrahlung is the most
significant qualitative difference between the LPM effect in
QCD and QED.

Parametrically, the rate for soft bremsstrahlung is then
\begin{subequations}
\label{eq:dGsoft0}
\begin{align}
   \left[ \frac{d\Gamma}{dx_\gamma} \right]^\LO_{e\to e\gamma}
   &\sim
   \frac{ \alpha \, P_{e\to e}(x_\gamma) }{ t_\form(x_\gamma) }
   \sim \frac{\alpha}{x_\gamma^{1/2}} \sqrt{ \frac{\qhat}{E} }
   = \frac{\alpha}{(1{-}\xe)^{1/2}} \sqrt{ \frac{\qhat}{E} }
   &
   \mbox{(QED, $x_\gamma \ll 1$)} ,
\label {eq:dGsoft0e}
\\
   \left[ \frac{d\Gamma}{dx_g} \right]^\LO_{q\to qg}
   &\sim
   \frac{ \alphas \, P_{q\to q}(x_g) }{ t_\form(x_g) }
   \sim \frac{\CF\alphas}{x_g^{3/2}} \sqrt{ \frac{\qhat}{E} }
   = \frac{\CF\alphas}{(1{-}\xq)^{3/2}} \sqrt{ \frac{\qhat}{E} }
   &
   \mbox{(QCD, $x_g \ll 1$)} .
\end{align}
\end{subequations}
Note the different power-law behavior in the soft limit, which arises
because of the different power-law behavior of the formation time,
which in turn arose because gluons interact much more directly
with a QCD medium than photons do with a QED medium.
Written in the language of net rates, (\ref{eq:dGsoft0}) is
\begin{subequations}
\label{eq:dGsoft}
\begin{align}
   \left[ \frac{d\Gamma}{d\xe} \right]^\LO_\uee
   &\sim \frac{\alpha}{(1{-}\xe)^{1/2}} \sqrt{ \frac{\qhat}{E} }
   &
   \mbox{(QED, $1{-}\xe \ll 1$)} ,
\label{eq:dGsoftqed}
\\
   \left[ \frac{d\Gamma}{d\xq} \right]^\LO_\uqq
   &\sim \frac{\Nc\alphas}{(1{-}\xq)^{3/2}} \sqrt{ \frac{\qhat}{E} }
   &
   \mbox{(QCD, $1{-}\xq \ll 1$)} .
\label{eq:dGsoftqcd}
\end{align}
\end{subequations}


\subsection {Soft bremsstrahlung limit: overlap corrections}

\begin {figure}[t]
\begin {center}
  \includegraphics[scale=0.6]{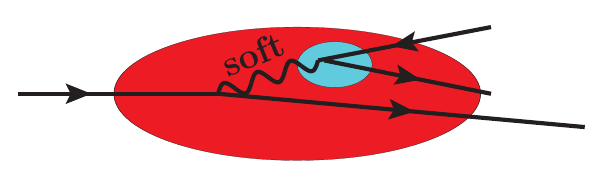}
  \caption{
     \label{fig:softoverlap}
     A qualitative depiction of
     QED bremsstrahlung overlapping with typical pair production
     in the soft photon limit.
     The red oval represents the long LO formation time for soft
     bremsstrahlung, which will be disrupted because of the pair
     production.
  }
\end {center}
\end {figure}

It's now easy to get a first idea of why overlap (NLO) corrections
can behave very differently from the LO bremsstrahlung rate in
QED.  The overlap corrections to the $e{\to}e$ net rate come
from the overlap of bremsstrahlung followed by either real or
virtual pair production, as depicted schematically in
fig.\ \ref{fig:softoverlap} for real production in the soft-photon limit.
As soon as the soft photon converts to a soft electron-positron pair,
however, the electron and positron
can interact directly with the QED plasma and so---like soft
gluons---they can receive and be easily deflected by kicks from
the plasma, reducing the collinearity of the entire
$e{\to}e\gamma{\to}e\E\Ebar$ process.  When the two splittings overlap,
the production of the soft $\E\Ebar$ pair therefore significantly
disrupts the collinearity needed for the LPM effect in the production of the
original $e{\to}e\gamma$ bremsstrahlung.  Based purely on the
analogy with LO soft gluon bremsstrahlung (\ref{eq:dGsoftqcd}),
one might then guess that this
conversion of a soft photon to an $\E\Ebar$ pair could
produce an overlap correction of size
\begin {align}
   \left[ \frac{d\Gamma}{d\xe} \right]^\NLO_\uee
   &\sim
     \Nf\alpha \times \frac{\alpha}{(1{-}\xe)^{3/2}} \sqrt{ \frac{\qhat}{E} }
   & \mbox{(QED, $1{-}\xe \ll 1$)},
\label {eq:NLOqed}
\\
\intertext{%
where the extra factor of $\Nf\alpha$ compared to (\ref{eq:dGsoftqcd})
comes from the vertex needed to make the $\E\Ebar$ pair.
The NLO/LO ratio (\ref{eq:Ratio}) of (\ref{eq:NLOqed}) to (\ref{eq:dGsoftqed})
would then behave like
}
   \Ratio(\xe) &\sim \frac{ \Nf\alpha }{1-\xe}
   \qquad
   & \mbox{(QED, $1{-}\xe \ll 1$)}\phantom{,}
\label {eq:Ratioqed}
\end {align}
and would blow up as $\xe \to 1$.
In contrast, QCD overlap effects, where the overlapping pair production
does not make the same qualitative change to LO bremsstrahlung,
would be
\begin {align}
   \left[ \frac{d\Gamma}{d\xq} \right]^\NLO_\uqq
   &\sim
     \Nf\alpha \times \frac{\Nc\alphas}{(1{-}\xq)^{3/2}} \sqrt{ \frac{\qhat}{E} }
   & \mbox{(QCD, $1{-}\xq \ll 1$)},
\\
\intertext{with}
   \Ratio(\xq) &\sim \Nf\alphas
   \qquad
   &\mbox{(QCD, $1{-}\xq \ll 1$)} .
\end {align}

The above parametric estimates are correct, neglecting logarithms.%
\footnote{
   For the QCD case, see eqs.\ (\ref{eq:fdef}), (\ref{eq:Rqq}) and
   (\ref{eq:fqq}).  For the QED case, see the analogous
   eqs.\ (3.5a), (3.6) and (3.10a) of ref.\ \cite{qedNfenergy}.
}
But it is desirable to better understand the origin of
(\ref{eq:NLOqed}) by giving a qualitative argument rather than
just a qualitative guess.


\subsubsection{\boldmath$\Nf\alpha \ll x_\gamma \ll 1$}

\begin {figure}[t]
\begin {center}
  \includegraphics[scale=0.6]{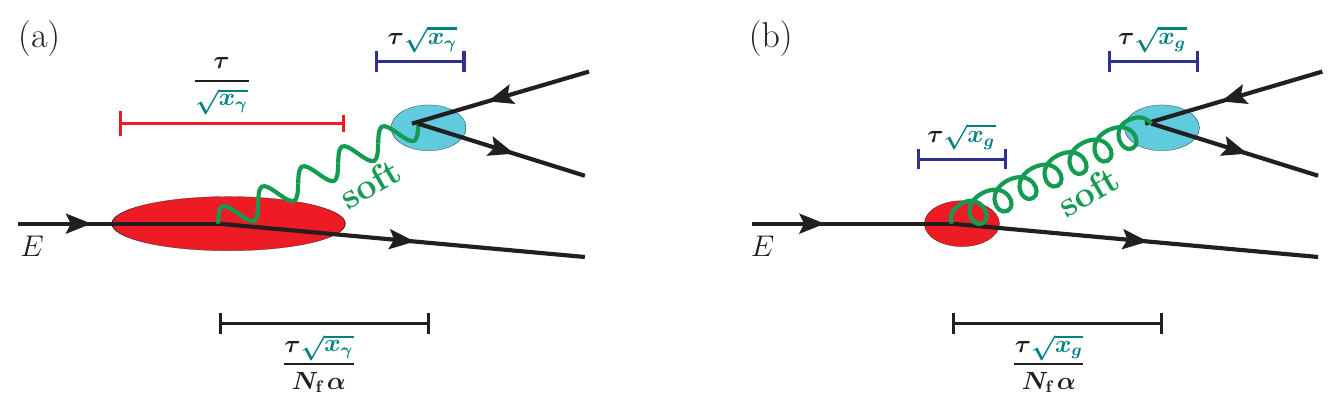}
  \caption{
     \label{fig:softNf}
     (a) Parametric summary of the relative size of typical formation lengths
     and distances between a soft QED bremsstrahlung and subsequent
     democratic pair production, contrasted with (b) the same for QCD.
     Above, $\tau$ represents the
     parametric scale $\tau \sim \sqrt{E/\qhat}$ for a democratic
     splitting of energy $E$.
     These figures may be contrasted with the first part of
     fig.\ \ref{fig:typicalNf_qcd}, where all splittings were democratic
     and the scale of the ``formation time'' there corresponds to
     ``$\tau$'' here.
  }
\end {center}
\end {figure}

In our analysis of overlap effects, we treat $\Nf\alpha$ as small and
picture overlapping formation times (or at least those that
cannot be absorbed into $\qhat$) as rare events. We only check
those assumptions \textit{a posteriori} when we numerically check the size of
final results, such as in table \ref{tab:chi2}.
In the spirit of this picture, fig.\ \ref{fig:softNf}a depicts
the typical QED scales involved for soft bremsstrahlung,
with formation time (\ref{eq:tformsoftqed}),
subsequently followed by democratic pair production.
(For comparison, fig.\ \ref{fig:softNf}b shows the QCD version.)
Whether in QED or QCD, democratic splittings from a parent of
energy $E$ have formation times of order $\sqrt{E/\qhat}$,
but the immediate parent of the pair production in the fig.\ \ref{fig:softNf}a
is the photon, whose energy is $x_\gamma E$, so that
$t_\form^{\rm pair} \sim \sqrt{x_\gamma E/\qhat}$ in the notation of
the figure.  The probability of pair production is parametrically
$\Nf\alpha$ per formation time, and so the typical distance the
photon travels before pair producing is of order $t_\form^{\rm pair}/\Nf\alpha$.

The qualitative features to take away from fig.\ \ref{fig:softNf}a
are that, in the soft photon limit, (i) the bremsstrahlung formation
time is parametrically large compared to the subsequent pair production
formation time, and (ii) overlap will be common unless the distance
between the bremsstrahlung and pair production is parametrically large
compared to the bremsstrahlung formation time, which requires
$\tau \sqrt{x_\gamma}/\Nf\alpha \gg \tau/\sqrt{x_\gamma}$ and so
$x_\gamma \gg \Nf\alpha$.  So, overall, fig.\ \ref{fig:softNf}a
depicts the situation
\begin {equation}
   \Nf\alpha \ll x_\gamma \ll 1 .
\label {eq:xgammaRange}
\end {equation}
The probability of overlap is parametrically
\begin {equation}
   \operatorname{Prob}({\rm overlap})
   \sim \frac{ \mbox{longer formation time} }
             { \mbox{typical separation} }
   \sim \frac{\Nf\alpha}{x_\gamma} = \frac{\Nf\alpha}{1{-}\xe} \,.
\label {eq:overlapprob}
\end {equation}
In the rare occasions that they \textit{do} overlap for (\ref{eq:xgammaRange}),
we should expect the separation between $e \to e\gamma$ and
$\gamma \to \E\Ebar$ to be of order the $e\to e\gamma$ formation length,
because it would be even more unlikely for the pair production to have
occurred much earlier than that.  The disruption to the formation of
the soft bremsstrahlung due to the creation of the more-easily scattered
$\E\Ebar$ can therefore be expected to have an $O(100\%)$ effect
(but not larger) on the $e \to e\gamma$ splitting for the events where
they overlap.  So we are led to estimate
\begin {align}
   \left[ \frac{d\Gamma}{d\xe} \right]^\NLO_\uee
   &\sim
   \left[ \frac{d\Gamma}{d\xe} \right]^\LO_\uee
   \times \operatorname{Prob}({\rm overlap}) \times O(100\%)
   \sim
   \left[ \frac{d\Gamma}{d\xe} \right]^\LO_\uee
     \times \frac{\Nf\alpha}{1{-}\xe} \,,
\end {align}
which is the same as the earlier guess (\ref{eq:NLOqed}).


\subsubsection{\boldmath$x_\gamma \ll \Nf\alpha$}

Nothing special about the scale $x_\gamma \sim \Nf\alphas$ was built into
the calculation in ref.\ \cite{qedNf} of rates
such as $[\Delta\,d\Gamma/d\xe\xE]_{e\to e\E\Ebar}^\NLO$ for
overlapping splitting.  So, as far as those calculations are concerned,
there's no reason not to qualitatively
discuss $x_\gamma \ll \Nf\alphas$ as well.
In that limit, the separation between the soft bremsstrahlung and the pair
production is short compared to the soft bremsstrahlung formation time,
and so the probability of overlap is
\begin {equation}
   \operatorname{Prob}({\rm overlap}) \simeq 100\%
\end {equation}
instead of (\ref{eq:overlapprob}).
But, once produced,
the creation of the more-easily scattered soft $\E$ and $\Ebar$
will quickly terminate the coherence of the original bremsstrahlung
process after a time of order the pair-production formation time.
This will effectively shorten the soft QED bremsstrahlung formation time
from the leading-order result (\ref{eq:tformsoftqed}) to
\begin {align}
  (t_\form)_{\rm eff} &\sim
  (\mbox{typical separation}) + (\mbox{pair formation time})
\\
  &\sim
  (\mbox{typical separation})
  \sim
  \frac{x_\gamma^{1/2}}{\Nf\alpha} \sqrt{ \frac{E}{\qhat} } .
\end {align}
Correspondingly,
the rate (\ref{eq:dGsoft0e}) for soft QED bremsstrahlung will then be changed
to
\begin {align}
   \left[ \frac{d\Gamma}{d\xe} \right]^\NLO_\uee
   &\sim
   \frac{ \alpha \, P_{e\to e}(x_\gamma) }{ (t_\form)_{\rm eff} }
   \sim
   \frac{\Nf\alpha^2}{ x_\gamma^{3/2} } \sqrt{ \frac{\qhat}{E} } ,
\end {align}
which also matches the earlier guess (\ref{eq:NLOqed}).

A precise calculation of the rate for $x_\gamma \lesssim \Nf\alpha$
in QED would likely require resumming higher-order effects.
But the analyses here for this case and the
$\Nf\alpha \ll x_\gamma \ll 1$ case provide a qualitative explanation for
(\ref{eq:Ratioqed}) and so for the QED behavior found
in fig.\ \ref{fig:ratio}.


\subsection{Logarithms and a more quantitative comparison}

In ref.\ \cite{qedNfenergy}, it was found that the power-law
behavior (\ref{eq:NLOqed}) of $[d\Gamma/d\xe]_\uee^\NLO$ in the
soft-photon limit was supplemented by a logarithm $\ln(1{-}\xe)$.
Moreover, the overlap rate could be calculated relatively simply to
leading order in this logarithm without
invoking any of the long and complicated formalism for computing
overlapping splitting rates.
Specifically, whenever one splitting has a parametrically small
LO formation time compared to the other, then one can compute
the overlap correction to leading-log order by treating the
splitting with the longer formation time as a vacuum-like
DGLAP initial radiation (or final-state fragmentation) correction
to the leading-order LPM formula for the other splitting.
Ref.\ \cite{qedNfenergy} found that%
\footnote{
   See eq.\ (B.7) of ref.\ \cite{qedNfenergy} and the discussion leading
   up to it, which invokes an earlier argument made in
   appendix B.1 of ref.\ \cite{seq}.
}
\begin {align}
   \left[ \frac{d\Gamma}{d\xe} \right]^\NLO_\uee
   &\approx
     -\frac{3\Nf\alpha^2}{8\pi} \,
     \frac{\ln(1{-}\xe)}{(1{-}\xe)^{3/2}}
     \sqrt{ \frac{\qhat}{E} }
   & \mbox{(QED, $1{-}\xe \ll 1$)} ,
\label{eq:qedlog}
\end {align}
where we use $\approx$ to indicate a leading-log approximation.
This approximation to the NLO net rate corresponds to the dotted
blue curve in the plot of the NLO/LO ratio in fig.\ \ref{fig:ratio}.
As can be seen, it roughly captures the large variation of that
ratio with $\xe$ and so one might expect it to capture a large
part of the overlap effects in QED that cannot be absorbed into $\qhat$.
If we use the recursion relations \cite{qedNfenergy}
for moments of the QED charge deposition distribution, analogous to
the initial-flavor discussion of section \ref{sec:flavor},
but use the approximation (\ref{eq:qedlog}) to the NLO net rate,
we find that the size of overlap corrections to
$\sigma/\lstop$ for charge or initial-flavor deposition would be
$-78\% \times \Nf\alpha$.%
\footnote{
  Useless but fun fact:
  Using the full LO rate for $e{\to}e$ but taking the approximation
  (\ref{eq:qedlog}) for the NLO rate, it is possible to perform all of
  the QED integrals analytically for this particular calculation.
  (See appendix E of ref.\ \cite{qedNfenergy} for the relevant integrals
  involving LO quantities.)  The result is
  \[
     \chi\alpha = -\left[
        \frac{8}{3\pi}
        + \frac{360(\pi-\pi\ln2-1)}{(525\pi-1472)}
     \right] \Nf\alpha ,
  \]
  which, when rounded, equals the $-78\% \times\Nf\alpha$ number quoted above.
}
By itself, the leading-log soft-photon approximation (\ref{eq:qedlog})
explains most of the $-85\% \times \Nf\alpha$ result quoted in
table \ref{tab:chi2}.
This gives support to our explanation that the dramatically
larger size of overlap effects in QED compared to QCD (at comparable
values of $N\alpha$) arises from
the qualitative mismatch between the behavior of soft photon
emission at leading order vs.\ NLO.


\section {Conclusion}
\label {sec:conclusion}

Our focus has been on the size of overlap effects for in-medium
shower development
that cannot be absorbed into an effective value of $\qhat$.
Our result here that such effects are small for
$\Nf{\gg}\Nc{\gg}1$ QCD
suggests that the small size found for
purely gluonic ($\Nf{=}0$) showers was not simply a numerical accident
peculiar to that particular calculation.
Equally important for confidence in the QCD results,
we believe we now understand why similar results
for QED are drastically larger for comparable values of $N\alpha$.
Overlap allows soft bremsstrahlung
photons to convert to a soft electron/positron pair while the
bremsstrahlung is still in progress, and a soft electron/positron pair
is easily scattered by the QED medium, significantly modifying the
LPM effect for the original soft bremsstrahlung.
In contrast, soft gluons already interact easily with the QCD medium
and so overlap does not make the same qualitative change.

\begin {figure}[tp]
\begin {center}
  \includegraphics[scale=0.6]{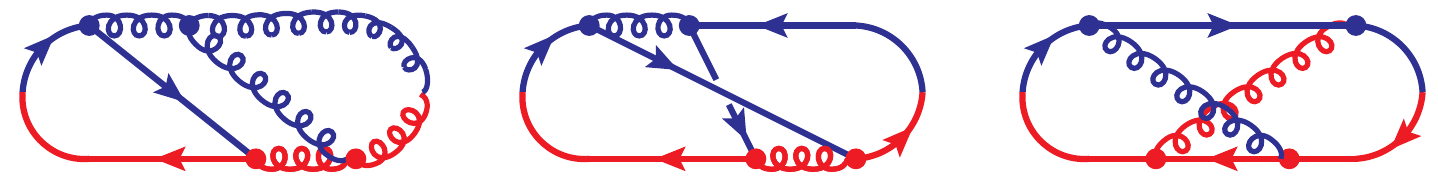}
  \caption{
     \label{fig:more}
     A few examples of additional interference diagrams that would have to be
     computed to generalize existing large-$\Nf$ and $\Nf{=}0$ results to
     any $\Nf$.
  }
\end {center}
\end {figure}

It would be interesting to check our qualitative conclusions for
$\Nf\sim\Nc$ as well, to make sure that nothing changes dramatically.
This would require computing several new interference diagrams, some
examples of which are shown in fig.\ \ref{fig:more}.
In the $\Nf\gg\Nc$ limit of this paper, we were able to ignore the soft
double logarithms that complicated the calculation of
$\qhat$-insensitive overlap effects for purely gluonic ($\Nf{=}0$) showers.
In the purely gluonic case, those soft double logarithms had to be
factorized and carefully absorbed into $\qhat$.
In the future,
studying $\Nf\sim\Nc$ could test whether that factorization might
somehow introduce larger overlap effects in showers equally composed of
quarks and gluons.

\begin{acknowledgments}

The work of Arnold and Elgedawy (while at U. Virginia) was supported,
in part, by the U.S. Department of Energy under Grant No.~DE-SC0007974.
The final phase of Arnold's work on this paper was supported in part by
the National Science Foundation under Grant No.~2412362.
Elgedawy's work at South China Normal University was supported by
Guangdong Major Project of Basic and Applied Basic Research No.\ 2020B0301030008
and by the
National Natural Science Foundation of China under Grant No.\ 12035007.
Arnold is grateful for the hospitality of the EIC Theory Institute
at Brookhaven National Lab for one of the months during which he was
working on this paper.
Iqbal thanks the CERN Theoretical Physics Department for their hospitality
and support during the time that this work was being completed.
We also thank John Collins for sharing his scripts that convert jaxodraw
files into encapsulated postscript files with transparent backgrounds,
which were used in the creation of fig.\ \ref{fig:largeNc0shaded}.

\end{acknowledgments}

\appendix

\section{Summary of Basic Rate Formulas}
\label{app:summary}

In this appendix, we gather our results for the basic rates of
eq.\ (\ref{eq:rates}), adapted as described in section \ref{sec:convert}
from the large-$\Nf$ QED results of ref.\ \cite{qedNf}.
We present the formulas in notation appropriate for
large-$\Nf$ QCD, where
\begin {align}
  \CA&=\Nc, &
  \CF&=\tfrac{\Nc}2 , &
  \tF&=\tfrac12 , &
  \dA&=\Nc^2 , &
  \dF&=\Nc  &
  &\mbox{($\Nc{\gg}1$ QCD)} . \\
\intertext{
  But the results for large-$\Nf$ QED can be recovered by instead using
}
  \CA&=0, &
  \CF&=1, &
  \tF&=1, &
  \dA&=1, &
  \dF&=1  &
  &\mbox{(QED)} ,
\end {align}
replacing $\alphas$ by electromagnetic $\alphaqed$,
and using the few explicitly designated QED versions of formulas below.

\begin {figure}[t]
\begin {center}
  \begin{picture}(440,370)(0,0)
    \put(5,200){\includegraphics[scale=0.4]{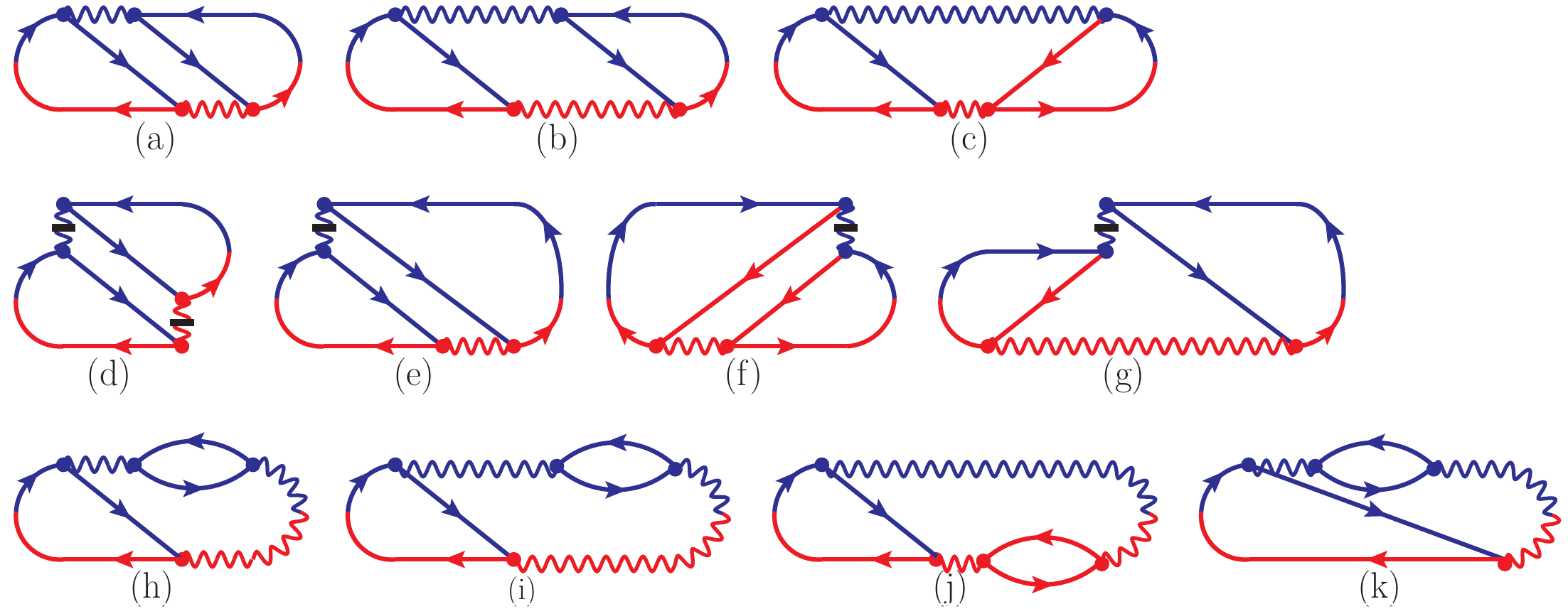}}
    \put(5,0){\includegraphics[scale=0.4]{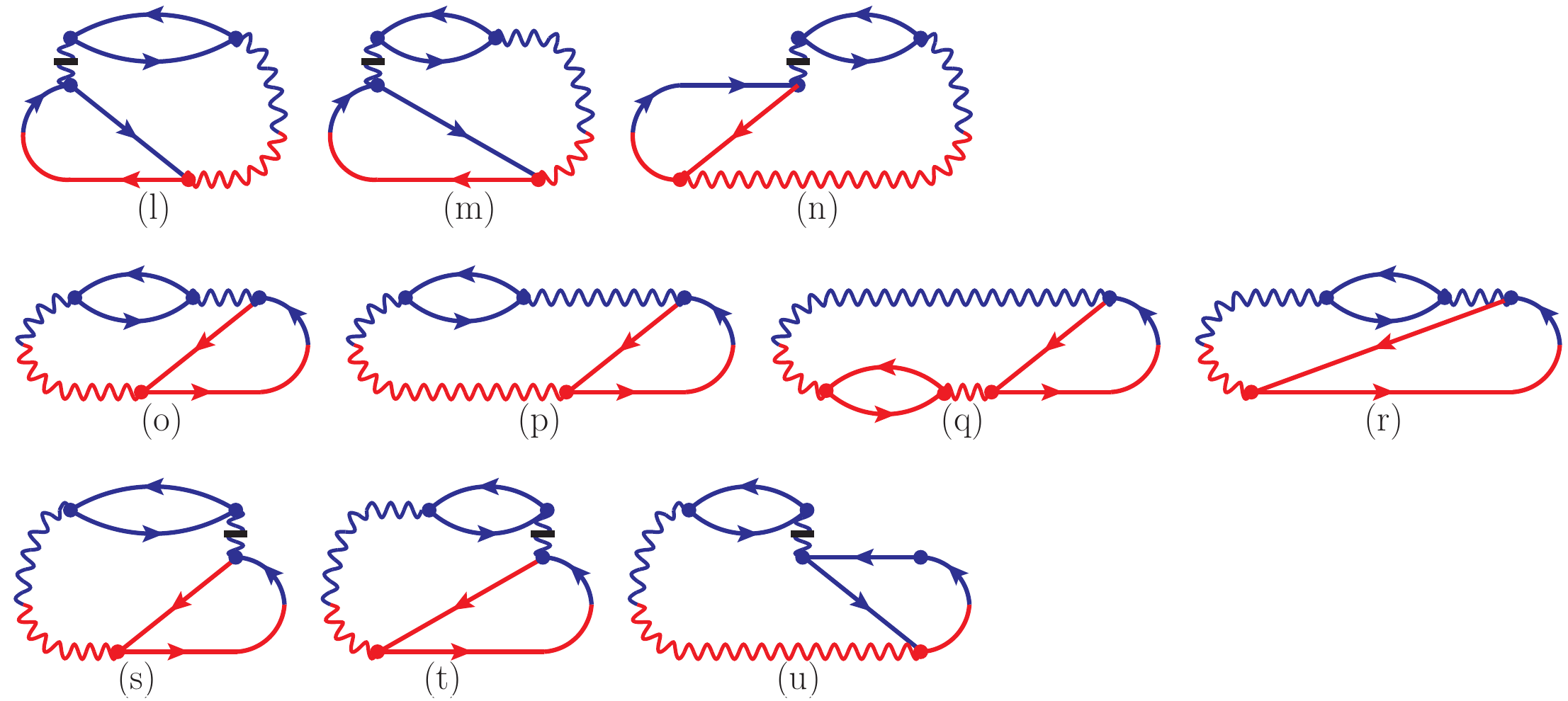}}
%
  \end{picture}
  \caption{
     \label{fig:qeddiags}
     The complete set of NLO time-ordered     
     diagrams needed to compute overlap effects in large-$\Nf$ QED,
     taken from ref.\ \cite{qedNf},
     with the caveat that $2\Re[\cdots]$ should be applied to add in the
     diagrams' complex conjugates.
     Reinterpreting the photon lines as gluon lines gives large-$\Nf$ QCD.
  }
\end {center}
\end {figure}

Many of the formulas below will refer to the contribution of specific
diagrams, as originally labeled in ref.\ \cite{qedNf}.
For reference, we reproduce the complete set of diagrams and labels
in fig.\ \ref{fig:qeddiags}.

In what follows, we will rewrite the $\xQ$ of the main text as
\begin {equation}
   \yQ \equiv \xQ
\end {equation}
just for the sake of (i) making it a little easier to avoid confusion
of $\xq$ and $\xQ$ and (ii) compatibility with the notation of
ref.\ \cite{qedNf}.



\subsection{Leading-order single-splitting rates}
\label {app:LO}


\subsubsection {\boldmath$q \to qg$}

In our notation, the leading-order $q{\to}qg$ rate is
\begin {equation}
   \left[ \frac{d\Gamma}{d\xq} \right]_{q\to qg}^{\rm LO}
   = \frac{\alphas}{\pi} \, P_{q\to q}(\xq) \, \Re(i\Omega_0)
   = \frac{\alphas}{\pi} \, P_{q\to g}(x_g) \, \Re(i\Omega_0)
\label {eq:LO}
\end {equation}
with $x_g = 1-\xq$,
\begin {equation}
   \Omega_0
   = \sqrt{ \frac{-i\qhatF}{2E} \left(
       \frac{1}{\xq} + \frac{\CA}{\CF(1{-}\xq)} - 1
     \right) } \,,
\end {equation}
and the relevant DGLAP splitting function
\begin {equation}
  P_{q\to q}(\xq) = P_{q\to g}(x_g)
  = \CF \frac{1 + \xq^2}{1-\xq} = \CF \frac{1 + (1-x_g)^2}{x_g} .
\end {equation}


\subsubsection {\boldmath$g \to \Q\Qbar$}

\begin {equation}
   \left[ \frac{d\Gamma}{d\yQ} \right]_{g\to\Q\Qbar}^{\rm LO}
   = \frac{\Nf\alphas}{\pi} \, P_{g\to q}(\yQ) \,\Re(i\Omega_0^{g\to\Q\Qbar})
\label {eq:LOpair2}
\end {equation}
with
\begin {equation}
   \Omega_0^{g\to\Q\Qbar}
   = \sqrt{ \frac{-i\qhatF}{2E} \left(
       \frac{1}{\yQ} + \frac{1}{1{-}\yQ} - \frac{\CA}{\CF}
     \right) }
\end {equation}
and
\begin {equation}
  P_{g\to q}(\yQ) = \tF[ \yQ^2 + (1-\yQ)^2 ].
\end {equation}


\subsection{Overlap corrections to real double splitting}
\label {app:realsummary}

The total result for the overlap correction to
real double splitting decomposes as
\begin {equation}
  \left[\frac{\Delta\,d\Gamma}{d\xq\,d\yQ}\right]_{q\to q\Q\Qbar} =
  \left[ \frac{\Delta\,d\Gamma}{d\xq\,d\yQ} \right]_{\rm seq}
  + \left[ \frac{d\Gamma}{d\xq\,d\yQ} \right]_{(I)}
  + \left[ \frac{d\Gamma}{d\xq\,d\yQ} \right]_{(II)} ,
\label {eq:total}
\end {equation}
where the pieces are given below and ``$I$'' indicates a
contribution involving an instantaneous 4-fermion interaction
(arising from longitudinally-polarized gluon exchange).


\subsubsection{Sequential diagrams}
\label {app:seqsummary}

The ``seq'' piece above is our result for diagrams (a--c) of
fig.\ \ref{fig:qeddiags}, which are
of a type referred to as ``sequential'' diagrams in earlier papers.
\begin {equation}
  \left[ \frac{\Delta\,d\Gamma}{d\xq\>d\yQ} \right]_\seq
   =
   2\Nf {\cal A}_\seqNf(\xq,\yQ)
\label {eq:dGammaseqNf}
\end {equation}
with%
\footnote{
  Note the factors of $\tF\CF$ needed for
  (\ref{eq:FseqNf}) and (\ref{eq:Aseqpole}) come from the
  formulas for $P_{q\to q}$ and $P_{g\to q}$.
}
\begin {equation}
   {\cal A}_\seqNf(\xq,\yQ)
   =
   {\cal A}^{\rm pole}_\seqNf(\xq,\yQ)
   + \int_0^{+\infty} d(\Delta t) \>
     \Bigl[
        2 \Re \bigl( B_\seqNf(\xq,\yQ,\Delta t) \bigr)
        + F_\seqNf(\xq,\yQ,\Delta t)
     \Bigr] ,
\end {equation}
\begin {align}
   B_\seqNf(\xq,\yQ,\Delta t) &=
       C_\seqNf(\hat x_1,\hat x_2,\hat x_3,\hat x_4,
               \bar\alpha,\bar\beta,\bar\gamma,\Delta t)
\nonumber\\
   &=
       C_\seqNf({-}1,\yQ,1{-}\xq{-}\yQ,\xq,
               \bar\alpha,\bar\beta,\bar\gamma,\Delta t) ,
\label {eq:Bseq}
\end {align}
\begin {equation}
   C_\seq = D_\seq - \lim_{\hat q\to 0} D_\seq ,
\end {equation}
\begin {align}
   D_\seq(x_1,&x_2,x_3,x_4,\bar\alpha,\bar\beta,\bar\gamma,\Delta t) =
\nonumber\\ &
   \tF\CF \frac{\alphas^2 M_\ix M_\fx^\seq}{32\pi^4 E^2} \, 
   ({-}x_1 x_2 x_3 x_4) \,
   \Omega_+\csc(\Omega_+\Delta t) \, \Omega_-\csc(\Omega_-\Delta t)
\nonumber\\ &\times
   \Bigl\{
     (\bar\beta Y_\yx^\seq Y_\xbx^\seq
        + \bar\alpha \Ybar_{\yx\xbx}^{\,\seq} Y_{\yx\xbx}^\seq) I_0^\seq
     + (\bar\alpha+\bar\beta+2\bar\gamma) Z_{\yx\xbx}^\seq I_1^\seq
\nonumber\\ &\quad
     + \bigl[
         (\bar\alpha+\bar\gamma) Y_\yx^\seq Y_\xbx^\seq
         + (\bar\beta+\bar\gamma) \Ybar_{\yx\xbx}^{\,\seq} Y_{\yx\xbx}^\seq
        \bigr] I_2^\seq
\nonumber\\ &\quad
     - (\bar\alpha+\bar\beta+\bar\gamma)
       (\Ybar_{\yx\xbx}^{\,\seq} Y_\xbx^\seq I_3^\seq+ Y_\yx^\seq Y_{\yx\xbx}^\seq I_4^\seq)
   \Bigl\}
   ,
\label {eq:Dseq}
\end {align}
\begin {align}
   F_\seq(\xq,\yQ,\Delta t) = &
   \frac{ \alphas^2 P_{q\to q}(\xq) P_{g\to q}(\frac{\yQ}{1-\xq}) }
        { 4\pi^2(1-\xq) }
   \Bigl[ 
      \Re\bigl(i(\Omega\sgn M)_\ix\bigr) \,
      \Re\bigl( \Delta t \, (\Omega_\fx^\seq)^2
                \csc^2(\Omega_\fx^\seq \, \Delta t) \bigr)
\nonumber\\ & \qquad
      +
      \Re\bigl(i(\Omega\sgn M)_\fx^\seq) \,
      \Re\bigl( \Delta t \, \Omega_\ix^2 \csc^2(\Omega_\ix \, \Delta t) \bigr)
   \Bigr] ,
\label {eq:FseqNf}
\end {align}
\begin {subequations}
\label {eq:Aseqpole}
\begin {multline}
   {\cal A}_\seq^{\rm pole}(\xq,\yQ)
   = - \frac{\alphas^2 \, P_{q\to q}(\xq) \, P_{g\to q}(\frac{\yQ}{1-\xq})}
          {4\pi^2(1-\xq)}
   \Re\Bigl[
     i (\Omega\sgn M)_\ix \, (1+\tfrac{i\pi}{2} \sgn M_\fx^\seq)
\\
     +
     i (\Omega\sgn M)_\fx^\seq \, (1+\tfrac{i\pi}{2} \sgn M_\ix)
   \Bigr]
\end {multline}
or equivalently
\begin {multline}
   {\cal A}_\seq^{\rm pole}(\xq,\yQ) =
   - \frac{\alphas^2 \, P_{q\to q}(\xq) \, P_{g\to q}(\frac{\yQ}{1-\xq})}
          {4\pi^2(1-\xq)}
   \Re\bigl[
     i (\Omega\sgn M)_\ix + i (\Omega\sgn M)_\fx^\seq
   \bigr]
\\ \times
   \bigl(1-\tfrac{\pi}{2} \sgn M_\ix \sgn M_\fx^\seq\bigr) ,
\end {multline}
\end {subequations}
and
\begin {equation}
   M_\ix = x_1 x_4 (x_1+x_4) E ,
   \qquad
   M_\fx^\seq = x_2 x_3 (x_2+x_3) E .
\label {eq:Mifseq}
\end {equation}
The $I_n^{\rm seq}$ above represent
\begin {subequations}
\label {eq:Iseq}
\begin {align}
   I_0^\seq &=
   \frac{4\pi^2}{[X_\yx^\seq X_\xbx^\seq - (X_{\yx\xbx}^\seq)^2]} \,,
\displaybreak[0]\\
   I_1^\seq &=
   - \frac{2\pi^2}{X_{\yx\xbx}^\seq}
   \ln\left( 1 - \frac{(X_{\yx\xbx}^\seq)^2}{X_\yx^\seq X_\xbx^\seq} \right) \,,
\displaybreak[0]\\
   I_2^\seq &=
   \frac{2\pi^2}{(X_{\yx\xbx}^\seq)^2}
     \ln\left( 1 - \frac{(X_{\yx\xbx}^\seq)^2}{X_\yx^\seq X_\xbx^\seq} \right)
   + \frac{4\pi^2}{[X_\yx^\seq X_\xbx^\seq - (X_{\yx\xbx}^\seq)^2]} \,,
\displaybreak[0]\\
   I_3^\seq &=
   \frac{4\pi^2 X_{\yx\xbx}^\seq}
        {X_\xbx^\seq[X_\yx^\seq X_\xbx^\seq - (X_{\yx\xbx}^\seq)^2]} \,,
\displaybreak[0]\\
   I_4^\seq &=
   \frac{4\pi^2 X_{\yx\xbx}^\seq}
        {X_\yx^\seq[X_\yx^\seq X_\xbx^\seq - (X_{\yx\xbx}^\seq)^2]} \,.
\end {align}
\end {subequations}
Here and in (\ref{eq:Dseq}), the $(X,Y,Z)^\seq$ are defined by 
\begin {subequations}
\label {eq:XYZseq}
\begin {align}
   \begin{pmatrix} X_\yx^\seq & Y_\yx^\seq \\ Y_\yx^\seq & Z_\yx^\seq \end{pmatrix}
   &\equiv
   \begin{pmatrix} |M_\ix|\Omega_\ix & 0 \\ 0 & 0 \end{pmatrix}
     - i a_\yx^{-1\top}
     \begin{pmatrix}
        \Omega_+ \cot(\Omega_+\,\Delta t) & 0 \\
        0 & \Omega_- \cot(\Omega_-\,\Delta t)
     \end{pmatrix}
     a_\yx^{-1} ,
\\
   \begin{pmatrix} X_\xbx^\seq & Y_\xbx^\seq \\ Y_\xbx^\seq & Z_\xbx^\seq \end{pmatrix}
   &\equiv
   \begin{pmatrix} |M_\fx^\seq|\Omega_\fx^\seq & 0 \\ 0 & 0 \end{pmatrix}
     - i (a_\xbx^\seq)^{-1\top}
     \begin{pmatrix}
        \Omega_+ \cot(\Omega_+\,\Delta t) & 0 \\
        0 & \Omega_- \cot(\Omega_-\,\Delta t)
     \end{pmatrix}
     (a_\xbx^\seq)^{-1} ,
\label {eq:XYZxbseq}
\\
   \begin{pmatrix} X_{\yx\xbx}^\seq & Y_{\yx\xbx}^\seq \\
                   \Ybar_{\yx\xbx}^\seq & Z_{\yx\xbx}^\seq \end{pmatrix}
   &\equiv
     - i a_\yx^{-1\top}
     \begin{pmatrix}
        \Omega_+ \csc(\Omega_+\,\Delta t) & 0 \\
        0 & \Omega_- \csc(\Omega_-\,\Delta t)
     \end{pmatrix}
     (a_\xbx^\seq)^{-1} ,
\end {align}
\end {subequations}
where the $a$'s and $\Omega$'s will be given below.
The quantities $(\bar\alpha,\bar\beta,\bar\gamma)$ in (\ref{eq:Bseq})
represent
various combinations of helicity-dependent DGLAP splitting functions
and are
\begin {equation}
   \begin{pmatrix}
      \bar\alpha \\ \bar\beta \\ \bar\gamma
   \end{pmatrix}_{\!\!q \to q\Q\Qbar}
   =
   \frac{1}{(1-\xq)^6}
   \left\{
     \begin{pmatrix} - \\ + \\ + \end{pmatrix}
       \frac{4}{|\xq\yQ\zQbar|}
   + \begin{pmatrix} + \\ + \\ - \end{pmatrix}
       \left[ \Bigl(1 + \frac{1}{\xq^2}\Bigr)
              \Bigl(\frac{1}{\yQ^2} + \frac{1}{\zQbar^2}\Bigr)
       \right]
   \right\}
   ,
\label {eq:abcNf}
\end {equation}
where
\begin {equation}
   \zQbar \equiv 1{-}\xq{-}\yQ .
\end {equation}
The 3-body frequencies needed in the above formulas are
\begin {equation}
   \Omega_\ix(x_1,x_2,x_3,x_4) =
     \sqrt{
       \frac{-i \qhatF}{2E}
       \Bigl(
          \frac{1}{x_1} + \frac{1}{x_4} - \frac{\CA}{\CF (x_1{+}x_4)}
       \Bigr)
     }
    \,,
\label {eq:Omegai2}
\end {equation}
\begin {equation}
   \Omega_\fx^\seq(x_1,x_2,x_3,x_4)
   =
     \sqrt{
       \frac{-i \qhatF}{2E}
       \Bigl(
          \frac{1}{x_2} + \frac{1}{x_3} - \frac{\CA}{\CF (x_2{+}x_3)}
       \Bigr)
     } .
\label{eq:Omegafseq4}
\end {equation}
We don't have neat packaging of the formulas for 4-body normal mode
frequencies $\Omega_\pm$ and matrices $a$ of normal modes
that covers both QCD and QED, so we
present them as separate cases.
Because of branch cut issues when later using these formulas with
front-end transformations, where variables such as $x_4$ may be negative,
it is important to avoid making algebraic simplifications like
$x_4/\sqrt{x_4} = \sqrt{x_4}$ in (\ref{eq:ayqcd}) and (\ref{eq:ayqed})
below.%
\footnote{
  As long as one uses a consistent convention for the square roots of
  negative numbers, e.g.\ they are always positive imaginary, the
  columns of $a_y$ resulting from
  eqs.\ (\ref{eq:ayqcd}) and (\ref{eq:ayqed}) will
  correctly give normal modes even in the case of front-end transformations.
  Consistently choosing the square roots of negative numbers to be
  negative imaginary would also work, since the overall sign of
  normal modes is a matter of convention and does not affect the
  combinations (\ref{eq:XYZseq}) that depend directly or indirectly
  on $a_y$.  However, changes like $x_4/\sqrt{x_4} \rightarrow \sqrt{x_4}$
  can incorrectly
  alter relative signs within a normal mode rather than just
  its overall sign.
}

\medskip
\noindent $\Nf{\gg}\Nc{\gg}1$ \textit{QCD}:

\begin {subequations}
\begin {equation}
  \Omega_+ =
    \sqrt{
       -\frac{i\qhatF}{2E} \left( \frac{1}{x_3} + \frac{1}{x_4} \right)
    } ,
  \qquad
  \Omega_- =
    \sqrt{
       -\frac{i\qhatF}{2E} \left( \frac{1}{x_1} + \frac{1}{x_2} \right)
    } ,
\label {eq:OmegaPMqcd}
\end {equation}
\begin {equation}
   a_y =
   \frac{1}{(x_1{+}x_4)}
   \begin{pmatrix}
       -x_3 & -x_2 \\
        \phantom{-}x_4 &  \phantom{-}x_1
   \end {pmatrix}
   \begin{pmatrix}
     \frac{1}{\sqrt{x_3 x_4 (x_3{+}x_4)E}} & 0 \\
     0 & \frac{1}{\sqrt{x_1 x_2 (x_1{+}x_2)E}}
   \end{pmatrix} .
\label {eq:ayqcd}
\end {equation}
\end {subequations}

\medskip
\noindent \textit{large-$\Nf$ QED}:

\begin {subequations}
\begin {equation}
   \Omega_+ =
   \sqrt{
     -\frac{i \qhatF}{2E}
     \Bigl( \frac{1}{x_1} + \frac{1}{x_2} + \frac{1}{x_3} + \frac{1}{x_4} \Bigr)
   }
   ,
   \qquad
   \Omega_- = 0 ,
\label {eq:OmegaPMqed}
\end {equation}
\begin {equation}
   a_\yx =
   \Bigl[
     (-x_1 x_2 x_3 x_4)
     \bigl(\frac{1}{x_1}+\frac{1}{x_2}+\frac{1}{x_3}+\frac{1}{x_4}\bigr)
     E
    \Bigr]^{-1/2}
   \begin {pmatrix}
      \frac{x_2 x_3}{(-x_1 x_2 x_3 x_4)^{1/2}} & ~~1\\[8pt]
      \frac{x_1 x_4}{(-x_1 x_2 x_3 x_4)^{1/2}} & ~~1
   \end{pmatrix}
   ,
\label {eq:ayqed}
\end {equation}
with the understanding that every $\Omega_- \cot(\Omega_-\,\Delta t)$
or $\Omega_- \csc(\Omega_-\,\Delta t)$ should be interpreted as
$1/\Delta t$ when $\Omega_- = 0$.
\end {subequations}

In both cases,
\begin {equation}
   a_\xbx^\seq =
   \begin{pmatrix} 0 & 1 \\ 1 & 0 \end{pmatrix} a_\yx .
\label {eq:abx}
\end {equation}


\subsubsection{Diagrams with one instantaneous vertex}
\label {app:Isummary}

The ``$(I)$'' piece of our result (\ref{eq:total}) is the contribution
from diagrams (e--g) of fig.\ \ref{fig:qeddiags}.
\begin {subequations}
\label {eq:dGIsummary}
\begin {equation}
   \left[ \frac{d\Gamma}{d\xq\>d\yQ} \right]_{(I)}
   =
   2\Nf \, {\cal A}_I(\xq,\yQ) ,
\label {eq:dGammaI}
\end {equation}
\begin {equation}
   {\cal A}_I(\xq,\yQ)
   \equiv
   \int_0^{\infty} d(\Delta t) \>
        2 \Re \bigl( B_I(\xq,\yQ,\Delta t) \bigr)
   ,
\end {equation}
\begin {align}
   B_I(\xq,\yQ,\Delta t) &=
       \frac{ 4 \bigl|\xq\yQ(1{-}\xq{-}\yQ)\bigr|^{1/2} }{ (1-\xq)^2 }
       \Bigl[
         D_I(\hat x_1,\hat x_2,\hat x_3,\hat x_4,\zeta,\Delta t)
\nonumber\\ & \hspace{10em}
         + D_I(-\hat x_3,-\hat x_4,-\hat x_1,-\hat x_2,\zeta,\Delta t)
      \Bigr]
\nonumber\\
   &=
       \frac{ 4 \bigl|\xq\yQ(1{-}\xq{-}\yQ)\bigr|^{1/2} }{ (1-\xq)^2 }
       \Bigl[
         D_I({-}1,\yQ,1{-}\xq{-}\yQ,\xq,\zeta,\Delta t)
\nonumber\\ & \hspace{10em}
         + D_I(-(1{-}\xq{-}\yQ),-\xq,1,-\yQ,\zeta,\Delta t)
       \Bigr]
   ,
\label {eq:BI}
\end {align}
\begin {multline}
   D_I(x_1,x_2,x_3,x_4,\zeta,\Delta t) =
\\
   - \tF\CF \frac{\alphas^2 M_\fx^\seq}{16 \pi^2 E} \,
   (-x_1 x_2 x_3 x_4)
   \zeta \,
   \Omega_+ \csc(\Omega_+\,\Delta t) \,\Omega_- \csc(\Omega_-\,\Delta t) \,
   \frac{Y_\xbx^\seq}{X_\xbx^\seq}
   \,,
\label {eq:DI}
\end {multline}
\end {subequations}
where
\begin {equation}
   \zeta =
   \frac{(1{+}|\xq|)\,(|\yQ|{+}|\zQbar|)}
        {(1-\xq)^3 |\xq\yQ\zQbar|^{3/2}} \,.
\label {eq:zeta}
\end {equation}
$M_\fx^\seq$, $\Omega_\fx^\seq$, and $\Omega_\pm$ are given here by
the previous general formulas (\ref{eq:Mifseq}), (\ref{eq:Omegafseq4}),
and (\ref{eq:OmegaPMqcd}) or (\ref{eq:OmegaPMqed}),
for use in (\ref{eq:XYZxbseq})
for $X_\xbx^\seq$ and $Y_\xbx^\seq$.

Later, we will need to refer separately to the contributions of the
three diagrams (e--g). The above formula
for $[d\Gamma/d\xq\,d\yQ]_{(I)}$ can be decomposed as
\begin {equation}
   \left[ \frac{d\Gamma}{d\xq\>d\yQ} \right]_{(I)}
   =
   2\Re \left[ \frac{d\Gamma}{d\xq\>d\yQ} \right]_{\rm(e)}
   +
   2\Re \left[ \frac{d\Gamma}{d\xq\>d\yQ} \right]_{\rm(f)}
   +
   2\Re \left[ \frac{d\Gamma}{d\xq\>d\yQ} \right]_{\rm(g)}
\end {equation}
where
\begin {align}
   2\Re \left[ \frac{d\Gamma}{d\xq\>d\yQ} \right]_{\rm(e)}
   &=
   \mbox{Eqs.\ (\ref{eq:dGIsummary}) using only the
         {\it first} $D_I$ term in (\ref{eq:BI});}
\label {eq:dGe}
\\
   2\Re \left[ \frac{d\Gamma}{d\xq\>d\yQ} \right]_{\rm(f)}
   &=
   \mbox{Eqs.\ (\ref{eq:dGIsummary}) using only the
         {\it second} $D_I$ term in (\ref{eq:BI});}
\label {eq:dGf}
\\
   2\Re \left[ \frac{d\Gamma}{d\xq\>d\yQ} \right]_{\rm(g)}
   &=
   0.
\label {eq:dGg}
\end {align}


\subsubsection{Diagrams with two instantaneous vertices}
\label {app:IIsummary}

The ``$(II)$'' piece of our result (\ref{eq:total}) is the contribution
from diagram (d) of fig.\ \ref{fig:qeddiags}, which gives%
\footnote{
  Eq.\ (\ref{eq:II}) above is $2\Re[\cdots]$ of the large-$\Nf$ QED to
  $\Nf{\gg}\Nc{\gg}1$ QCD generalization of
  the first line of eq.\ (E.21) of ref.\ \cite{qedNf}.
  Unlike the QED case shown there and in eq.\ (A.35) of ref.\ \cite{qedNf},
  we do not know how to do the integral
  analytically for the QCD case.
}
\begin {multline}
   \left[\frac{d\Gamma}{d\xq\,d\yQ}\right]_{(II)}
   =
   2\Re \left[\frac{d\Gamma}{d\xq\,d\yQ}\right]_{\rm(d)}
\\
   =   
   - \tF\CF \frac{2\Nf \alphas^2}{\pi^2} \,
   \frac{\xq\yQ\zQbar}{(1-\xq)^4}
   \int_0^\infty d(\Delta t) \,
   2\Re
   \left[
     \Omega_+ \csc(\Omega_+\,\Delta t) \, \Omega_- \csc(\Omega_-\,\Delta t)
     - \frac{1}{(\Delta t)^2}
   \right]
  .
\label {eq:II}
\end {multline}
This completes the set of formulas needed to numerically evaluate
$\Delta[d\Gamma/d\xq\,d\yQ]_{q\to q\Q\Qbar}$.


\subsection{NLO corrections to single splitting \boldmath$q\to qg$}

We will decompose the contributions of figs.\ \ref{fig:qeddiags}(h--n) to
single splitting $q \to qg$ as
\begin {equation}
  \left[ \frac{\Delta\,d\Gamma}{d\xq} \right]_{q\to qg}^{\rm NLO}
  =
  2\Re \biggl\{
    \left[ \frac{\Delta\,d\Gamma}{d\xq} \right]_{\rm(h+i+j)}
    +
    \left[ \frac{d\Gamma}{d\xq} \right]_{\rm(k)}
    +
    \left[ \frac{d\Gamma}{d\xq} \right]_{\rm(l)}
    +
    \left[ \frac{d\Gamma}{d\xq} \right]_{\rm(m)}
    +
    \left[ \frac{d\Gamma}{d\xq} \right]_{\rm(n)}
  \biggr\} .
\end {equation}
By back-end transformation \cite{qedNf},
\begin {align}
   2 \Re
   \left[ \frac{\Delta\,d\Gamma}{d\xq} \right]_{\rm (h+i+j)}
   &= - \int_0^{1-\xq} d\yQ \>
        \left[\frac{\Delta\,d\Gamma}{d\xq\,d\yQ}\right]_{\rm seq}
   ,
\\
   2 \Re
   \left[ \frac{d\Gamma}{d\xq} \right]_{\rm (l)} \quad
   &= - \int_0^{1-\xq} d\yQ \>
         2\Re\left[\frac{d\Gamma}{d\xq\,d\yQ}\right]_{\rm(e)}
   ,
\\
   2 \Re
   \left[ \frac{d\Gamma}{d\xq} \right]_{\rm (n)} \quad
   &= 0 ,
\end {align}
where the integrands on the right-hand side are specified in
(\ref{eq:dGammaseqNf}), (\ref{eq:dGe}) and implicitly (\ref{eq:dGg}).
By combined front- and back-end transformation \cite{qedNf},
\begin {equation}
   2\Re
   \left[ \frac{d\Gamma}{d\xq} \right]_{\rm (m)}
   = + \int_0^{1-\xq} d\yQ \>
   \left\{
     2\Re \left[ \frac{d\Gamma}{d\xq\,d\yQ} \right]_{\rm (e)}
     \mbox{with}~
     (\xq,\yQ,E) \to
     \Bigl(
      \frac{-\yQ}{\zQbar} \,,\,
      \frac{-\xq}{\zQbar} \,,\,
      \zQbar E\Bigr)
   \right\}
\end {equation}
with $2\Re[d\Gamma/d\xq\,d\yQ]_{\rm(e)}$ again given by (\ref{eq:dGe}).

Finally, diagram (k) gives
\begin {multline}
  2\Re\left[ \frac{d\Gamma}{d\xq} \right]_{\rm(k)}
  =
  2\Re\biggl\{ 
  -
  \frac{\beta_0\alphas}{2}
  \left[ \frac{d\Gamma}{d\xq} \right]_{x\bar x} 
  \biggl(
    \ln\Bigl( \frac{\mu^2}{(1{-}\xq)E\Omega_\ix\sgn M_\ix} \Bigr)
    + \gammaE
    - 2\ln2
    + \tfrac53
  \biggr)
  \biggr\}
\\
  +
  \int_0^{1-\xq} d\yQ \>
    2\Re\left[ \frac{d\Gamma}{d\xq\,d\yQ} \right]_{xyy\bar x}^{(\rm subtracted)} ,
\label {eq:dGk}
\end {multline}
where the coefficient of the 1-loop renormalization group $\beta$-function
for $\alphas$ is (in the large-$\Nf$ limit)
\begin {equation}
   \beta_0 = \frac{2\tF\Nf}{3\pi}
\label {eq:beta02}
\end {equation}
and
\begin {equation}
  \left[ \frac{d \Gamma}{d\xq} \right]_{x\bar x}
  = \frac{\alphas}{2\pi} \, P_{q\to q}(\xq) \, i\Omega_\ix \sgn M_\ix
\label {eq:xxd2}
\end {equation}
is the $x\bar x$ diagram%
\footnote{
  $x\bar x$ refers to the naming convention for
  individual rate diagrams in fig.\ 31 of ref.\ \cite{qedNf}.
}
from which $2\Re[\cdots]$ gives the
leading-order bremsstrahlung rate (\ref{eq:LO}).
The ``subtracted'' rate above is
\begin {align}
  \left[\frac{d\Gamma}{d\xq\,d\yQ}\right]_{xyy\bar x}^{\rm(subtracted)} =
  - & \tF\CF \frac{\Nf\alphas^2 M_\ix^2}{16\pi^4 E^2} \,
  ({-}\hat x_1 \hat x_2 \hat x_3 \hat x_4)
  \int_0^\infty d(\Delta t)
  \Biggl[
\nonumber\\ &
   \Omega_+\csc(\Omega_+\Delta t) \, \Omega_-\csc(\Omega_-\Delta t)
\nonumber\\ &\quad
   \times\Bigl\{
     (\bar\beta Y_\yx^2
        + \bar\gamma \Ybar_{\yx\yx'} Y_{\yx\yx'}) I_0^\new
     + (2\bar\alpha+\bar\beta+\bar\gamma) Z_{\yx\yx'} I_1^\new 
\nonumber\\ &\quad\quad
     + \bigl[
         (\bar\alpha+\bar\gamma) Y_\yx^2
         + (\bar\alpha+\bar\beta) \Ybar_{\yx\yx'} Y_{\yx\yx'}
        \bigr] I_2^\new
\nonumber\\ &\quad\quad
     - (\bar\alpha+\bar\beta+\bar\gamma)
       (\Ybar_{\yx\yx'} Y_\yx I_3^\new + Y_\yx Y_{\yx\yx'} I_4^\new)
   \Bigl\}
\nonumber\\ &
   - (2\bar\alpha+\bar\beta+\bar\gamma)
     \frac{\hat x_2\hat x_3}{\hat x_1\hat x_4} \, {\cal D}_2^{(\bbI)}
  \Biggr]
\end {align}
and
\begin {equation}
   {\cal D}_2^{(\bbI)}(\Delta t) =
   2\pi^2
   \left[
     \frac{\ln(2i\Omega_\ix \,\Delta t\sgn M_\ix)}{(\Delta t)^2}
     - i\Omega_\ix^3\,\Delta t \csc^2(\Omega_\ix\, \Delta t) \sgn M_\ix
   \right] .
\label {eq:D2Isummary}
\end {equation}
Here the $I^{\rm new}_n$ are the same as the
$I^\seq_n$ of (\ref{eq:Iseq}) except that the $(X,Y,Z)^\seq$ there are
replaced by
\begin {subequations}
\begin {align}
   \begin{pmatrix} X_\yx^\new & Y_\yx^\new \\ Y_\yx^\new & Z_\yx^\new \end{pmatrix}
   &\equiv
   \begin{pmatrix} |M_\ix|\Omega_\ix & 0 \\ 0 & 0 \end{pmatrix}
     - i a_\yx^{-1\top}
     \begin{pmatrix}
        \Omega_+ \cot(\Omega_+\,\Delta t) & 0 \\
        0 & \Omega_- \cot(\Omega_-\,\Delta t)
     \end{pmatrix}
     a_\yx^{-1} ,
\\
   \begin{pmatrix} X_{\yx'}^\new & Y_{\yx'}^\new
                   \\ Y_{\yx'}^\new & Z_{\yx'}^\new \end{pmatrix}
   &\equiv
   \begin{pmatrix} X_\yx^\new & Y_\yx^\new \\ Y_\yx^\new & Z_\yx^\new \end{pmatrix} ,
\\
   \begin{pmatrix} X_{\yx\yx'}^\new & Y_{\yx\yx'}^\new \\[2pt]
                   \Ybar_{\yx\yx'}^\new & Z_{\yx\yx'}^\new \end{pmatrix}
   &\equiv
     - i a_\yx^{-1\top}
     \begin{pmatrix}
        \Omega_+ \csc(\Omega_+\,\Delta t) & 0 \\
        0 & \Omega_- \csc(\Omega_-\,\Delta t)
     \end{pmatrix}
     a_\yx^{-1} .
\end {align}
\end {subequations}
The $M$'s, $\Omega$'s and $a$'s are as in section
\ref{app:seqsummary} with $(x_1,x_2,x_3,x_4)$ set
to
$
  (\hat x_1,\hat x_2,\hat x_3,\hat x_4)=
  (-1,\yQ,1{-}\xq{-}\yQ,\xq)
$.
The only reason that the
factors of $\sgn M_\ix$ in (\ref{eq:dGk}--\ref{eq:D2Isummary}) are necessary
is to accommodate the transformation of diagram (k) that will later
be used to evaluate diagram (r).

The specific additive constants shown in (\ref{eq:dGk})
assume that the coupling $\alphas$ used in the leading-order calculation
(\ref{eq:LO}) of single splitting
is $\alphas(\mu)$ in the modified minimal subtraction ($\overline{\rm MS}$)
scheme.
To use a different renormalization scheme, one should convert
(\ref{eq:dGk}) accordingly, but nothing else would change.


\subsection{NLO corrections to single splitting \boldmath$g\to\Q\Qbar$}

We decompose the contributions of figs.\ \ref{fig:qeddiags}(o--u) to
single splitting $g\to\Q\Qbar$ as
\begin {equation}
  \left[ \frac{\Delta\,d\Gamma}{d\yQ} \right]_{g\to\Q\Qbar}^{\rm NLO}
  =
  2\Re \biggl\{
    \left[ \frac{\Delta\,d\Gamma}{d\yQ} \right]_{\rm(o+p+q)}
    +
    \left[ \frac{d\Gamma}{d\yQ} \right]_{\rm(r)}
    +
    \left[ \frac{d\Gamma}{d\yQ} \right]_{\rm(s)}
    +
    \left[ \frac{d\Gamma}{d\yQ} \right]_{\rm(t)}
    +
    \left[ \frac{d\Gamma}{d\yQ} \right]_{\rm(u)}
  \biggr\} .
\end {equation}
By front-end transformation \cite{qedNf},
modified as in section \ref{sec:frontend} of the current paper,
\begin {align}
   2 \Re
   \left[ \frac{\Delta\,d\Gamma}{d\yQ} \right]_{\rm (o+p+q)}
   &= - \frac{\dF}{\dA} \Nf \int_0^1 d\xq
   \left\{
     \left[\frac{\Delta\,d\Gamma}{d\xq\,d\yQ}\right]_{\rm seq}
     \mbox{with (\ref{eq:frontend1b}) below}
   \right\}
   ,
\\
   2 \Re
   \left[ \frac{d\Gamma}{d\yQ} \right]_{\rm (s)} \quad
   &= - \frac{\dF}{\dA} \Nf \int_0^1 d\xq \>
   \left\{
         2\Re\left[\frac{d\Gamma}{d\xq\,d\yQ}\right]_{\rm(f)}
     \mbox{with (\ref{eq:frontend1b}) below}
   \right\}
   ,
\\
   2 \Re
   \left[ \frac{d\Gamma}{d\yQ} \right]_{\rm (u)} \quad
   &= 0 ,
\end {align}
where
\begin {equation}
     (\xq,\yQ,E) \to
     \Bigl(\frac{-\xq}{1-\xq}\,,\,\frac{\yQ}{1-\xq}\,,\,(1{-}\xq)E\Bigr)
     .
\label {eq:frontend1b}
\end {equation}
By front- and back-end transformation \cite{qedNf},
\begin {equation}
   2\Re
   \left[ \frac{d\Gamma}{d\yQ} \right]_{\rm (t)}
   \!= \frac{\dF}{\dA} \Nf \int_0^1 d\xq
   \left\{
     2\Re\left[ \frac{d\Gamma}{d\xq\,d\yQ} \right]_{\rm (f)}
     \mbox{with}~
   (\xq,\yQ,E) \to
   \Bigl(\frac{-(1{-}\yQ)}{\yQ},\frac{\xq}{\yQ},\yQ E\Bigr)
   \right\} \!.
\end {equation}
Finally, the result (\ref{eq:dGk}) for diagram (k) can
be transformed to%
\footnote{
  See eq.\ (4.40) of ref.\ \cite{qedNf}.
  Also see appendix H of ref.\ \cite{qedNf}
  for technicalities on getting the first
  term in (\ref{eq:dGr}) from the transformation of (\ref{eq:dGk}).
  Also, we have not bothered to write any general $\sgn M$ factors in
  (\ref{eq:dGr}--\ref{eq:yybar}) that would allow this result
  to in turn be transformed
  back again to diagram (k).  Instead, here we have just specialized to
  the specific values of $\sgn M$ of diagram (r).
}
\begin {align}
  2&\Re\left[ \frac{d\Gamma}{d\yQ} \right]_{\rm(r)}
  \!=
  2\Re\biggl\{ 
  -
  \frac{\beta_0\alphas}{2}
  \left[ \frac{d\Gamma}{d\yQ} \right]_{y\bar y}^{g\to\Q\Qbar} 
  \biggl(
    \ln\Bigl( \frac{\mu^2}{E\Omega_0^{g\to\Q\Qbar}} \Bigr)
    - i\pi
    + \gammaE
    - 2\ln2
    + \tfrac53
  \biggr)
  \biggr\}
\nonumber\\ &
  +
  \frac{\dF}{\dA} \Nf \int_0^1 d\xq
   \left\{
     2\Re\left[ \frac{d\Gamma}{d\xq\,d\yQ} \right]_{xyy\bar x}^{(\rm subtracted)}
     \mbox{with}~
   (\xq,\yQ,E) \to
   \Bigl(\frac{-\yQ}{1{-}\yQ},\frac{\xq}{1{-}\yQ},(1{-}\yQ)E\Bigr)
   \right\} ,
\label {eq:dGr}
\end {align}
where
\begin {equation}
  \left[ \frac{d \Gamma}{d\yQ} \right]_{y\bar y}
  = \frac{\Nf\alphas}{2\pi} \, P_{g\to q}(\yQ)
         \, i\Omega_0^{g\to\Q\Qbar}
\label {eq:yybar}
\end {equation}
is the $y\bar y$ diagram from which $2\Re[\cdots]$ gives the
leading-order pair production rate (\ref{eq:LOpair2}).


\section{\boldmath$\ln x$ or $\ln(1{-}x)$ behavior in eqs.\ (\ref{eq:fits})
         for $f_{i\to j}(x)$}
\label{app:logs}

This appendix follows appendix B of ref.\ \cite{qedNfenergy}, which should
be read first.  That appendix discussed
$\ln x$ and $\ln(1{-}x)$ behavior of the functions $f_\uij(x)$ in the case
of large-$\Nf$ QED.
Here, we merely add to that discussion to explain where
large-$\Nf$ QCD is different from large-$\Nf$ QED.


\subsection{\boldmath$q \to q\Q\Qbar$}

\begin {figure}[t]
\begin {center}
  \includegraphics[scale=0.7]{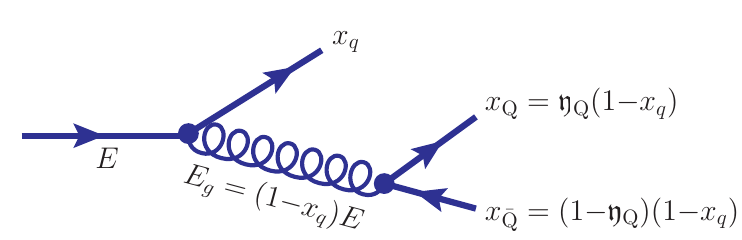}
  \caption{
     \label {fig:qQQnotation2}
     $q \to q\Q\Qbar$ process
  }
\end {center}
\end {figure}

We start by looking at $f_{q\to q}^\real(x)$ (\ref{eq:fqqReal}) for
real, overlapping splitting $q\to q\Q\Qbar$.  The notation we use is
shown in fig.\ \ref{fig:qQQnotation2}, where
$E_g = (1{-}\xq)E$ and $\yfrakQ \equiv \xQ/x_g = \xQ/(1{-}\xq)$.
The formation times for LO splittings
$q\to qg$ and $g\to\Q\Qbar$ with these energies are parametrically
\begin {subequations}
\label {eq:tformsReal}
\begin {align}
  t_{\rm form}^{q\to qg} &\sim \sqrt{ \frac{\xq\xg E}{\qhat} }
  = \sqrt{ \frac{\xq(1{-}\xq)E}{\qhat} } \,,
\\
  t_{\rm form}^{g\to\Q\Qbar} &\sim
    \sqrt{ \frac{\yfrakQ\yfrakQbar E_g}{\qhat} }
  = \sqrt{ \frac{\yfrakQ(1{-}\yfrakQ)(1{-}\xq)E}{\qhat} } \,.
\end {align}
\end {subequations}

\subsubsection{\boldmath$\xq\to 1$}

\begin {figure}[t]
\begin {center}
  \includegraphics[scale=0.3]{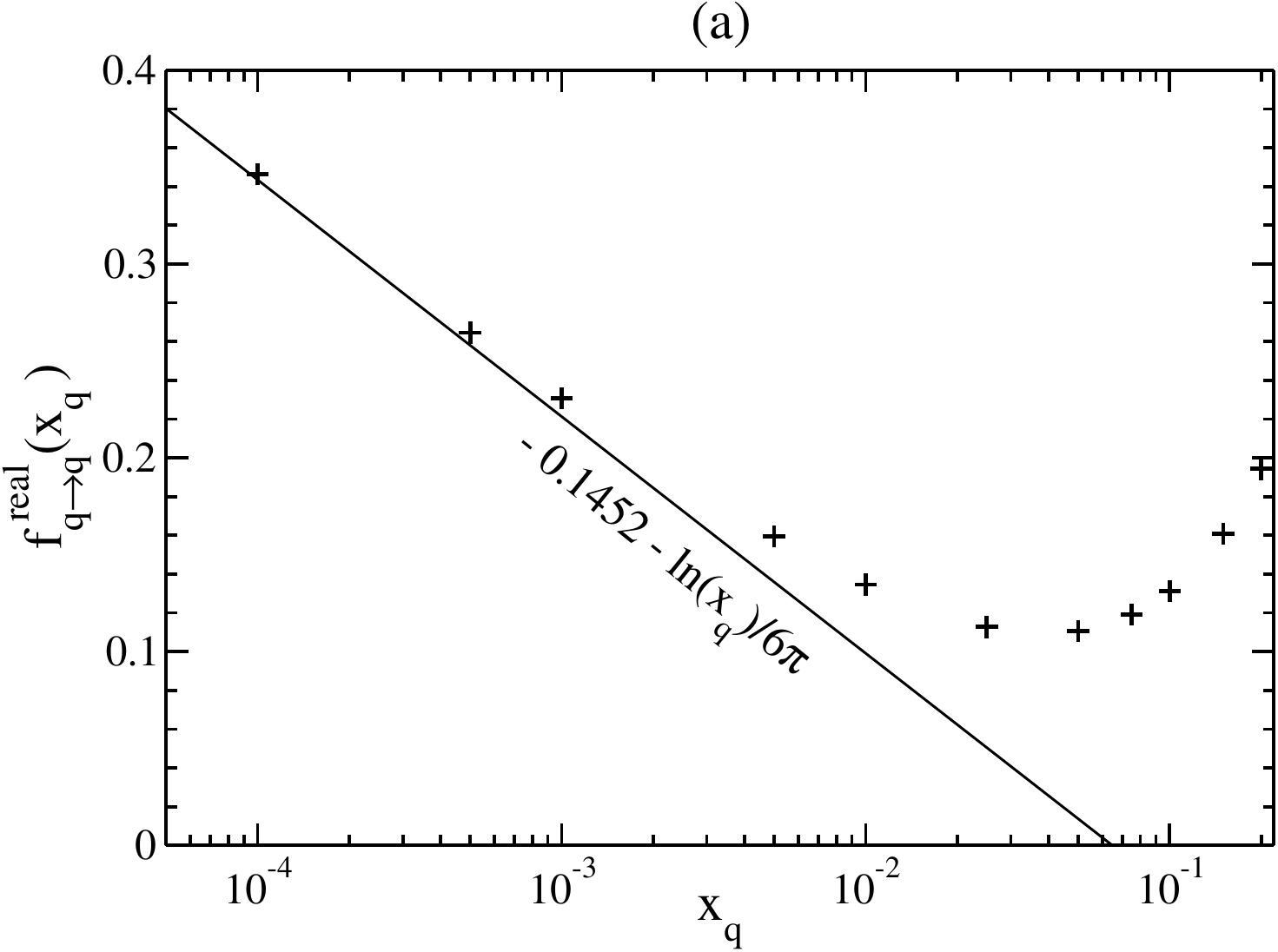}
  ~~
  \includegraphics[scale=0.3]{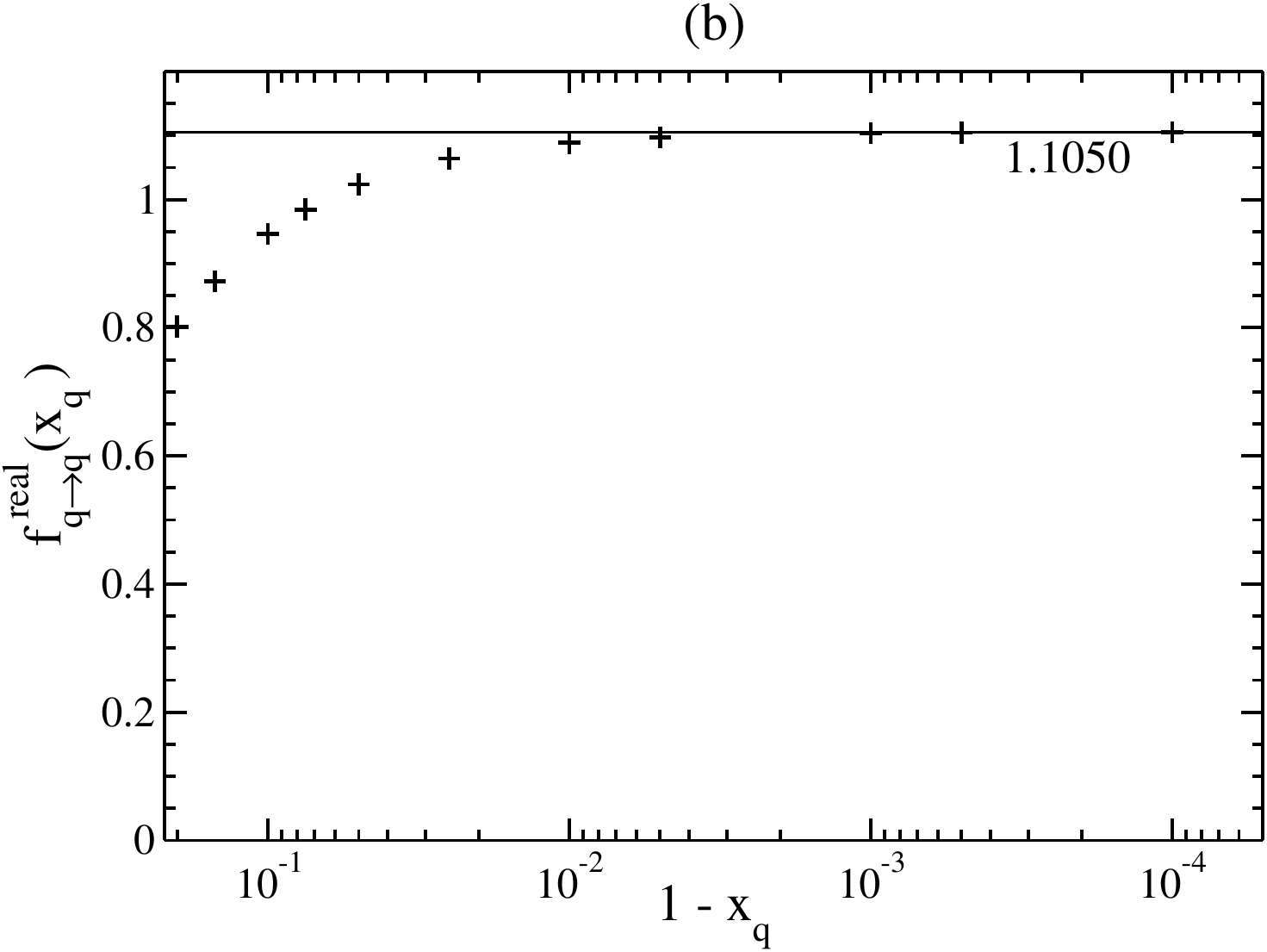}
  \caption{
     \label{fig:REALlog}
     Log-linear plots of numerical results for $f_{q\to q}^{\rm real}(\xq)$
     vs.\ (a) $\xq$ and (b) $1{-}\xq$.
     The numerical data points ($+$)
     are taken from table \ref{tab:dGnet}.
     The points are compared to (a) lines whose slopes on these plots
     correspond to the leading-log analytic results
     $-\frac{1}{6\pi} \ln\xq$ and
     (b) a constant (signifying no logarithmic behavior).
     Note that the horizontal axis in both plots is oriented so that
     $\xq\to 0$ is toward the left and $\xq\to 1$ is toward the right. 
  }
\end {center}
\end {figure}

In the limit $\xq \to 1$, (\ref{eq:tformsReal}) gives
\begin {equation}
   t_\form^{g\to\Q\Qbar} \sim t_\form^{q \to qg}
\label {eq:hierarchy1}
\end {equation}
for democratic pair production (neither $\yfrakQ$ nor $1{-}\yfrakQ$ small).
This is in stark contrast to the QED case where
$t_\form^{\gamma\to\E\Ebar} \ll t_\form^{e \to e\gamma}$ in the same limit.
It was this hierarchy that was responsible for
the $\ln(1{-}\xe)$ behavior of
$f_{e\to e}^\real(\xe)$.  Without a hierarchy, we do not expect such
logarithms.
(Formation time hierarchies generated when $\yfrakQ \to 0$ or
$\yfrakQ \to 1$ do not matter because $\yfrakQ$ is integrated
in the calculation of $f_{q\to q}^\real(\xq)$, just as in the calculation
of the $\ln(1{-}\xe)$ behavior in appendix B.1.1 of
ref.\ \cite{qedNfenergy}.)
We verify the absence of logarithmic behavior numerically in the
log-linear plot of fig.\ \ref{fig:REALlog}b, which compares our
numerical results for $f^\real_{q\to q}(\xq)$ to a constant as $x{\to}1$.
The constant used in the plot
is the value of our earlier fit (\ref{eq:fqqReal}) evaluated at
$x{=}1$.

\subsubsection{\boldmath$\xq\to 0$}

In the limit $\xq \to 0$ with $\yfrakQ$ held fixed, (\ref{eq:tformsReal})
gives
\begin {equation}
   t_\form^{g\to\Q\Qbar} \gg t_\form^{q \to qg} ,
\label {eq:hierarchy0}
\end {equation}
which is the same hierarchy that was found for QED.
Following the exact same analysis as in appendix B.1.2 of
ref. \cite{qedNfenergy} leads to exactly the same $\ln\xq$ term
for QCD, shown in our eq.\ (\ref{eq:fqqReal}) here.
We check the logarithmic behavior by verifying that the slope
of numerical results in the log-linear plot of fig.\ \ref{fig:REALlog}a
approaches the coefficient predicted for the logarithm in the
limit $\xq\to 0$.
Specifically, the line drawn for comparison shows the limiting
$\xq{\to}0$ behavior of (\ref{eq:fqqReal}),
\begin {equation}
   f_{q\to q}^\real(x) = -\tfrac{1}{6\pi} \ln x + \mbox{constant}
   \qquad \mbox{for $x \to 0$}.
\end {equation}


\subsubsection{\boldmath$\xQ\to 0$}

The limit $\xQ\to 0$ is dominated by $x_g \to 0$ with
$\xQ \sim \xQbar$ as in
(\ref{eq:softlim}).  We then get no hierarchy of scales, as in
(\ref{eq:hierarchy1}), and so expect no $\ln\xQ$ term
in $f_\uqQ(\xQ)$.  This is different than the behavior for QED.
The absence of a logarithm is verified numerically in
fig.\ \ref{fig:fotherLog}a.
Similarly, we expect no $\ln\xQbar$ term in $f_\uqQbar(\xQbar)$.


\subsubsection{\boldmath$\xQ\to 1$}

\begin {figure}[t]
\begin {center}
  \includegraphics[scale=0.3]{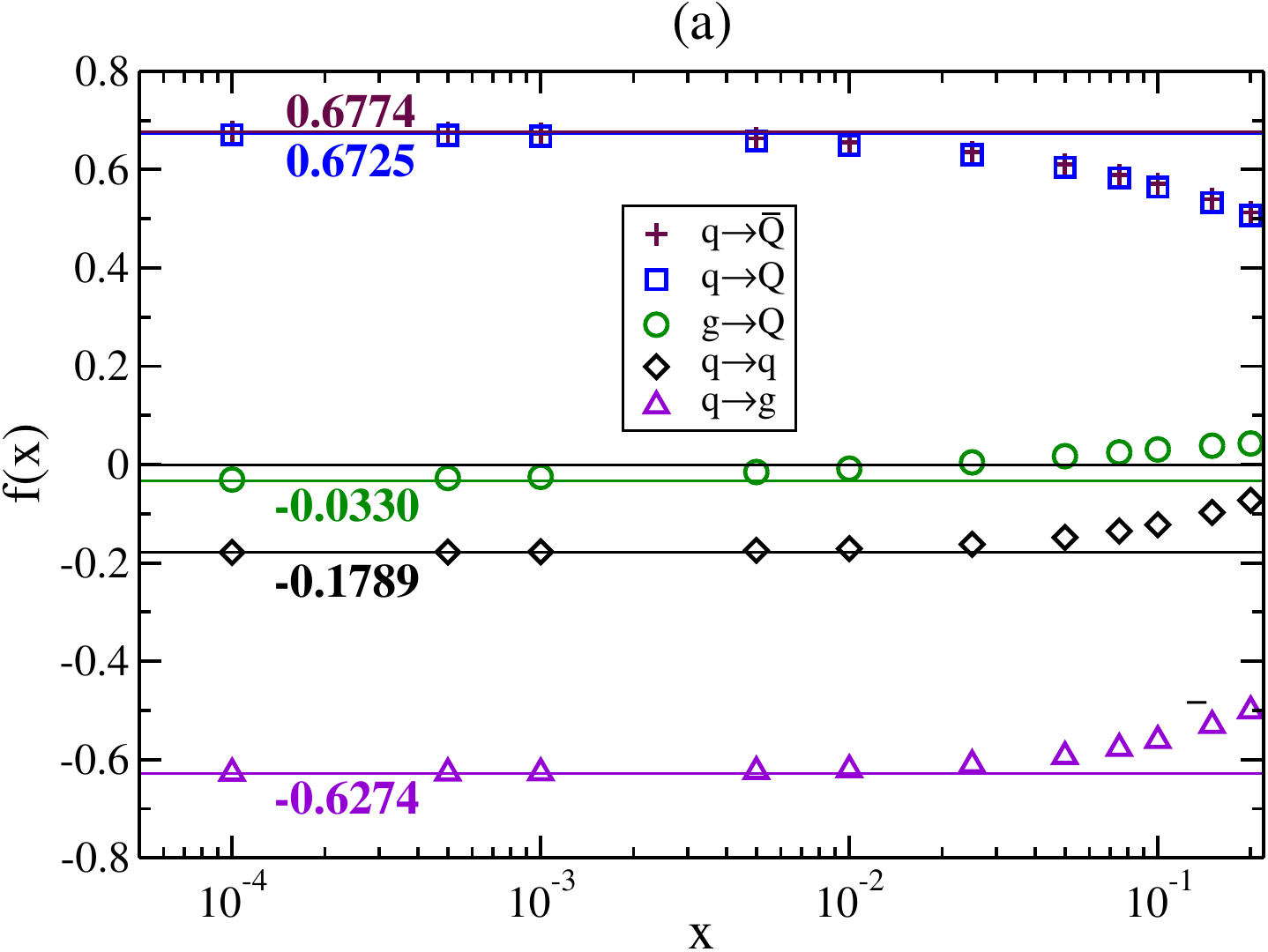}
  ~~
  \includegraphics[scale=0.3]{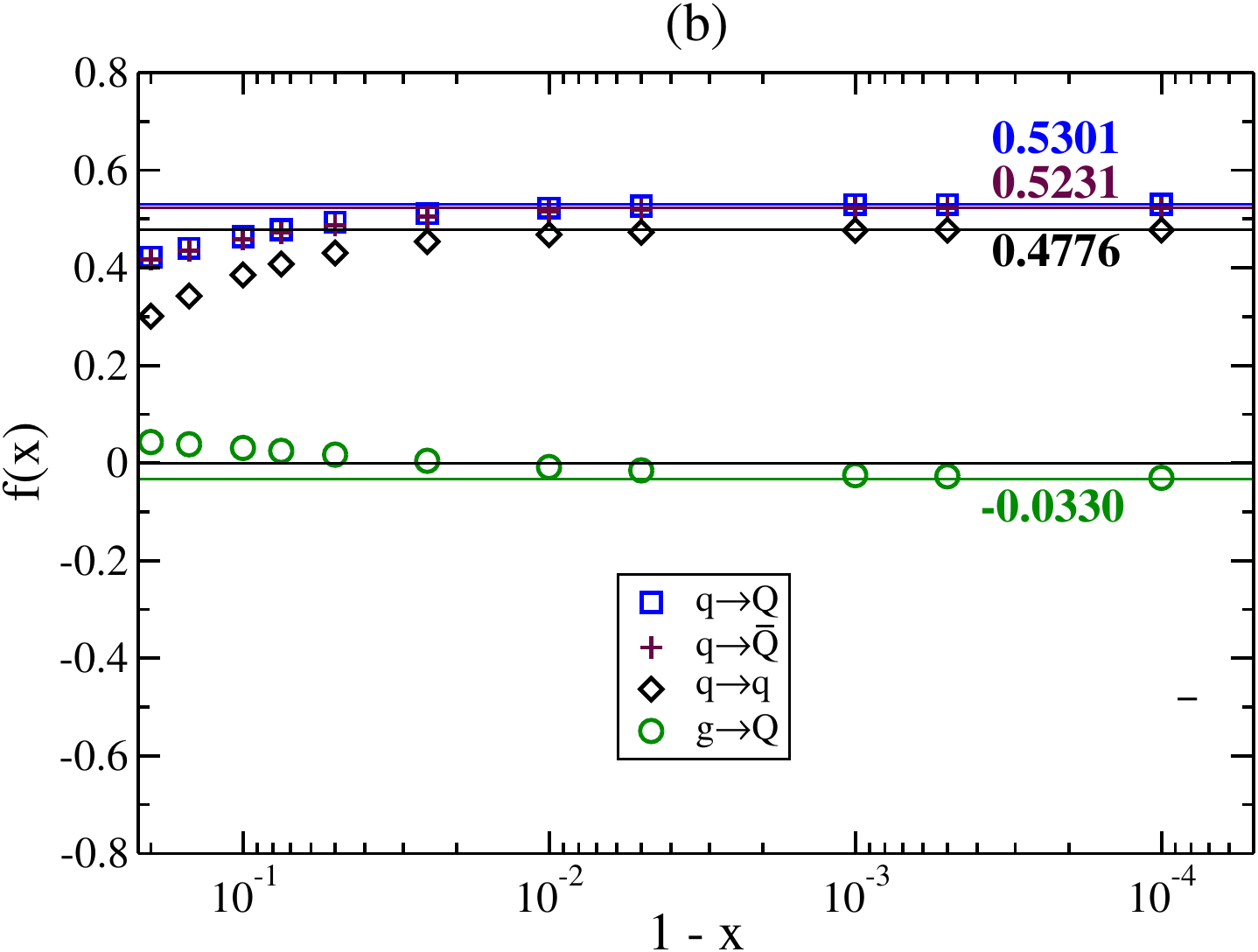}
  \caption{
     \label{fig:fotherLog}
     Like fig.\ \ref{fig:REALlog} but here for a variety of $f_\uij(x)$
     that show no log behavior for (a) $x\to0$ or (b) $x\to 1$.
     Note that $g{\to}\Q$ data (which corresponds to NLO $g\to\Q\Qbar$)
     are mirror reflections of each other in the two plots because
     of charge conjugation symmetry.
  }
\end {center}
\end {figure}

This limit did not generate any $\ln(1{-}\xE)$ for
$f_{\underline{e\to\E}}(\xE)$ in QED.
The discussion of that case in ref.\ \cite{qedNfenergy}
is unaltered for QCD, which we've checked numerically in
fig.\ \ref{fig:fotherLog}b.

\subsection{Virtual diagrams}

Like ref.\ \cite{qedNfenergy}, we do not have a method for deducing
\textit{ab initio} the logarithmic behavior of $f_\uij$'s that contain
virtual diagrams, and we rely on numerics to identify which limits
lack any logarithmic terms $\ln x$ or $\ln(1{-}x)$.
These are shown for various cases of $f_\uqq$, $f_\uqg$ and $f_\ugQ$
in fig.\ \ref{fig:fotherLog}.

The one limit of an $f_\uij$ that is not shown in the figure is
the $x{\to}1$ limit of $f_\uqg(x)$.  But
this limit is already determined by using
the relations (\ref{eq:fdecomposeAll}) to rewrite the
$x{\to}1$ limit of $f_\uqg$ in terms
of the $x{\to}0$ limit of $f_{q\to q}^\virt = f_\uqq - f_{q\to q}^\real$.
Our previous results for the limits of $f_\uqq$ and $f_{q\to q}^\real$
then give
\begin {equation}
   f_\uqg(x) \simeq \tfrac{1}{6\pi} \ln(1{-}x) + {\rm constant}
   \qquad \mbox{for $x \to 1$},
\end {equation}
as in (\ref{eq:fqg}).


\section{Qualitative picture for IR sensitivity of
         quark number deposition}
\label{app:phi}

In this appendix, we give a very qualitative argument about why
the fermion-number deposition distribution $\phi_q(z)$ is IR sensitive
in large-$\Nf$ QCD.


\subsection{The picture}

\begin {figure}[t]
\begin {center}
  \includegraphics[scale=0.8]{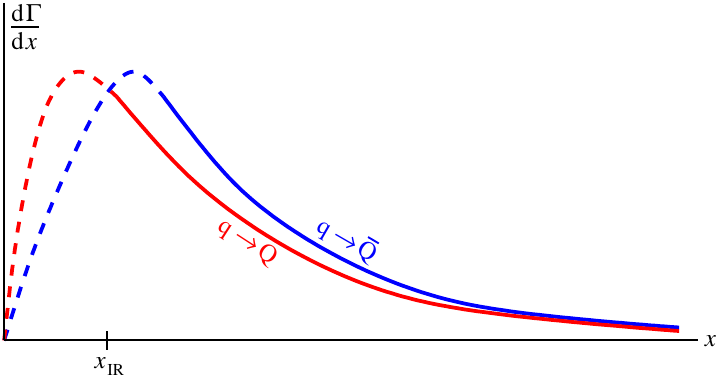}
  \caption{
     \label{fig:qQQbarRates}
     A schematic picture qualitatively comparing the behavior of
     $[d\Gamma/dx]^\net_\uqQ$ and $[d\Gamma/dx]^\net_\uqQbar$ in
     the small-$x$ limit.  The tails fall like $x^{-3/2}$.
     Details of this figure should not be taken seriously except
     for the takeaway that $\Qbar$'s are statistically produced
     with a little more energy than $\Q$'s in
     overlapping $q{\to}qg{\to}q\Q\Qbar$.
  }
\end {center}
\end {figure}

The staring point is the fact, apparent in table \ref{tab:dGnet},
that
\begin {equation}
  \left[ \frac{d\Gamma}{dx} \right]^\net_\uqQ
  \not=
  \left[ \frac{d\Gamma}{dx} \right]^\net_\uqQbar
\end {equation}
even in the small $x$ limit.
As discussed in section \ref{sec:IRunsafe}, these rates arise solely
from overlapping $q{\to}qg{\to}q\Q\Qbar$, and both rates scale like
$x^{-3/2}$ for small $x$ in the $\qhat$ approximation.
A qualitative picture of the situation for
$x \ll 1$ is shown by the solid curves in fig.\ \ref{fig:qQQbarRates}.
The blue solid curve for $q{\to}\Qbar$ is drawn slightly above the red
solid curve for $q{\to}\Q$ because table \ref{tab:dGnet} shows that
$q{\to}\Qbar$ is the slightly larger rate at small $x$.

The $x^{-3/2}$ small-$x$ behavior of the rates
no longer applies when the $\qhat$ approximation breaks down
for $x\lesssim\xIR$ given by (\ref{eq:xIR}).
The total integrated rate
\begin {equation}
  \Gamma_\uqQ \equiv \int_0^1 dx \> 
    \left[ \frac{d\Gamma}{dx} \right]^\net_\uqQ ,
\end {equation}
corresponding to the area beneath the red curve,
would diverge in the $\qhat$ approximation because of the
$x^{-3/2}$ behavior, but will be finite once
IR physics is properly accounted for, as \textit{schematically} depicted by
the dashed red line in the figure.
(We have made no attempt to think through the details of the shape
of the $x\ll\xIR$ behavior.)
The total integrated rate $\Gamma_\uqQ$ for making a $\Q$
in overlapping $q{\to}qg{\to}q\Q\Qbar$ must
be the same as that for making a $\Qbar$
(since one can't be made without the other), and so the areas under the
red and blue curves in fig.\ \ref{fig:qQQbarRates} must be equal.
This means that the red curve lies consistently to the left of the blue
curve (as in the figure), and so, on average, the $\Q$ produced in overlapping
$q{\to}qg{\to}q\Q\Qbar$ will carry a bit less energy that then
$\Qbar$ that is produced with it.

\begin {figure}[t]
\begin {center}
  \includegraphics[scale=1.0]{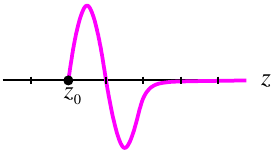}
  \caption{
     \label{fig:charge0}
     A schematic picture of the deposition of fermion number from
     a soft $\Q\Qbar$ created at $z{=}z_0$ by overlapping
     $q{\to}qg{\to}q\Q\Qbar$.  Details of this picture should not
     be taken seriously except for the fact that, in this particular
     situation,
     $\Qbar$ deposition occurs
     (statistically) a little further away from the pair creation
     than $\Q$ deposition does.
  }
\end {center}
\end {figure}

Having slightly more energy, the $\Qbar$ will travel slightly further
through the medium than the $\Q$ before depositing its fermion number.
A $\Q\Qbar$ pair has no net fermion number and so, on net, will deposit
none.  But the fact that (statistically) the $\Qbar$ goes a little
further means that, as a function of time,
the $\Q\Qbar$ pair will (statistically) first
deposit positive fermion number, with the canceling deposition of negative
fermion number by $\Qbar$ slightly later.
Fig.\ \ref{fig:charge0} shows
a schematic picture of the fermion number deposited by a soft $\Q\Qbar$
pair that was produced at some position $z{=}z_0$
by overlapping $q{\to}qg{\to}q\Q\Qbar$.
The most common $\Q\Qbar$ energies will be the
softest ones ($x\sim\xIR$).  Quarks with energies of order $\xIR E$
(where $E$ is the energy of the original parent $q$) will stop in
a distance of order
\begin {equation}
  \Delta z \sim \xIR^{1/2} \ell_0 ,
\end {equation}
where
$\ell_0$ is the length scale (\ref{eq:ell0}) of a shower initiated by
a particle of energy $E$.  Crudely, one may think of the tick marks
in fig.\ \ref{fig:charge0} as counting off intervals with sizes
of order $\Delta z$.

\begin {figure}[t]
\begin {center}
  \includegraphics[scale=0.7]{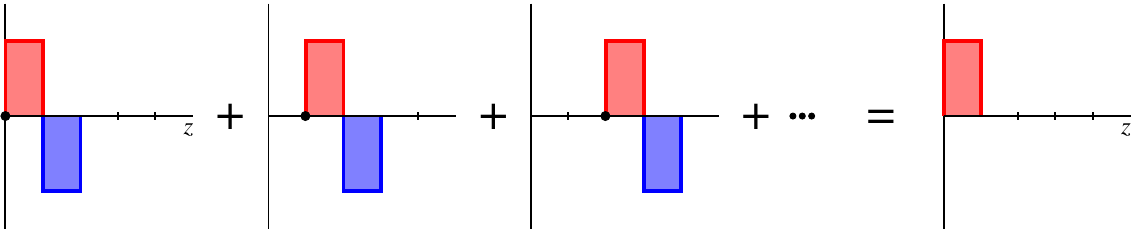}
%
  \caption{
     \label{fig:charge}
     The left-hand side of the above equation schematically represents
     the possibilities that a single, overlapping $q{\to}qg{\to}q\Q\Qbar$
     (represented by the black dot)
     might occur at different early times $z_0\ge 0$, and the right-hand side
     represents the resulting average fermion-number deposition $\phi_q(z)$
     at early times ($0 \le z \ll \ell_0$).
  }
\end {center}
\end {figure}

To provide a simple visualization for the following argument, let's
simplify and abstract fig.\ \ref{fig:charge0} (at various positions
$z_0$) by the graphs
on the left-hand side of fig.\ \ref{fig:charge}.  In this figure,
$z{=}0$ represents
the starting position of the original high-energy $q$ of energy $E$.
Since overlapping $q{\to}qg{\to}q\Q\Qbar$ is a NLO effect, and since
we formally expand to first order in $\Nf\alphas$ in our treatment of
shower development, there will only be one overlapping splitting in the
entire development of the shower, and that one NLO splitting
can happen at any random time.  We will focus on the case where
it happens early in the shower, at some $z \ll \ell_0$.
The first three graphs in fig.\ \ref{fig:charge} crudely represent
the cases where it happens at $z\simeq 0$, $z\simeq\Delta z$, and
$z\simeq2\Delta z$.  These have equal probability, and so we
should add them up (imagining that probabilities are already included
in the normalization of the vertical axis) to get the final,
statistically averaged, fermion-number deposition distribution for
$z \ll \ell_0$.  As can be seen visually in our cartoon
(fig.\ \ref{fig:charge}), there
is a cancellation, in all intervals beyond the first, 
between the chance that (i) a $\Q$ emitted
previously deposits its fermion number there and (ii) a $\Qbar$
emitted even earlier deposits its fermion number there.
The only place where this cancellation cannot happen (for $z \ll \ell_0$) is the
first interval of size $\Delta z$.  So, the average fermion-number distribution
deposited by the $\Q\Qbar$ pairs is peaked at very early positions
$z \sim \Delta z \ll \ell_0$.

Of course, the times when the overlapping $q{\to}qg{\to}q\Q\Qbar$ might
have happened are continuous, not discrete, and a curve like
fig.\ \ref{fig:charge0} is also not the same thing as the square pulse
curves shown in fig.\ \ref{fig:charge}.  But one will get the same
qualitative result by averaging any curve like fig.\ \ref{fig:charge0}
continuously over the location, from $z{=}0$ onward, of its starting point
$z_0$.

Parametrically, how large is the short-time fermion-number deposition
represented by the right-hand side of fig.\ \ref{fig:charge}?
The rate for soft overlapping $q{\to}qg{\to}q\Q\Qbar$
with $x\sim \xIR$ is
\begin {subequations}
\label {eq:IRestimates}
\begin {equation}
  \operatorname{rate}(\xIR)
  \sim \xIR \times \left[\frac{d\Gamma}{dx}(\xIR)\right]^\NLO_\uqQ
  \sim \frac{\Nf\alphas}{\xIR^{1/2}\ell_0} \,.
\end {equation}
The chance that it happens in the first interval $\Delta z$ is
\begin {equation}
  \operatorname{Prob}(\xIR)
  \sim \operatorname{rate}(\xIR) \times \Delta z
  \sim \Nf\alphas .
\end {equation}
\end {subequations}
When it does happen, the fermion number deposited in the first interval
is $O(1)$.  So, on average, the far right-hand side of fig.\ \ref{fig:charge}
represents very early deposition of fermion number
of order $\Nf\alphas \ll 1$.

\begin {figure}[t]
\begin {center}
  \includegraphics[scale=0.7]{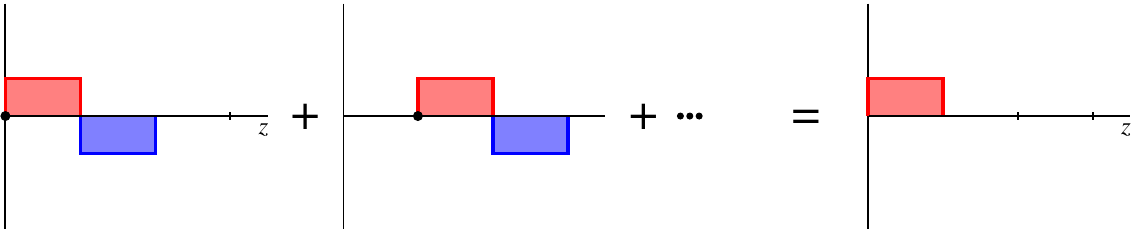}
  \caption{
     \label{fig:charge2}
     Like fig.\ \ref{fig:charge} but representing the smaller probabilities
     of overlapping $q{\to}qg{\to}q\Q\Qbar$ when the $\Q\Qbar$ pair
     has enough energy that the $\Q$'s and $\Qbar$'s take twice as
     long to stop.
  }
\end {center}
\end {figure}

This is certainly an IR-sensitive feature of the deposition distribution
$\phi_q(z)$ at NLO.  However, the story so far does not explain the
logarithmic enhancement to NLO corrections that we found in
section \ref{sec:IRlog}.  That's because the discussion so far has
focused only on the most common case of overlapping $q{\to}qg{\to}q\Q\Qbar$,
which is the softest possible pair production ($x\sim\xIR$).
But now imagine rarer $q{\to}qg{\to}q\Q\Qbar$
with $x\sim x_1$ where $\xIR \ll x_1 \ll 1$.
Though still soft compared to the original $q$, the $\Q$ and $\Qbar$
will now have more energy than before and so will go further than
before.  The analog of fig.\ \ref{fig:charge} will now look
like fig.\ \ref{fig:charge2} (where we've made the distances only
twice as long so that the figure still fits easily on the page).
All of the above estimates (\ref{eq:IRestimates}) still
go through, but with the scale $\xIR$ replaced by the scale $x_1$.
So $x \sim x_1$ contributes an additional $\Nf\alphas$ to early deposition
of fermion number; it's just more spread out than the $x\sim\xIR$
contribution.  Roughly speaking, there should be a contribution of
$\Nf\alphas$ for every distinct choice of scale $x \sim x_1$
with $\xIR \ll x_1 \ll 1$, which means a logarithmic enhancement
$\approx \ln(1/\xIR)$.

Allowing for contributions from all $x$ scales, the early deposition
of fermion number from soft, overlapping $q{\to}qg{\to}q\Q\Qbar$ will
be
\begin {equation}
   \phi_q(z) \sim \frac{\Nf\alphas}{z}
   \quad \mbox{for $\Delta z \ll z \ll \ell_0$} .
\label {eq:phiq}
\end {equation}

The integral of $\phi_q(z)$ over early times $z \ll \ell_0$ is
parametrically $\Nf\alphas \ln(1/\xIR)$, which is the early-time
fermion number deposition.
By conservation of fermion number,
$O\bigl( \Nf\alphas\log)$ fermion number deposited early in the shower
$(z \ll \ell_0)$ means that fermion number
$1 - O\bigl( \Nf\alphas\log)$ must be deposited at $z \sim \ell_0$.
One crude way to picture this is to imagine that we suddenly turned
off the production of $\Q\Qbar$ pairs at some large time in
figs.\ \ref{fig:charge} and \ref{fig:charge2}.  Then there would be
a deposition (on average) of negative fermion number by $\Qbar$ at the very end,
in addition to the deposition $1$ by the original quark $q$ that
initiated the shower.

Like the main text, we will not attempt to sort out the case where
$\Nf\alphas \ln(1/\xIR) \gg 1$.


\subsection{Relation to moments \boldmath$\langle z^n \rangle$}

In section \ref{sec:IRunsafe},
the infrared logarithm arose when calculating
the recursion relation for moments $\langle z^n \rangle$.
We now discuss how that is related to the form (\ref{eq:phiq}) of
the early time deposition.

But first, we examine more
thoroughly how the infrared logarithm affected
the recursion relation for $\langle z^n \rangle$.
The logarithm arose in the calculation of ${\cal M}_n$ for the specific
case $n{=}1$ of (\ref{eq:Mn}).  The recursion relation (\ref{eq:znphi})
gives
\begin {equation}
  \langle z^n \rangle_{\phi} =
  \frac{ n! }{ {\cal M}_n {\cal M}_{n-1} \cdots {\cal M}_1 } \,,
\label {eq:znphi2}
\end {equation}
where all factors are infrared-safe except for ${\cal M}_1$.
So, the IR sensitivity of all of the
$\langle z^n \rangle$ arise from
the same common overall factor
\begin {equation}
  \frac{1}{ {\cal M}_1 }
  \propto
  \frac{1}{ 1 + O\bigl(\Nf\alphas\ln(1/\xIR)\bigr) }
  =
  1 - O\bigl(\Nf\alphas\ln(1/\xIR)\bigr) .
\label {eq:M1}
\end {equation}

To help understand this result, we now qualitatively argue for a common
factor like (\ref{eq:M1}) starting instead from the early-time behavior
(\ref{eq:phiq}) of $\phi_q(z)$.
Imagine calculating the moments
\begin {equation}
  \langle z^n \rangle = \int_0^\infty dz \> z^n \phi_q(z)
\label {eq:znjunk}
\end {equation}
directly from $\phi_q(z)$ for all $n{\ge}1$.
The early-time behavior (\ref{eq:phiq})
generates no infrared
logarithm from the early-time ($z {\ll}\ell_0$)
portion of the integrals (\ref{eq:znjunk}).
The IR-sensitive corrections to moments instead arises
indirectly from
the fact that early-time deposition
(\ref{eq:phiq}) leaves only
total fermion number $1 - O\bigl(\Nf\alphas\log\bigr)$ remaining
to be deposited at late times $z \sim \ell_0$.
That means that $\phi_q(z)$ will be proportional to
$1 - O\bigl(\Nf\alphas\log\bigr)$ at late times,
which introduces a common IR-sensitive correction to
the $z{\sim}\ell_0$ part of the integrals (\ref{eq:znjunk})
for the moments $\langle z^n \rangle$.


\section{\boldmath$\xQ{\leftrightarrow}\xQbar$ behavior of overlapping
  $q{\to}qg{\to}q\Q\Qbar$ in soft gluon limit}
\label{app:QQbar}

In section \ref{sec:WhatQQbar}, we reported that one can check numerically
that the diagrams contributing to the rate
$[\Delta\,d\Gamma/d\xq\,\xQ]_{q\to q\Q\Qbar}$
are all individually symmetric or anti-symmetric under
\begin {equation}
   \xQ \leftrightarrow \xQbar = 1{-}\xq{-}\xQ
\label {eq:swapQQbar2}
\end {equation}
in the limit
\begin {equation}
  \xQ \sim \xQbar \ll 1 ,
\label {eq:softlim2}
\end {equation}
and that the diagrams which are anti-symmetric are the
rate diagrams of fig.\ \ref{fig:phidiags}, which are
characterized by having a single longitudinal gluon exchange.
[In the full set of diagrams shown in fig.\ \ref{fig:qeddiags},
the ones that contribute to real double splitting
$q{\to}qg{\to}q\Q\Qbar$ are diagrams (a--g), those that have a single
longitudinal exchange are (e--g), and diagram (g) evaluates to zero.]
The purpose of this appendix is to briefly sketch an
explanation of these symmetry properties.
We will compare and contrast this situation
to the discussion of large-$\Nf$ QED.


\subsection{4-body Hamiltonian}

\subsubsection{QED}

Let's start with QED.  The critical starting point for the symmetry
argument in appendix A of ref.\ \cite{qedNfenergy} was the effective 2-particle
Hamiltonian for 4-body $(\bar e,\E,\Ebar,e)$ evolution, which was
\begin {equation}
   \frac{P_{41}^2}{2 x_1 x_4(x_1{+}x_4) E}
   + \frac{P_{23}^2}{2 x_2 x_3(x_2{+}x_3) E}
   - \frac{i\qhat}{4} (x_1{+}x_4)^2 \left( \C_{41} - \C_{23} \right)^2
  \qquad \mbox{(QED)}
\end {equation}
in the basis $(\C_{41},\C_{23})$ most convenient for the evaluation of
large-$\Nf$ diagrams.
Above, $(x_1,x_2,x_3,x_4) = (-1,\xE,\xEbar,\xe)$.
This Hamiltonian is exactly symmetric under the exchange of the
values of $x_2$ and $x_3$, which corresponds to (\ref{eq:swapQQbar2}).
From there, ref.\ \cite{qedNfenergy} argued that all
large-$\Nf$ QED diagrams were symmetric under
exchanging $\xE \leftrightarrow \xEbar$.


\subsubsection{\boldmath$\Nc{\gg}1$ QCD}

For $\Nc{\gg}1$ QCD, we earlier presented the effective 2-particle
Hamiltonian for 4-body $(\bar q,\Q,\Qbar,q)$ evolution in
eqs.\ (\ref{eq:V4C12C34}) and (\ref{eq:H4two})
but so far only in the basis $(\C_{34},\C_{12})$.
We then found the normal modes and only then switched to the
$(\C_{41},\C_{23})$ basis convenient to the evaluation of the
large-$\Nf$ diagrams.  Here, let us instead make this change
of basis%
\footnote{
  By cyclic permutation of the particle numbers (1,2,3,4) to (4,1,2,3),
  eqs.\ (\ref{eq:changebasis}) and (\ref{eq:S}) give
  $\C_{34} = (-x_2\C_{23} - x_1\C_{41})/(x_3+x_4)$ and
  $\C_{12} = (x_3\C_{23} + x_4\C_{41})/(x_3+x_4)$.
  (This is equivalent to, but much quicker than,
  explicitly computing ${\cal S}^{-1}$ and then
  tediously simplifying using $x_1{+}x_2{+}x_3{+}x_4 = 0$.)
}
directly to the Hamiltonian (\ref{eq:H4two}), which
then becomes
\begin {equation}
  {\cal H}_4
    = \frac{P_{41}^2}{2x_1 x_4(x_1{+}x_4)E} + \frac{P_{23}^2}{2x_2 x_3(x_2{+}x_3)E}
    - V_4(\C_{41},\C_{23})
\label {eq:H4two2}
\end {equation}
with
\begin {equation}
  V_4(\C_{41},\C_{23}) =
    -\frac{i\qhatF}{4}
    \bigl[
      (x_2^2{+}x_3^2) C_{23}^2
        + 2(x_1 x_2{+}x_3 x_4) \C_{23}\cdot\C_{41} + (x_1^2{+}x_4^2) C_{41}^2
    \bigr]
 .
\label {eq:V4two0}
\end {equation}
The kinetic terms (which are the same as in the QED case)
are symmetric under $x_2 \leftrightarrow x_3$, but the potential is not.
But now remember that
\begin {equation}
  (x_1,x_2,x_3,x_4) = (-1,\xQ,\xQbar,\xq) = (-1,\xQ,\xQbar,1{-}\xQ{-}\xQbar) .
\end {equation}
Then, taking the soft limit (\ref{eq:softlim2}) while avoiding any
particular assumption about the relative sizes of $C_{23}$ and $C_{41}$,
\begin {equation}
  V_4(\C_{41},\C_{23}) \simeq
    -\frac{i\qhatF}{4}
    \bigl[
      (x_2^2{+}x_3^2) C_{23}^2 + 2(x_3 {-} x_2) \C_{23}\cdot\C_{41} + 2 C_{41}^2
    \bigr]
 .
\label {eq:V4two}
\end {equation}
This soft limit is still not symmetric under $x_2 \leftrightarrow x_3$
because the $(x_3 {-} x_2) \C_{23}\cdot\C_{41}$ term is anti-symmetric.
But we can make the following useful observation regarding the
4-body potential:
\begin {quote}
  In the soft limit (\ref{eq:softlim2}), exchanging $x_2 \leftrightarrow x_3$
  is equivalent to leaving $x_2$ and $x_3$ unchanged but
  replacing $\C_{23}$ by $\C_{32} = -\C_{23}$.
\end {quote}
For consistency, the corresponding conjugate momentum
$\P_{23}$ should then also be replaced
by $\P_{32} = -\P_{23}$, but this has no effect on the Hamiltonian
(\ref{eq:H4two2}).


\subsubsection{\boldmath$\Nc{\gg}1$ QCD with less algebra}

There's an alternate way to arrive at the last conclusion with less algebra
but a little more qualitative argument about scales in the soft limit.
Start from the original form (\ref{eq:Vfour0}) of the 4-body potential
in terms of the four transverse positions $\b_i$:
\begin {equation}
   V_4(\b_1,\b_2,\b_3,\b_4)
   = -\frac{i\qhatF}{4} \left[
        (\b_2{-}\b_1)^2 + (\b_4{-}\b_3)^2
     \right] .
\label {eq:Vfour0b}
\end {equation}
In the soft limit $x_2 \sim x_3 \ll 1$ of (\ref{eq:softlim2}),
particles 1 and 4 will have
relatively high energy and so collisions with the medium will not
deflect them significantly compared to particles 2 and 3.
So the sizes of $\b_1$ and $\b_4$ will be negligible compared to the
sizes of $\b_2$ and $\b_3$.  So we may approximate (\ref{eq:Vfour0b})
as
\begin {equation}
   V_4(\b_1,\b_2,\b_3,\b_4)
   \simeq -\frac{i\qhatF}{4} ( b_2^2 + b_3^2 ) .
\end {equation}
The corresponding kinetic terms of the quantum mechanics
problem, coming from the $p_\perp^2$ terms in the high-energy
expansion of $|\p_i| \simeq p_{z,i} + p_{\perp,i}^2/p_{z,i}$,
then give a Hamiltonian
\begin {equation}
   H_4
   \simeq \frac{p_{\perp,2}^2}{2x_2E} + \frac{p_{\perp,3}^2}{2x_3 E}
     -\frac{i\qhatF}{4} ( b_2^2 + b_3^2 )
\end {equation}
in this language.  For that Hamiltonian, exchanging the values of
$x_2$ and $x_3$ is equivalent to exchanging $(\b_2,\p_{\perp,2})$
with $(\b_3,\p_{\perp,3})$.
Because $\C_{ij} \equiv (\b_i{-}\b_j)/(x_i{+}x_j)$
(with conjugate momentum $\P_{ij} = x_j\p_{\perp,i} - x_i\p_{\perp,j}$),
that's equivalent to negating $\C_{23}$ (and $\P_{23}$), as found previously.


\subsection{Diagrams}

Let's start with (the QCD version of) the diagram of fig.\ \ref{fig:qeddiags}a.
Following the related analysis of large-$\Nf$ QED in appendix A of
ref.\ \cite{qedNfenergy}, we note that the formula for this diagram was
originally derived from a formula of the form%
\footnote{
  Specifically, see eq.\ (A.3) of ref.\ \cite{qedNfenergy}.
  That's taken from eq.\ (E.1) of ref.\ \cite{seq}, with the QED modifications
  described in appendices E.1 and E.2 of ref.\ \cite{qedNf},
  followed here by the QCD modifications described in this paper.
}
\begin {align}
   \mbox{(splitting}&~\mbox{amplitude factors from the vertices)}
   \times
   \frac{\Nf\alpha^2}{(x_1{+}x_4)^2}
   \int_{\rm times} \int_{\B',\B''}
\nonumber\\ &\times
   \grad_{\B'''}
   \langle\B''',t'''|\B'',t''\rangle
   \Bigr|_{\B'''=0}
\nonumber\\ &\times
   \grad_{\C''_{41}}
   \grad_{\C'_{23}}
   \langle\C''_{41},\C''_{23},t''|\C'_{41},\C'_{23},t'\rangle
   \Bigr|_{\C_{41}''=0=\C_{23}'; ~ \C_{23}''=\B''; ~ \C_{41}'=\B'}
\nonumber\\ &\times
   \grad_{\B}
   \langle\B',t'|\B,t\rangle
   \Bigr|_{\B=0} .
\label {eq:seqform}
\end {align}
Above, the $\langle\C''_{41},\C''_{23},t''|\C'_{41},\C'_{23},t'\rangle$
is the propagator associated with the 4-body Hamiltonian ${\cal H}_4$.
The other two $\langle \cdots \rangle$'s are similar factors for
the initial and final 3-particle evolution, with time
running from right to left in the formula ($t < t' < t'' < t'''$).
The derivatives $\grad$ are position-space versions of transverse
momentum factors associated with the vertices. 
We will not need to discuss the somewhat complicated details of how
the splitting-amplitude vertex factors contract
those derivatives with each other.

By our previous discussion of ${\cal H}_4$, exchanging
$\xQ \leftrightarrow \xQbar$ in the soft limit (\ref{eq:softlim2})
will be equivalent to replacing
$\C_{23}$ by $-\C_{23}$ in the 4-body propagator:
\begin {equation}
   \langle\C''_{41},\C''_{23},t''|\C'_{41},\C'_{23},t'\rangle
   \to
   \langle\C''_{41},-\C''_{23},t''|\C'_{41},-\C'_{23},t'\rangle .
\label {eq:4change}
\end {equation}
The Hamiltonian for the initial 3-body evolution is
\begin {equation}
   {\cal H}_{3,i}
   = \frac{P_\perp^2}{2M_\ix} + \tfrac12 M_\ix \Omega_\ix^2 B^2
\end {equation}
with $M_\ix$ and $\Omega_\ix$ given by (\ref{eq:MLambdai0}),
which only depend on $\xQ$ and $\xQbar$ through the symmetric
combination $x_g = \xQ{+}\xQbar = 1{-}\xq$, and so
$\langle\B',t'|\B,t\rangle$ is unchanged.
Similarly, the Hamiltonian for the final 3-body evolution is
\begin {equation}
   {\cal H}_{3,f}
   = \frac{P_\perp^2}{2M_\fx^\seq} + \tfrac12 M_\fx^\seq (\Omega_\fx^\seq)^2 B^2
\end {equation}
with $M_\fx^\seq$ and $\Omega_\fx^\seq$ given by (\ref{eq:MLambdaf}),
which again are symmetric in $\xQ \leftrightarrow \xQbar$.
So $\langle\B''',t'''|\B'',t''\rangle$ is also unchanged.

Now focus on the first two factors of the integrand in
(\ref{eq:seqform}).  The only change is (\ref{eq:4change}), and so
those two factors becomes
\begin {align}
   \grad_{\B'''} &
   \langle\B''',t'''|\B'',t''\rangle
   \Bigr|_{\B'''=0}
\nonumber\\ &\times
   \grad_{\C''_{41}}
   \grad_{\C'_{23}}
   \langle\C''_{41},-\C''_{23},t''|\C'_{41},-\C'_{23},t'\rangle
   \Bigr|_{\C_{41}''=0=\C_{23}'; ~ \C_{23}''=\B''; ~ \C_{41}'=\B'}
\label {eq:seqform2}
\end {align}
and nothing else changes.
Mathematically, $\C'_{23}$ and $\C''_{23}$
are just dummy variable names in the above expression since at the end
they are set to zero and $\B''$ respectively, and so we could call them
anything.  So switch variables by defining new symbols $\C'_{23}$ and
$\C''_{23}$ that are the negative of what they were previously.
The factors (\ref{eq:seqform2}) are therefore equal to the expression
\begin {align}
   -
   \grad_{\B'''} &
   \langle\B''',t'''|\B'',t''\rangle
   \Bigr|_{\B'''=0}
\nonumber\\ &\times
   \grad_{\C''_{41}}
   \grad_{\C'_{23}}
   \langle\C''_{41},\C''_{23},t''|\C'_{41},\C'_{23},t'\rangle
   \Bigr|_{\C_{41}''=0=\C_{23}'; ~ \C_{23}''=-\B''; ~ \C_{41}'=\B'}
\label {eq:seqform3}
\end {align}
where the overall minus sign came from the change of variable
on $\grad_{\C'_{23}}$, and $\C_{23}''$ is now set to $-\B''$ instead
of to $\B''$.  Next, change integration variables $\B'' \to -\B''$
in (\ref{eq:seqform}).  By parity invariance in the transverse plane,
$\langle\B''',t'''|\B'',t''\rangle$ would be invariant under parity
but $\grad_{\B'''} \langle\B''',t'''|\B'',t''\rangle \Bigr|_{\B'''=0}$ will
negate.  The factors (\ref{eq:seqform3}) are then equal to
what they were in (\ref{eq:seqform}) before we exchanged $\xQ$ and $\xQbar$.
Overall, we picked up one minus sign from the $\grad_{\C'_{23}}$ and
a canceling minus sign from the $\grad_{\B'''}$ (because $\B'''$ was
connected to $\B''$ which was connected to $\C''_{23}$).
The takeaway is that (\ref{eq:seqform}) is symmetric under
$\xQ \leftrightarrow \xQbar$ in the soft limit.

Now consider instead diagram (e) of fig.\ \ref{fig:qeddiags}, which has
a single longitudinal gluon exchange.  The structure analogous to
(\ref{eq:seqform}) is%
\footnote{
  See eq.\ (E24) of ref.\ \cite{qedNf}.
}
\begin {align}
   \mbox{(splitting}&~\mbox{amplitude factors from the vertices)}
   \times
   \frac{\Nf\alpha^2}{(x_1{+}x_4)^2}
   \int_{\rm times} \int_{\B''}
\nonumber\\ &\times
   \grad_{\B'''}
   \langle\B''',t'''|\B'',t''\rangle
   \Bigr|_{\B'''=0}
\nonumber\\ &\times
   \grad_{\C''_{41}}
   \langle\C''_{41},\C''_{23},t''|\C'_{41},\C'_{23},t'\rangle
   \Bigr|_{\C_{41}''=0=\C_{23}'; ~ \C_{23}''=\B''; ~ \C_{41}'=0}
\end {align}
with $t' < t'' < t'''$.  If we follow the same steps as we did for
(\ref{eq:seqform2}), we will only get one minus sign from
$\xQ \leftrightarrow \xQbar$ (in the soft limit) because
there is no $\grad_{\C'_{23}}$ in this expression.
That's because longitudinal gluon exchange does not come with
a factor of transverse momentum.

Similarly, diagram (f) works out to be anti-symmetric.
Diagrams (b--c) do not involve any region of 4-body evolution, which is
what caused potential difficulties, and so they are symmetric.
Diagram (g) evaluates to zero and so is irrelevant.
Diagram (d), which has two longitudinal gluon exchanges, has the form%
\footnote{
   See eq.\ (3.11) of ref.\ \cite{4point}, adapted to longitudinal gluon
   exchange between fermion lines
   as discussed in appendix E.3.1 of ref.\ \cite{qedNf}.
}
\begin {align}
   \mbox{(splitting}&~\mbox{amplitude factors from the vertices)}
   \times
   \frac{\Nf\alpha^2}{(x_1{+}x_4)^2}
\nonumber\\ &\times
   \langle\C''_{41},\C''_{23},t''|\C'_{41},\C'_{23},t'\rangle
   \Bigr|_{\C_{41}''=0=\C_{23}'; ~ \C_{23}''=0; ~ \C_{41}'=0}
   .
\end {align}
There are no vertices that generate transverse momentum factors
and so no transverse position gradients.  So this diagram is
symmetric.


\subsection{More details regarding rate formulas}

For any reader wishing to see how the symmetry and anti-symmetry
manifest in the final rate formulas given in appendix
\ref{app:realsummary}, we offer an observation here about
the symmetry of some of the more complicated elements of
those formulas.
Refs.\ \cite{seq,qedNf} define a variety of symbols
by writing the harmonic oscillator propagator for the 4-body evolution
in the form%
\footnote{
  Our (\ref{eq:XYZexp}) is not shown explicitly in refs.\ \cite{seq,qedNf}
  but the explicit version may be found in eq.\ (3.25) of ref.\ \cite{1overN}.
  In the earlier paper \cite{seq} the argument, appearing in appendix E.2
  of that paper, proceeded by analogy with section 5.3 of ref.\ \cite{2brem}
  and skips the explicit formula.  The analogous formula is
  eq.\ (5.41) of ref.\ \cite{2brem}.
}
\begin {subequations}
\label{eq:4prop}
\begin {multline}
  \langle\C_{41}^\xbx,\C_{23}^\xbx,t_\xbx|\C_{41}^\yx,\C_{23}^\yx,t_\yx\rangle
  \propto
\\
  \exp\Biggl[
     - \frac12
     \begin{pmatrix} \C_{41}^\yx \\ \C_{23}^\yx \end{pmatrix}^{\!\!\top} \!
       \begin{pmatrix}
          \calX_\yx^\seq & Y_\yx^\seq \\
          Y_\yx^\seq & Z_\yx^\seq 
       \end{pmatrix}
       \begin{pmatrix} \C_{41}^\yx \\ \C_{23}^\yx \end{pmatrix}
     -
     \frac12
     \begin{pmatrix} \C_{23}^\xbx \\ \C_{41}^\xbx \end{pmatrix}^{\!\!\top} \!
       \begin{pmatrix}
          \calX_\xbx^\seq & Y_\xbx^\seq \\
          Y_\xbx^\seq & Z_\xbx^\seq
       \end{pmatrix}
       \begin{pmatrix} \C_{23}^\xbx \\ \C_{41}^\xbx \end{pmatrix}
\\
     +
     \begin{pmatrix} \C_{41}^\yx \\ \C_{23}^\yx \end{pmatrix}^{\!\!\top} \!
       \begin{pmatrix}
          X_{\yx\xbx}^\seq & Y_{\yx\xbx}^\seq \\
          \Ybar_{\yx\xbx}^\seq & Z_{\yx\xbx}^\seq
       \end{pmatrix}
       \begin{pmatrix} \C_{23}^\xbx \\ \C_{41}^\xbx \end{pmatrix}
   \Biggr] ,
\label {eq:XYZexp}
\end {multline}
where the $\calX$'s
are related to the $X$'s of refs.\ \cite{seq,qedNf} by
\begin {align}
   X_\yx^\seq &= |M_\ix|\Omega_\ix + \calX_\yx^\seq ,
\\
   X_\xbx^\seq &= |M_\fx^\seq| \Omega_\fx^\seq + \calX_\xbx^\seq .
\end {align}
\end {subequations}
Combined with (\ref{eq:4prop}),
our previous result that soft $\xQ \leftrightarrow \xQbar$ is equivalent
to $\C_{23} \to -\C_{23}$
means that
\begin {equation}
\begin {array}{cl}
   (X_\yx, Z_\yx; \, X_\xbx, Z_\xbx; \, Y_{\yx\xbx}, \Ybar_{\yx\xbx})^\seq &
   \mbox{are even under $\xQ\leftrightarrow\xQbar$} ,
\\
   (Y_\yx; \, Y_\xbx; \, X_{\yx\xbx}, Z_{\yx\xbx})^\seq &
   \mbox{are \kern2pt\underline{odd}\kern4.5pt
             under $\xQ\leftrightarrow\xQbar$}
\end {array}
\end {equation}
in the soft limit (\ref{eq:softlim2}).
These even/odd properties may be used to see the symmetry or
anti-symmetry of formulas like (\ref{eq:Dseq}) and (\ref{eq:DI})
without drilling down into the complicated detailed formulas for
the $(X,Y,Z)'s$.


\end {document}